\documentclass[10pt,twocolumn,letterpaper]{article}

\usepackage[arxiv,pagenumbers]{cvpr}      
\usepackage[table,dvipsnames]{xcolor}
\usepackage{bm}
\usepackage{amsthm}
\usepackage{multirow}
\usepackage{caption}
\usepackage{subcaption}

\usepackage{pifont}
\newcommand{\xmark}{\ding{55}}

\usepackage[dvipsnames]{xcolor}

\definecolor{cvprblue}{rgb}{0.21,0.49,0.74}
\usepackage[pagebackref,breaklinks,colorlinks,citecolor=cvprblue]{hyperref}
\usepackage{longtable}

\newcommand{\subpara}[1]{\vspace{0.4em} \noindent \textbf{#1.}}
\newcommand{\tychetrain}{Tyche-TS}
\newcommand{\tychetest}{Tyche-IS}

\def\paperID{1079} 
\def\confName{CVPR}
\def\confYear{2024}

\title{Tyche: Stochastic In-Context Learning for\\Medical Image Segmentation}

\author{
Marianne Rakic\\
CSAIL MIT\\
Broad Institute\\
{\tt\small mrakic@mit.edu}
\and
Hallee E. Wong\\
CSAIL MIT \& MGH\\
\and
Jose Javier Gonzalez Ortiz\\
MosaicML DataBricks\\ 
\& MIT \\
\and 
Beth Cimini\\
Broad Institute\\
\and
John Guttag\\
CSAIL MIT\\
\and
Adrian V. Dalca\\
CSAIL MIT\\
\& HMS, MGH\\
}

\begin{document}
\maketitle
\begin{abstract}
Existing learning-based solutions to medical image segmentation have two important shortcomings. First, for most new segmentation task, a new model has to be trained or fine-tuned. This requires extensive resources and machine-learning expertise, and is therefore often infeasible for medical researchers and clinicians. Second, most existing segmentation methods produce a single deterministic segmentation mask for a given image. In practice however, there is often considerable uncertainty about what constitutes the \textit{correct} segmentation, and different expert annotators will often segment the same image differently. We tackle both of these problems with \emph{Tyche}, a model that uses a context set to generate stochastic predictions for previously unseen tasks without the need to retrain. Tyche differs from other in-context segmentation methods in two important ways. (1) We introduce a novel convolution block architecture that enables interactions among predictions. (2) We introduce in-context test-time augmentation, a new mechanism to provide prediction stochasticity. When combined with appropriate model design and loss functions, Tyche can predict a set of plausible diverse segmentation candidates for new or unseen medical images and segmentation tasks without the need to retrain. Code available at: \url{https://github.com/mariannerakic/tyche/}.
\end{abstract}    
\begin{figure}
    \centering
    \includegraphics[width=0.46\textwidth]{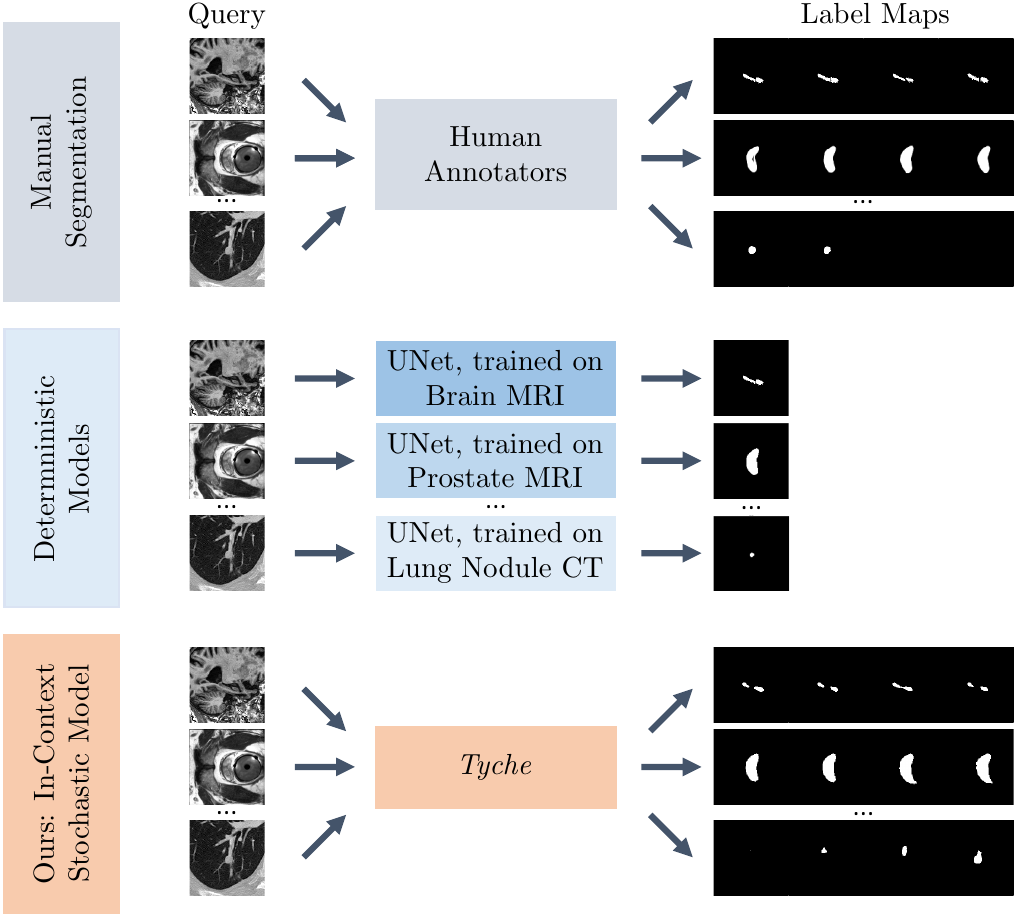}
    \caption{\textbf{Tyche: the first in-context stochastic segmentation framework.} Human annotators (top) can handle a wide variety of tasks, and different annotators often produce differing segmentations. Existing automated methods (middle) are typically task-specific and provide only one segmentation per image. \emph{Tyche} (bottom) can capture the disagreement among annotators across many modalities and anatomies without retraining or fine-tuning.}
    \label{fig:teaser}
\end{figure}

\section{Introduction}
\label{sec:intro}

Segmentation is a core step in medical image analysis, for both research and clinical applications. However, current approaches to medical image segmentation fall short in two key areas. First, segmentation typically involves training a new model for each new modality and biomedical domain, which quickly becomes infeasible given the resources and expertise available in biomedical research and clinical environments. Second, models most often provide a single solution, whereas in many cases, the target image contains ambiguous regions, and there isn't a \textit{single} correct segmentation. This ambiguity can arise from noisy or low contrast images, variation in the task definition, or human raters' interpretations and downstream goals~\cite{becker2019variability,joskowicz2019inter}. Failure to take this ambiguity into account can affect downstream analysis, diagnosis, and treatment.

Recent work tackles these issues separately. \emph{In-context learning} methods generalize to unseen medical image segmentation tasks, employing an input \textit{context} or \textit{prompt} to guide inference~\cite{butoi2023universeg,wang2023seggpt,wang2023images}.  These methods are deterministic and predict a \textit{single} segmentation for a given input image and task.

Separately, stochastic or probabilistic segmentation methods output multiple plausible segmentations at inference, reflecting the task uncertainty~\cite{baumgartner2019phiseg, kohl2018probabilistic, monteiro2020stochastic}. Each such model is trained for a specific task, and can only output multiple plausible segmentations at inference for that task. Training or fine-tuning a model for a new task requires technical expertise and computational resources that are often unavailable in biomedical settings. 

We present \emph{Tyche}, a framework for stochastic in-context medical image segmentation (Figure \ref{fig:teaser}). \emph{Tyche} includes two variants for different settings. The first, \emph{\tychetrain} (Train-time Stochasticity), is a system explicitly designed to produce multiple candidate segmentations. The second, \emph{\tychetest} (Inference-time Stochasticity), is a test time solution that leverages a pretrained deterministic in-context model.

\emph{Tyche} takes as input the image to be segmented (target), and a \textit{context set} of image-segmentation pairs that defines the task. This enables the model to perform unseen segmentation tasks upon deployment, omitting the need to train new models. \emph{\tychetrain} learns a \textit{distribution of possible label maps}, and predicts a set of plausible stochastic segmentations. \emph{\tychetrain} encourages \textit{diverse} predictions by enabling the internal representations of the different predictions to interact with each other through a novel convolutional mechanism, a carefully chosen loss function and noise as an additional input. In \emph{\tychetest}, we show that applying test time augmentation to both the target and context set in combination with a trained in-context model leads to competitive segmentation candidates.

We make the following contributions. 
\begin{itemize}
    \item We present the first solution for probabilistic segmentation for in-context learning. We develop two variants to our framework: \emph{\tychetrain} that is trained to maximize the quality of the best prediction, and \emph{\tychetest}, that can be used straightaway with an existing in-context model. 
    \item For \emph{\tychetrain}, we introduce a new mechanism, \emph{SetBlock}, to encourage diverse segmentation candidates. Simpler than existing stochastic methods, \emph{\tychetrain} predicts all the segmentation candidates in a single forward pass.
    \item Through rigorous experiments and ablations on a set of twenty unseen medical imaging tasks, we show that both frameworks produce solutions that outperform existing in-context and interactive segmentation benchmarks, and can match the performance of specialized stochastic networks trained on specific datasets.
\end{itemize}

\section{Related Work}
\label{sec:related}

Biomedical segmentation is a widely-studied problem, with recent methods dominated by UNet-like architectures~\cite{ronneberger_u-net_2015,badrinarayanan2017segnet,isensee2021nnu}. These models tackle a wide variety of tasks, such as different anatomical regions, different structures to segment within a region, different image modalities, and different image settings. With most methods, a new model has to be trained or fine-tuned for each combination of these. Additionally, most models don't take into account image ambiguity, and provide a single deterministic output.

\subpara{Multiple Predictions}
Uncertainty estimation can help users decide how much faith to put in a segmentation~\cite{czolbe2021segmentation} and guide downstream tasks. Uncertainty is often categorized into aleatoric, uncertainty in the data, and epistemic, uncertainty in the model~\cite{der2009aleatory,kendall2017uncertainties}. In this work, we focus on aleatoric uncertainty. Medical images are also heteroscedastic in that the degree of uncertainty varies across the image. 

Different strategies exist to capture uncertainty. One can assign a probability to each pixel~\cite{kendall2015bayesian,hong2021hypernet,islam2021spatially,larrazabal2021maximum}, or use contour strategies and difference loss functions to predict the largest and smallest plausible segmentations~\cite{lambert2023triadnet, ye2023confidence}. These strategies however do not capture the correlations across pixels. To address this, some methods generate multiple plausible label maps given an image~\cite{monteiro2020stochastic, baumgartner2019phiseg, kohl2018probabilistic, kohl2019hierarchical, wolleb_diffusion_2021}. To achieve this, one can directly model pixel correlations, such as through a multivariate Gaussian distribution (with low rank) covariance~\cite{monteiro2020stochastic}, or more complex distributions~\cite{bhat2022generalized}. Alternatively, various frameworks combine potentially hierarchical representations for UNet-like architectures with variational auto-encoders~\cite{kohl2018probabilistic,baumgartner2019phiseg,kohl2019hierarchical}. More recently, diffusion models have been used for ensembling~\cite{wolleb_diffusion_2021} or to produce stochastic segmentations~\cite{rahman2023ambiguous,zbinden2023stochastic}. Some methods explicitly model the different annotators to capture ambiguity~\cite{hu2023inter,nichyporuk2022rethinking,schmidt2023probabilistic,tanno2019learning}. But these methods do not apply to our framework where the number of annotators and their characteristics are unknown.  

Most of the models above involve sophisticated modeling or lengthy runtimes, and need to be trained on each segmentation task. In \emph{Tyche}, we build on intuition across these methods, but combine a more efficient mechanism with an in-context strategy to predict segmentation candidates. 

\subpara{In-context Learning}
Few-Shot frameworks use a small set of examples to generalize to new tasks~\cite{ding2023few, li2022prototypical, nguyen2019feature, shen2022q,pandey2023robust, seo2022task, zhang2019canet}, sometimes by fine-tuning an existing pretrained model~\cite{finn2017model, nichol2018first, vinyals2016matching, snell2017prototypical}. 
In-context learning segmentation methods (ICL) use a small set of examples directly as input to infer label maps for a task~\cite{butoi2023universeg,balavzevic2023towards, kim2023universal,wang2023seggpt,wang2023seggpt,kim2023universal}. This enables them to generalize to new tasks. For example, UniverSeg uses an enhanced UNet-based architecture to generalize to medical image segmentation tasks unseen during training~\cite{butoi2023universeg}. We build on these ideas to enable segmentation of new tasks without the need to re-train, but expand this paradigm to model stochastic segmentations.

\subpara{Test Time Augmentation}
The test-time augmentation (TTA) strategy uses perturbations of a test input and ensembles the resulting predictions. Existing TTA frameworks model accuracy~\cite{shanmugam2021better,melanieTAAL,sinha2019variational,kim2021task}, robustness~\cite{cohen2023boosting}, and estimates of uncertainty~\cite{matsunaga2017image,ayhan2022test}. Test-time augmentation has  been applied to diverse anatomies and modalities including brain MRI and retinal fundus~\cite{amiri2020two,ayhan2022test,huang2021style,moshkov2020test,ResNetTTAColon,wang2019automatic}. Prior work has formalized the variance of a model's predictions over a set of input transformations as capturing aleatoric uncertainty~\cite{TTA2019aleatoric,wang2019automatic,ayhan2022test}. 

\emph{Tyche}'s use of TTA is distinct from prior work. Instead of ensembling segmentations over perturbations of a test input or pixel-wise estimates of uncertainty, \emph{Tyche} extends TTA to the in-context setting and uses the individual TTA predictions to model uncertainty.

\begin{figure}
    \centering
    \includegraphics[width=0.46\textwidth]{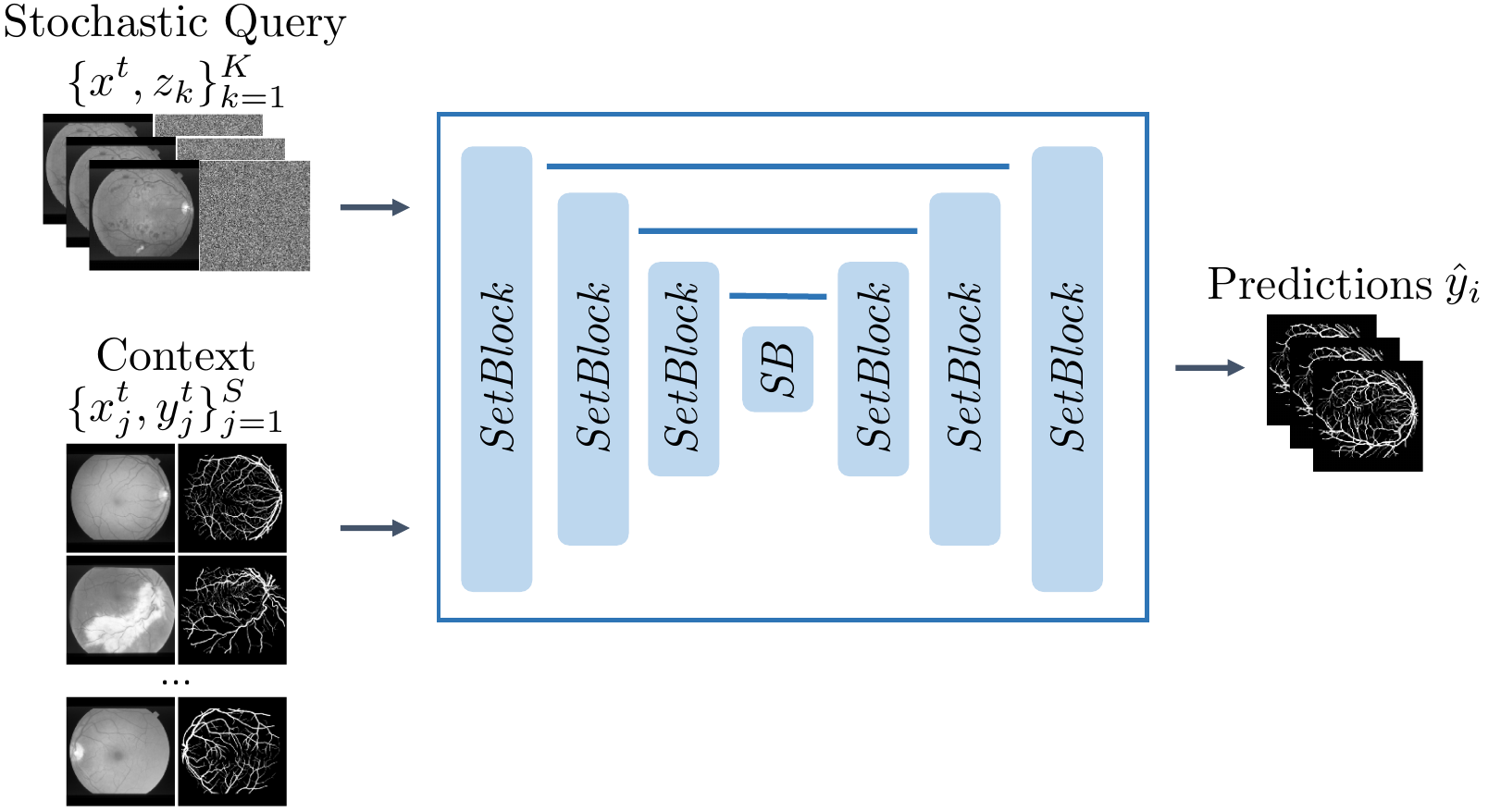}
    \caption{\textbf{Tyche Model Schematic.} The target $x^t$, context set ${(x^t_j, y^t_j)}_{j=1}^S$, and noise images $\{z_k\}_{k=1}^K$ are inputs to the network. The architecture employs UNet-like levels, but uses \emph{SetBlocks} that enable interactions between the context set and the target segmentation candidates.}
    \label{fig:method:network}
\end{figure}

\begin{figure}
    \centering
    \includegraphics[width=0.43\textwidth]{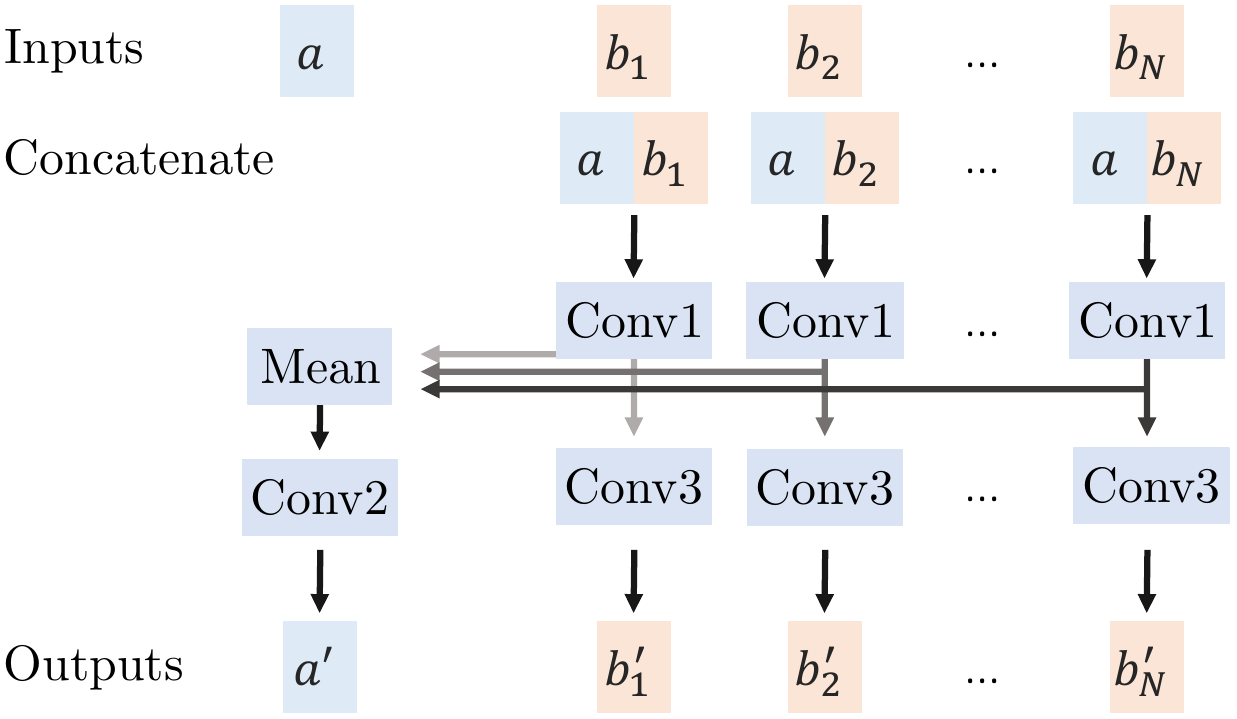}
    \caption{\textbf{CrossBlock Mechanism} The CrossBlock involves interactions between a single feature and a set of features and outputs new feature for the target and new features for each.}
    \label{fig:method:xblock}
\end{figure}
\begin{figure}
    \centering
    \includegraphics[width=0.47\textwidth]{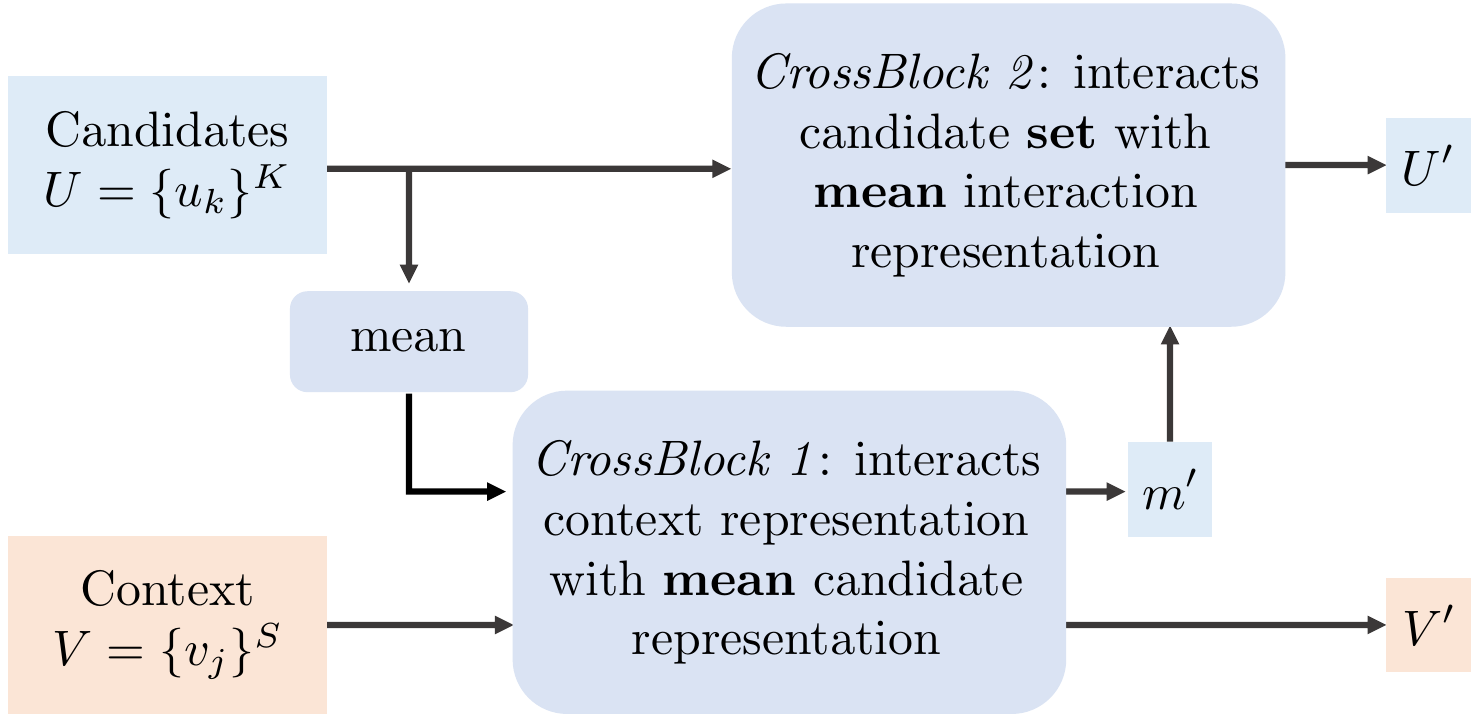}
    \caption{\textbf{SetBlock Mechanism}. \emph{SetBlock} enables interactions between the \textbf{set} of features from the context set and the \textbf{set} of features from the prediction candidates. It outputs two sets of features, one for the context and one for the prediction candidates.}
    \label{fig:method:setblock}
\end{figure}

\section{Method}
\label{sec:methods}

For segmentation task $t$, let $\{(x^t_j, y^t_j)\}_{j=1}^N$ be a dataset with images $x^t$ and label maps $y^t$. Typical segmentation models learn a different function $\hat{y}^t = g_{\theta^t}(x^t)$ with parameters $\theta^t$ for each task~$t$, where $\hat{y}^t$ is a single segmentation map prediction. 

We design \emph{Tyche} as an in-context learning (ICL) model using a \textit{single} function for all tasks:
\begin{equation}
    \hat{y}_k^t = f_\theta(x^t, z_k, \mathcal{S}^t).
    \label{eq:tyche_model}
\end{equation}
This function, with \textit{global} parameters $\theta$, captures a \textit{distribution of label maps} $\{\hat{y}^{t}_k\}_{k=1}^K$, given target $x^t$, context set \mbox{$\mathcal{S}^t=\{x^t_j, y^t_j\}_{j=1}^S$} defining task \textit{t}, and noise $z_k \sim \mathcal{N}(\boldsymbol{0}, \mathbb{I})$. We use this modelling strategy in two ways: we either explicitly train a network to approximate the model $f_\theta(\cdot)$ in \emph{\tychetrain}, or design a test-time strategy to approximate $f_\theta(\cdot)$ using an existing (pretrained) deterministic in-context network in \emph{\tychetest}.

\subsection{\tychetrain}

In \emph{\tychetrain}, we explicitly train a neural network for $f_\theta(\cdot)$ that can make different predictions given the same image input $x^t$ but different noise channels $z_k$. We model interaction between predictions, and employ a loss that encourages diverse solutions (Figure \ref{fig:method:network}).

\subsubsection{Neural Network}
We use a convolutional architecture focused on interacting representations of sets of flexible sizes using a modified version of the usual UNet structure~\cite{ronneberger_u-net_2015}.

\subpara{Inputs} \emph{\tychetrain} takes as input the target $x^t$, a set of K Gaussian noise channels $z_k$, and a context set, $\mathcal{S}^t$. 

\subpara{Layers} Each level of the UNet takes as input a set of K candidate representations and S context representations. We design each level to encourage communication between the intermediate elements of the sets, and between the two features of the segmentation candidates. The size $K$ is flexible and can vary with iterations. 

\subpara{SetBlock} We introduce a new operation called \emph{SetBlock}, which interacts the candidate representations $U=\{u_i\}_{i=1}^K$, with the context representations $V=\{v_i\}_{i=1}^S$, illustrated in Figure \ref{fig:method:setblock}. We use the \textit{CrossBlock}~\cite{butoi2023universeg} as a building block for this new layer. The $\text{CrossBlock}(u, V) \rightarrow (u^\prime, V^\prime)$ compares an image representation $u$ to a context set representation $V$ through convolutional and averaging operations, and outputs a new image representation $u'$ and a new set representation $V'$ (Figure \ref{fig:method:xblock}). $\text{SetBlock}(U, V) \rightarrow  (U', V')$ builds on $\text{CrossBlock}$ and performs a set to set interaction of the entries of $U$ and $V$: 
\begin{align} 
    \bar{u} &= 1/m {\textstyle\sum}_{i=1}^m u_i\\
    \bar{u}', V' &= \text{CrossBlock}(\bar{u}, V)\\
    u_i' &= \text{Conv}_m\left(u_i || \bar{u}'\right), \quad i=1, \ldots, K\,\\
    u_i' &= \text{Conv}_u\left(u_i'\right),\quad i=1, \ldots,\, K\\
    v_i' &= \text{Conv}_v\left(v_i\right),\quad i=1, \ldots,\, S,
\end{align}
where $||$ is the concatenation operation along the feature dimension. The CrossBlock interacts the context representation with the \textbf{mean} candidate. The $\textit{Conv}_m$ step communicates this result to all candidate representation. $\textit{Conv}_u$ and $\textit{Conv}_v$ then update all representations. All convolution operations include a non-linear activation function.

\subsubsection{Best candidate Loss}
Typical loss function compute the loss of a single prediction relative to a single target, but \emph{\tychetrain} produces multiple predictions and has one or more corresponding label maps.%
We optimize 
\begin{equation}
\small
\mathcal{L}(\theta; \mathcal{T}) = 
\mathbb{E}_{t\in \mathcal{T}}\left[\mathbb{E}_{(x^t, y^t_r), \mathcal{S}^t}\left[ \mathcal{L}_{seg}\left(\{\hat{y}_k\}, y^t_r\right)\right]\right], 
\end{equation}
with
\begin{equation}
\small
\mathcal{L}_{seg}  (\{\hat{y}_k\}, y)= \, \min_{k} \; \mathcal{L}_{Dice}\left(y_k, y\right),
\end{equation}
where  $y_r^t$ is a segmentation from rater $r$, and $\mathcal{L}_{Dice}$ is a weighted sum of soft Dice loss~\cite{milletari2016v} and categorical crossentropy. By only back-propagating through the best prediction among $K$ candidates, the network is encouraged to produce diverse solutions~\cite{charpiat2008automatic, guzman2012multiple, kirillov2023segment, li2018interactive}. 

\subsubsection{Training Data}
We employ a large dataset of single- and multi-rater segmentations across diverse biomedical domains. We then use data augmentation~\cite{butoi2023universeg}, as described in \ref{sup:tes:info}. 

We add synthetic multi-annotator data by modelling an image as the average of four blobs representing four raters (Figure \ref{fig:sup:blob}). Each blob is white disk $b_i$ deformed by a random smoothed deformation field $\phi_i$. The synthetic image is a noisy weighted sum of raters: $\sum_{i=1}^4 w_i (b_i \circ \phi_i)$ where $\circ$ represents the spacial warp operation.

\subsubsection{Implementation Details}
We use a UNet-like architecture of 4 \emph{SetBlock} layers for the encoder and decoder, with 64 features each and Leaky ReLU as activation function. We use the Adam optimizer and a learning rate of 0.0001. At training, we have a fixed number of candidates per sample $K_\text{tr}=8$. At inference, we consider different numbers of candidates. %

\subsection{\tychetest}
In \emph{\tychetest}, we first train (or use an existing trained) \textit{deterministic} in-context segmentation system: 
$$\hat{y}^t = h_\theta(x^t, \mathcal{S}^t).$$
We then introduce a \textit{test-time} in-context augmentation strategy to provide stochastic predictions: 
\begin{align}
    \hat{y}^t_k &= f_\theta (x^t, z_k, \mathcal{S}^t)\\
                &= h_\theta(aug(x^t, z_k, \mathcal{S}^t)),
\end{align}
where $\tilde{x}^t, \tilde{\mathcal{S}}^{t} = aug(x^t, z_k, \mathcal{S}^t)$ is an augmentation function.

\subsubsection{Augmentation Strategy}
Test time augmentation for single task networks $y=g_t(x)$ applies different transforms to an input image $x$:
\begin{equation}
\small
   \tilde{x}_k = a_{\phi}(x, z_k),%
\end{equation}
where $\phi$ are augmentation parameters and $z_k$ is a random vector. A final prediction is then obtained by combining several predictions of augmented images. Most commonly, the combining function averages the predictions: 
\begin{equation}
    y = \frac{1}{k} \sum_k g_t (\tilde{x}_k),
\end{equation}
where the sum operates pixel-wise.

We introduce in-context test-time augmentation (ICTTA) prediction as another mechanism to generate diverse stochastic predictions. %

We apply augmentation to both the test target $x^t$ and the context set $\mathcal{S}^t$:%
\begin{align}
\small
    (\tilde{x_i}^t, y^t_i) = (a_{\phi} (x^t_i), y^t ) \\
    \tilde{\mathcal{S}}^t = \{a_\phi\left(x_j^t\right), y_j^t\}_{j=1}^S.
\end{align}

We repeat this process $K_i$ times to obtain $K_i$ stochastic predictions: 
\begin{align}
    \hat{y}_k = f_\theta(\tilde{x_i}^t, z_k, \tilde{\mathcal{S}^t})
\end{align}
We only apply intensity based transforms, to avoid the need to invert the segmentations back. We apply Gaussian noise, blurring and pixel intensity inversion. We detail the specific augmentations in \ref{sup:aug:training}.

\begin{figure*}
    \centering
    \includegraphics[width=.9\textwidth]{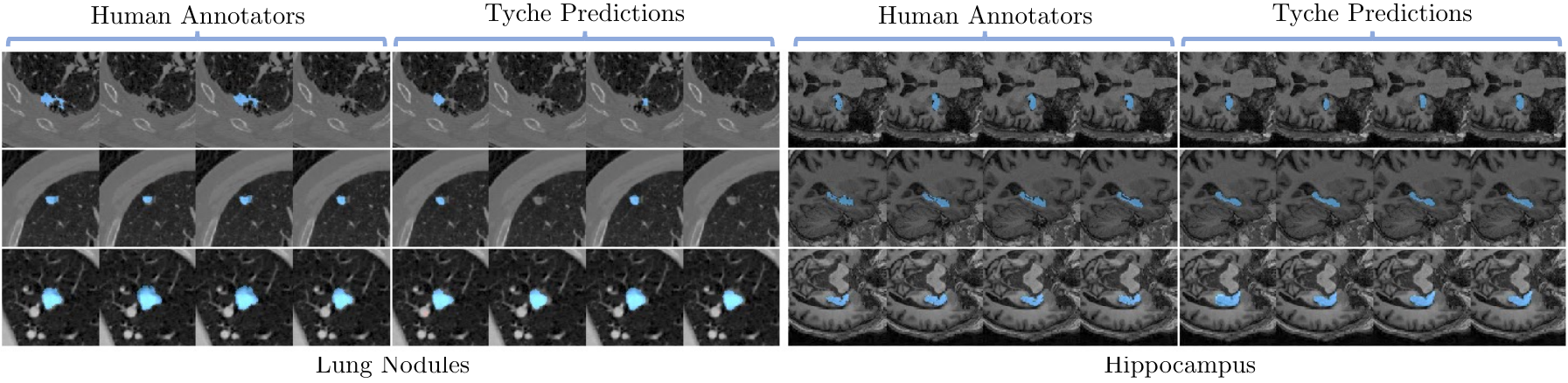}
    \caption{\textbf{Visualization of predictions for three different samples}, 1 per row. Left: LIDC-IDRI. Right: Hippocampus dataset. The leftmost columns are raters' annotations. The 4 last columns are model predictions. \emph{Tyche} provides a set of prediction that is diverse and matches the raters, for tasks unseen at training time.}
    \label{fig:visual}
\end{figure*}

\begin{figure*}
\centering
    \includegraphics[width=0.9\textwidth]{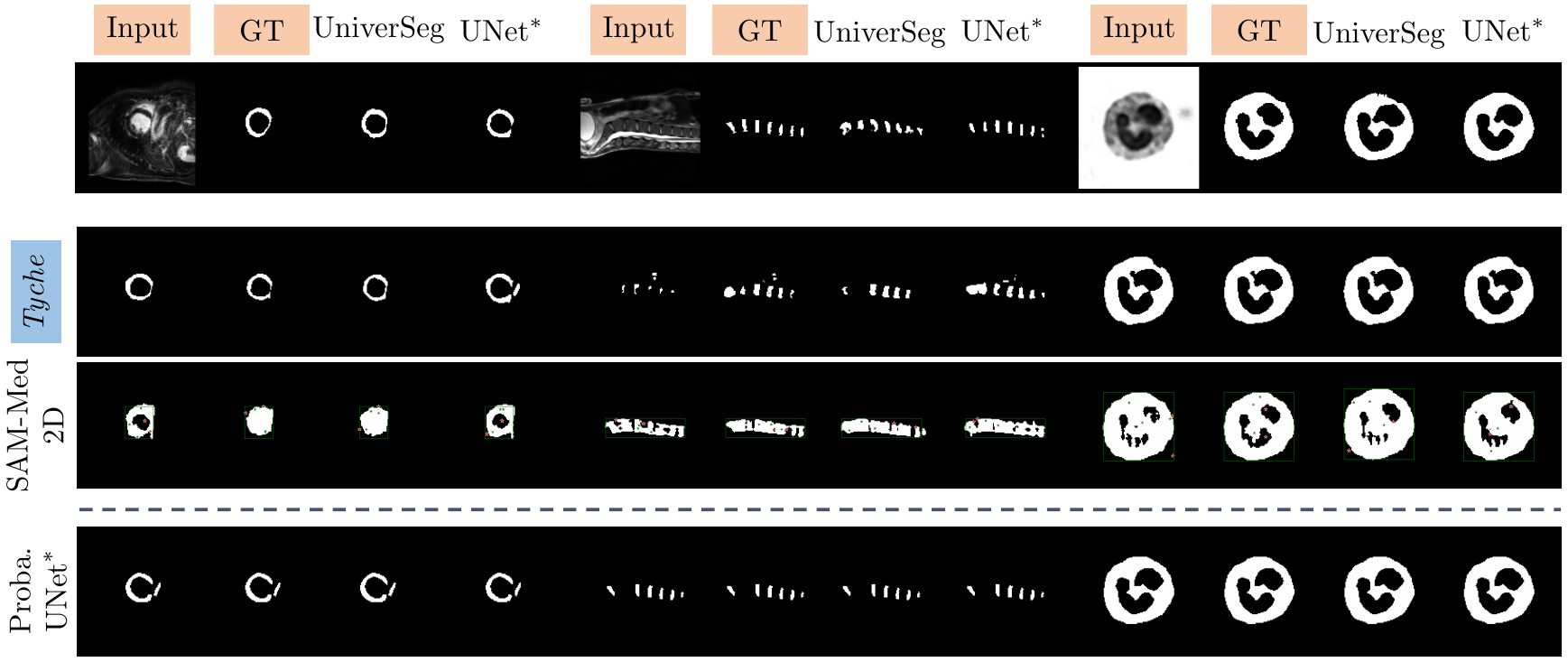}
    \caption{\textbf{Single annotator visualization for different models.} We show three example images that show very different corresponding segmentation. \emph{Tyche} can output plausible segmentation for single annotator data with varying degrees of variability in the segmentation. Methods with an asterisk are upper baselines. }
    \label{fig:exp:visualsr}
\end{figure*}

\section{Experimental Setup}
\subsection{Data}
We evaluate our method using a large collection of biomedical and synthetic datasets. Most datasets include a single manual segmentation for each example, while a few have several raters per image. 

\subpara{Data Splits}
We partition each dataset into development, validation, and test splits. We assign each dataset to an \textit{in-distribution} set (\textit{I.D.}) or an \textit{out-of-distribution} set (\textit{O.D.}). We train exclusively on the development splits of the \textit{I.D.} datasets, and use the validation splits of the \textit{I.D.} datasets to tune parameters. We use the validation splits of the \textit{O.D.} datasets for final model selection. We report results on the test splits of the \textit{O.D.} datasets. 

For each use case, we sample the context from each dataset's corresponding development set. As a result, the network doesn't see any of the \textit{O.D.} datasets at training time.%

We distinguish between single annotator data and multi-annotator data.

\subpara{Single-Annotator Data} For single annotator data, we build on MegaMedical used in recent publications~\cite{butoi2023universeg,wong2023scribbleprompt}  and employ a collection of 73 datasets, of different public biomedical domains and different modalities~\cite{ACDC,PanDental,cDemris,CAMUS,CoNSeP, CHAOS2021Challenge,OCTA500, Rose, IDRID, DRIVE,CheXplanation,LUNA, SCD,SpineWeb, VerSe,WBC, BBBC003,butoi2023universeg,ssTEM, MMOTU,ISBI_EM, PanNuke,ISIC,LiTS,NCI-ISBI,KiTS,AMOS,CHAOSdata2019,SegTHOR,BTCV,Promise12,I2CVB, Word, MSD, DukeLiver, CT2US,CT_ORG,SegThy,NerveUS,BRATS,MCIC,ISLES,WMH, BrainDevelopment, OASIS-data, PPMI, LGGFlair,PAXRay,HipXRay, ToothSeg,BUID,BUSIS}. MegaMedical spans a variety of anatomies and modalities, including brain MRI, cardiac ultrasound, thoracic CT and dental X-ray. We also use synthetic data involving simulated shapes, intensities, and image artifacts~\cite{butoi2023universeg,hoffmann2021synthmorph}. The  single-annotator datasets used for out-of-domain (\textit{O.D.}) testing are: PanDental~\cite{PanDental}, WBC~\cite{WBC}, SCD~\cite{SCD},  ACDC~\cite{ACDC}, and SpineWeb~\cite{SpineWeb}.

\subpara{Multi-Annotator Data} For multi-annotator \textit{I.D.} data, we use four datasets from Qubiq~\cite{qubiq}: Brain Growth, Brain Lesions, Pancreas Lesions, and Kidney. We also simulate a multi-rater dataset consisting of random shapes (blobs). For the \textit{O.D.} multi-annotator data we use four datasets. One contains hippocampus segmentation maps on brain MRIs from a large hospital. We crop the volumes around the hippocampus~\cite{kohl2018probabilistic} to focus on the areas where the raters disagree. The second is a publicly available lung nodule dataset, LIDC-IDRI~\cite{armato2011lung}. This dataset is notable for the substantial inter-rater variability. It contains 1018 thoracic CT scans, each annotated by 4 annotators from a pool of 12 annotators. Finally, we also use retinal fundus images, STARE~\cite{STARE}, annotated by 2 raters, and prostate data from the MICCAI 2021 QUBIQ challenge~\cite{qubiq}, annotated by 6 raters on two tasks. Single and multi-annotator combined, our \textit{O.D.} group contains 20 tasks unseen at training time (some datasets have several tasks).

\subsection{Evaluation}
We evaluate our method by analysing individual prediction quality and distribution of predictions, both qualitatively and quantitatively. We also examine model choices through an ablation study.

A main use case of stochastic segmentation is to propose a small set of segmentations to a human rater, who can select the most appropriate one for their purpose. For this scenario, a model can be viewed as good if at least one prediction matches what the rater is looking for. We thus employ the best candidate Dice metric. 

In the multi-annotator setting, we evaluate using both best candidate Dice score, also called maximum Dice score, as well as Generalized Energy Distance (GED)\cite{bellemare2017cramer,salimans2018improving,szekely2013energy}. GED is commonly used in the stochastic segmentation literature to asses the difference between the distribution of predictions and the distribution of annotations~\cite{kohl2018probabilistic,baumgartner2019phiseg,monteiro2020stochastic,rahman2023ambiguous,zbinden2023stochastic}. GED has limitations, such as rewarding excessive prediction diversity~\cite{rahman2023ambiguous}. Let $\mathcal{Y}$ and $\hat{\mathcal{Y}}$ be the set of annotations, GED is defined as:
\begin{equation}
\small
    D^2_{GED}(\mathcal{Y}, \hat{\mathcal{Y}}) = 2\mathbb{E}\left[d(p,\hat{p} )\right]-\mathbb{E}\left[d(p, p')\right]-\mathbb{E}\left[d(\hat{p}, \hat{p}')\right],
\end{equation}
where $p,p'\sim \mathcal{Y}$, $\hat{p}, \hat{p}' \sim \hat{\mathcal{Y}}$ and $d(\cdot, \cdot)$ is a distance metric. We use the Dice score~\cite{dice1945measures}.

\subsection{Benchmarks}
\emph{Tyche} is the first method to produce stochastic segmentation predictions in-context. Consequently, we compare \emph{Tyche} to existing benchmarks, each of which achieves only a subset of our goals.%

\begin{table}[]
    \centering
   \rowcolors{2}{white}{gray!15}
    \begin{tabular}{lccc}
    & In-Context & Stochastic & Automatic\\
    \hline
    SENet     & \checkmark &  & \checkmark \\
    UniverSeg     & \checkmark &  & \checkmark \\
    SegGPT     & \checkmark &  & \checkmark \\
    Prob. UNet & & \checkmark & \checkmark \\
    PhiSeg & & \checkmark &  \checkmark\\
    CIMD & & \checkmark &  \checkmark\\
    SAM-based & \checkmark & \checkmark & \\ \hline
    \textbf{Tyche} & \checkmark & \checkmark & \checkmark\\\hline
    \end{tabular}
    \caption{\textbf{Summary of evaluated methods used and their properties.} Only \emph{Tyche} is both stochastic and in-context, and does not require user interaction.}
    \label{tab:benchmark:summary}
\end{table}

\subpara{In-Context Methods} We compare to deterministic frameworks that can leverage a context set: a few-shot method, SENet~\cite{roy2020SEnet}, and two in-context learning (ICL) methods, UniverSeg~\cite{butoi2023universeg} and SegGPT~\cite{wang2023seggpt}. We train UniverSeg and SENet with the same data split strategies and the same sets of augmentation transforms as for \emph{Tyche}. For SegGPT, we use the public model, trained on a mix of natural and medical images. Figure \ref{fig:sup:OfficialvsRetraineduniverseg} in the Supplemental Material shows that UniverSeg trained with additional data outperforms its public version.

\subpara{Stochastic Upper Bounds}
We compare to task-specialized probabilistic segmentation methods that are trained-on and perform well on specific datasets. We independently train Probabilistic UNet~\cite{kohl2018probabilistic}, PhiSeg\cite{baumgartner2019phiseg} and CIDM, a recent diffusion network~\cite{rahman2023ambiguous}, on each of the 20 held-out tasks. For each task, we train three model variants: no augmentation, weak augmentation, and as much augmentation as for the \emph{Tyche} targets. For each benchmark variant, we train on a \textit{O.D.} development split and select the model that performs best on the corresponding \textit{O.D.} validation split. We then compare these benchmarks to \emph{Tyche} on the held-out \textit{O.D.} test splits.

These models are explicitly optimized for the datasets on which they are evaluated, unlike \emph{Tyche}, which does not use those datasets for training. Since these are trained, tuned and evaluated on the \textit{O.D.} datasets splits, something we explicitly aim to avoid in the problem set up as it is not easily done in many medical settings, they serve as upper bounds on performance.

\subpara{Interactive Segmentation Methods} We compare to two interactive methods: SAM~\cite{kirillov2023segment} and SAM-Med2D~\cite{cheng_sam-med2d_2023}. These methods can provide multiple segmentations, but, unlike \emph{Tyche}, require human interaction, which is outside our scope. SAM also has a functionality to segment all elements in an image, however less optimized for medical imaging. We assume that the SAM-based models have access to the same information as the ICL methods: several image-segmentation pairs as context to guide the segmentation task. We fine-tune SAM using our \textit{I.D.} development datasets. To replace the human interaction, we provide a bounding box, the average context label map, and 10 clicks, 5 positive and 5 negative as input. With SAM-Med2D, we use a bounding box, and several positive and negative clicks as input. For both SAM and SAM-Med2D, we generate clicks and bounding box from the average context label map.

 We use one iteration of interaction, and sample different plausible segmentation candidates by sampling different sets of clicks and different averaged context sets.

Table \ref{tab:benchmark:summary} summarizes the features of all the methods. Additional information on the benchmarks is provided in the Supplemental Material.

\begin{table*}
\centering
\rowcolors{2}{white}{gray!15}
    \begin{tabular}{l|l|ccccc}
\multicolumn{2}{l}{$GED^2$ ($\downarrow$)}         &  Hippocampus & LIDC-IDRI & Prostate Task 1 & Prostate Task 2 & STARE \\\hline
\cellcolor{white}    &SAM & $0.57 \pm 0.02$ & $0.90 \pm 0.01$ & $0.20 \pm 0.03$ & $0.31 \pm 0.06$ & $0.89 \pm 0.06$ \\
\multirow{-2}{*}{\cellcolor{white} Interactive}    &SAM-Med2d & $0.93 \pm 0.02$ & $1.01 \pm 0.01$ & $0.80 \pm 0.09$ & $0.78 \pm 0.11$ & $1.52 \pm 0.05$ \\\hline
\cellcolor{white} & \textbf{\tychetest} & $\bm{0.21 \pm 0.01}$ & $0.41 \pm 0.01$ & $0.12 \pm 0.02$ & $0.20 \pm 0.05$ & $0.73 \pm 0.03$ \\
\multirow{-2}{*}{\cellcolor{white} \shortstack{I-C \& Stochastic \\ (Ours)}} & \textbf{\tychetrain} & $0.22 \pm 0.01$ & $\bm{0.40 \pm 0.01}$ & $\bm{0.09 \pm 0.02}$ & $\bm{0.15 \pm 0.03}$ & $\bm{0.62 \pm 0.03}$ \\\hline\hline
\cellcolor{white}    &PhiSeg & $0.14 \pm 0.01$ & $0.33 \pm 0.01$ & $0.12 \pm 0.01$ & $0.17 \pm 0.05$ & $1.22 \pm 0.02$ \\
\cellcolor{white}    & ProbaUNet & $0.13 \pm 0.01$ & $0.51 \pm 0.01$ & $0.08 \pm 0.01$ & $0.18 \pm 0.05$ & $0.76 \pm 0.06$ \\
\multirow{-3}{*}{\cellcolor{white} \shortstack{Stochastic\\ Upper Bound}}    & CIDM & $0.17 \pm 0.01$ & $0.42 \pm 0.01$ & $0.14 \pm 0.02$ & $0.26 \pm 0.04$ & $0.87 \pm 0.05$ \\\hline
    \end{tabular}
    \caption{\textbf{Generalized Energy Distance} for different models with a context size of 16 for in-context methods and a number of predictions set to 8. Lower is better. \emph{Tyche} outperforms interactive and in-context baselines, and matches stochastic upper bounds.}
    \label{exp:tab:GED}
\end{table*}

\begin{table*}
\centering
\rowcolors{2}{white}{gray!15}
    \begin{tabular}{l|l|ccccc}
\multicolumn{2}{l}{Max Dice ($\uparrow$)}          &  Hippocampus & LIDC-IDRI & Prostate Task 1 & Prostate Task 2 & STARE \\\hline
\cellcolor{white}   & UniverSeg & $0.84 \pm 0.01$ & $0.67 \pm 0.01$ & $0.91 \pm 0.01$ & $0.88 \pm 0.03$ & $0.51 \pm 0.02$ \\
    & SegGPT & $0.10 \pm 0.01$ & $0.68 \pm 0.01$ & $0.94 \pm 0.01$ & $0.89 \pm 0.03$ & $0.02 \pm 0.01$ \\
\multirow{-3}{*}{\cellcolor{white} In-Context}     & SENet & $0.68 \pm 0.01$ & $0.00 \pm 0.00$ & $0.83 \pm 0.02$ & $0.83 \pm 0.02$ & $0.30 \pm 0.03$ \\\hline
\cellcolor{white}    &SAM & $0.71 \pm 0.01$  & $0.55 \pm 0.01$ & $0.90 \pm 0.01$ & $0.85 \pm 0.03$ & $0.50 \pm 0.03$ \\
\multirow{-2}{*}{\cellcolor{white} Interactive}    &SAM-Med2d & $0.52 \pm 0.01$ & $0.42 \pm 0.01$ & $0.62 \pm 0.04$ & $0.64 \pm 0.06$ & $0.21 \pm 0.03$ \\\hline

\cellcolor{white}  & \textbf{\tychetest} & $0.87 \pm 0.01$ & $0.90 \pm 0.00$ & $0.94 \pm 0.01$ & $0.91 \pm 0.01$ & $0.52 \pm 0.03$ \\
\multirow{-2}{*}{\cellcolor{white} \shortstack{I-C \& Stochastic \\ (Ours)}} & \textbf{\tychetrain} & $\bm{0.88 \pm 0.01}$ & $\bm{0.91 \pm 0.00}$ & $\bm{0.95 \pm 0.01}$ & $\bm{0.93 \pm 0.01}$ & $\bm{0.60 \pm 0.02}$ \\\hline\hline

\cellcolor{white}    &PhiSeg & $0.88 \pm 0.00$ & $0.91 \pm 0.00$ & $0.93 \pm 0.01$ & $0.91 \pm 0.02$ & $0.15 \pm 0.01$ \\
\cellcolor{white}    & ProbaUNet & $0.91 \pm 0.00$ & $0.86 \pm 0.01$ & $0.95 \pm 0.00$ & $0.91 \pm 0.03$ & $0.59 \pm 0.02$ \\
\multirow{-3}{*}{\cellcolor{white} \shortstack{Stochastic\\ Upper Bound}}    & CIMD & $0.84 \pm 0.01$ & $0.92 \pm 0.00$ & $0.93 \pm 0.01$ & $0.87 \pm 0.02$ & $0.41 \pm 0.04$ \\ \hline

    \end{tabular}
    \caption{\textbf{Best candidate Dice score} for different models with a context size of 16 for in-context methods and a number of predictions set to 8. Higher is better. \emph{Tyche} outperforms interactive and in-context baselines, and matches stochastic upper bounds.}
    \label{exp:tab:maxdice}
\end{table*}

%
\subsection{Experiments}
We evaluate all models on the multi-annotator and single-annotator \textit{O.D.} data. We then analyze the \emph{Tyche} variants individually and perform an ablation study on each to validate parameter choices. Finally, we compare the GPU inference runtimes and model parameters. 

In the Supplemental Material, we analyze further the noise given as input, the context set, the number of predictions, the \emph{SetBlock} and the candidate loss. We also provide additional performance metrics and per dataset results. We also compare the performance of \emph{Tyche} and PhiSeg in a few-shot setting. Finally, we provide additional visualizations.

\subpara{Inference Setting}
We use a fixed context of 16 image-segmentation pairs, because existing in-context learning systems show minimal improvements beyond this size~\cite{butoi2023universeg}. Because there is variability in performance depending on the context sampled, we sample 5 different context sets for each datapoint and average performance. Similarly, for the stochastic upper bounds and interactive methods, we do 5 rounds of sampling $K_i$ samples.%


\section{Results}
\subsection{Comparison to Benchmarks}
\subpara{Multi-Annotator O.D. Data}
We evaluate on the datasets where \textit{multiple annotations} exist for each sample. Figure \ref{fig:visual} shows that, for both the lung nodules and the hippocampus datasets, \emph{Tyche} predictions are diverse and capture rater diversity, even though these datasets are out-of-domain. Tables \ref{exp:tab:GED} and \ref{exp:tab:maxdice} show that both versions of \emph{Tyche} outperform the interactive and deterministic benchmarks on all datasets except for Prostate Task 1, on which SegGPT has similar performance. Using a paired Student t-test, we find that \emph{\tychetrain} outperforms \emph{\tychetest} in terms of maximum Dice score, with $p<10^{-10}$. We find no statistical difference between the two methods in terms of GED. 

\begin{figure}
    \centering
    \includegraphics[width=0.48\textwidth]{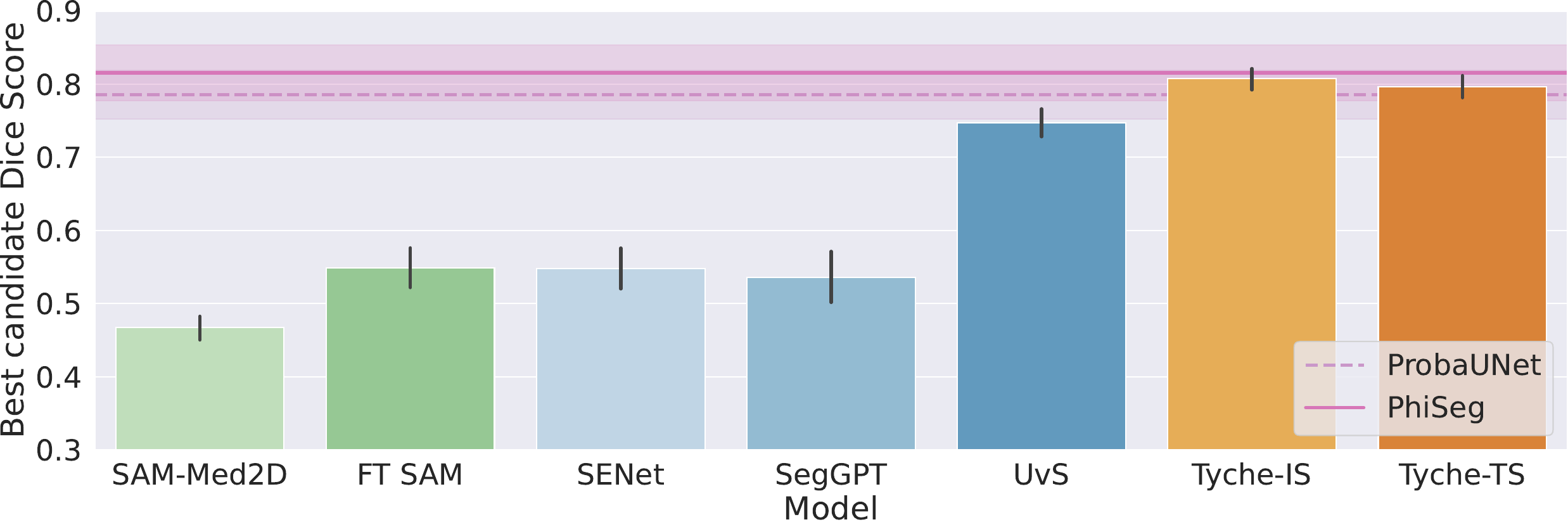}
    \caption{\textbf{Best candidate Dice Score for single annotator data aggregated per task.} %
    \emph{Tyche} outperforms the in-context and interactive segmentation benchmarks, and approaches the stochastic upper bounds. Error bars represent the 95\% confidence interval.}
    \label{fig:exp:BDice:MM}
\end{figure}
%

\subpara{Single-Annotator Data}
Figure \ref{fig:exp:visualsr} shows examples of predictions for \emph{Tyche} and the corresponding benchmarks for the single-annotator datasets. \emph{Tyche} produces a more diverse set of candidates than its competitors. Figure \ref{fig:exp:BDice:MM} compares all plausible models in terms of aggregate best candidate Dice score, except for CIDM, which underperformed for single-annotator data with a mean best candidate Dice score of: $0.673\pm0.032$. For clarity, we only present the full Figure in the Supplemental Material. \emph{Tyche} performs better than the deterministic and interactive frameworks, and similarly to Probabilistic UNet, one of the upper bound benchmarks that is trained on the O.D. data. A paired Student t-test shows that \emph{\tychetest} produces statistically higher GED ($p=0.044$) than \emph{\tychetrain}, but we find no statistical difference in terms of best candidate Dice. We hypothesize that \emph{\tychetest} is competitive because of the implicit annotator characterization provided by the context.

\subsection{Tyche Analysis}
We analyze \emph{Tyche} variants and study the influence of two important parameters: the number of inference-time candidate predictions $K_i$ and the size of the context set $\|\mathcal{S}\|$. 

\begin{figure}
    \centering
    \includegraphics[width=0.48\textwidth]{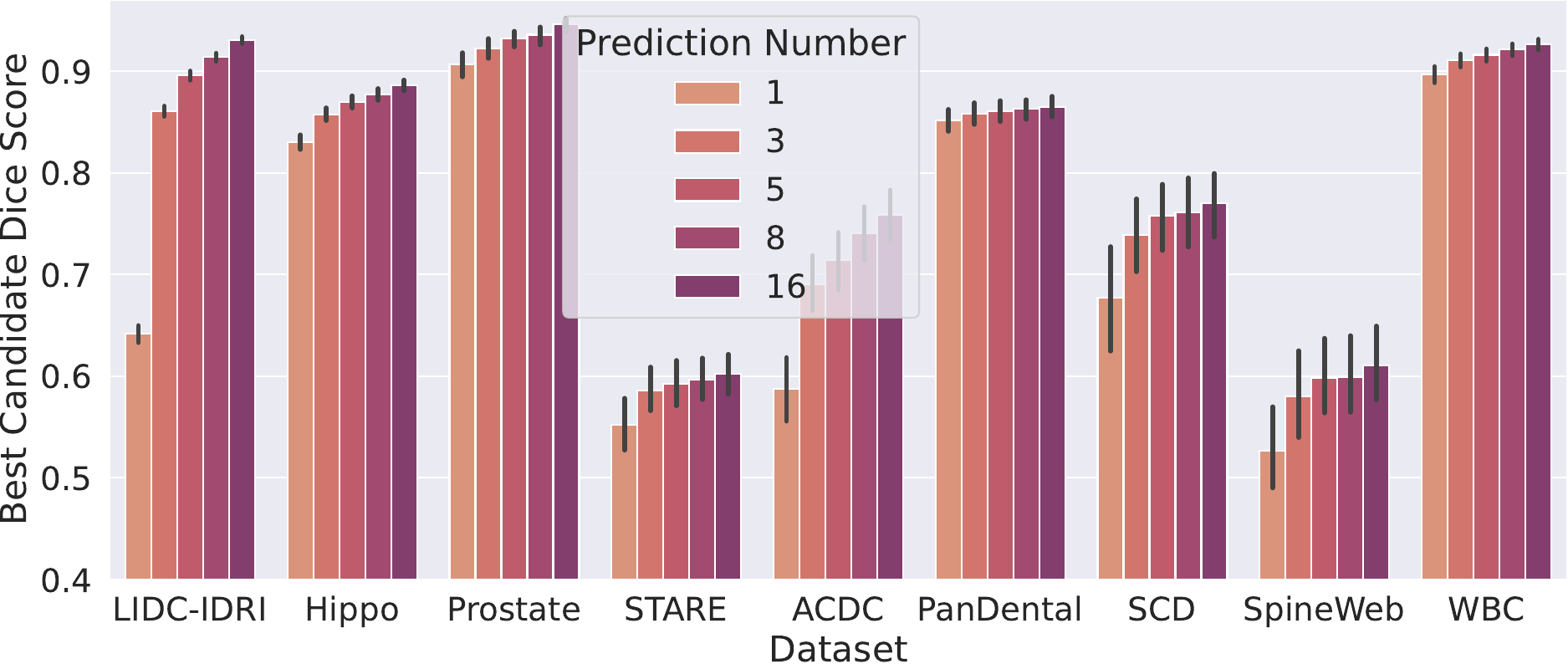}
    \caption{\textbf{Best candidate Dice Score as the number of candidate prediction increases.} The largest improvements are usually obtained for a small number of predictions. The error bars represent the 95\% confidence interval.}
    \label{fig:exp:kistudy}
\end{figure}

\subpara{Influence of the number of prediction $K_i$}
We study how the number of predictions impacts the best candidate Dice score, keeping the context size constant. Figure \ref{fig:exp:kistudy} shows that for \emph{\tychetrain}, the best candidate Dice score rises with the number of predictions, but with diminishing returns.

\begin{figure}
    \centering
    \includegraphics[width=0.48\textwidth]{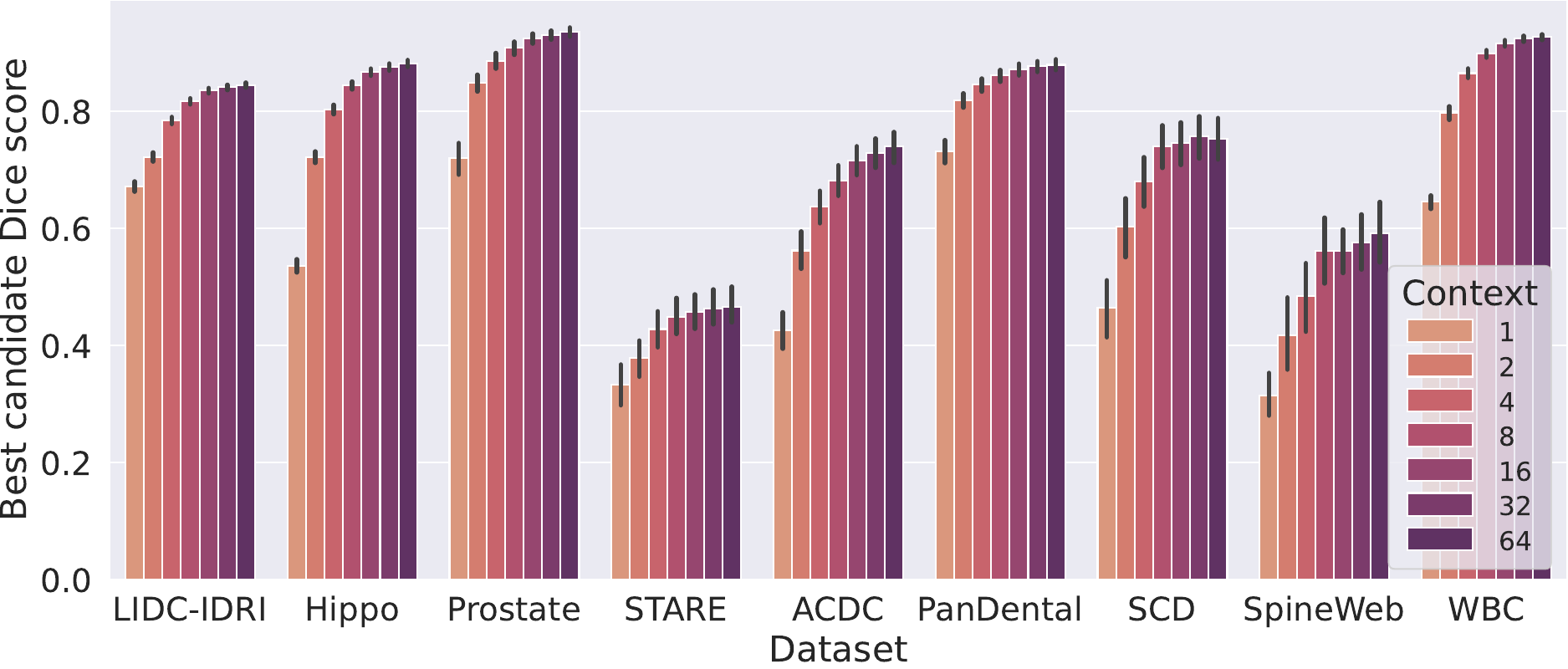}
    \caption{\textbf{Best candidate Dice Score per dataset as context size increases.} A context size of 16 is already large enough to obtain a reasonable best candidate Dice. The error bars represent the 95\% confidence interval.}
    \label{fig:exp:ssstudy}
\end{figure}

\subpara{Influence of context size $\|\mathcal{S}\|$}
Figure \ref{fig:exp:ssstudy} shows that \emph{\tychetrain} is capable of leveraging the increased context size to improve its best candidate Dice score, and that a context size of 16 is sufficient to achieve most of the gain. 



\begin{table}
\centering
\rowcolors{2}{white}{gray!15}
\begin{tabular}{cccccc}
      Blob & SetBlock & Std  & Max. DSC($\uparrow$)  & $GED^2$($\downarrow$)\\\hline
       \xmark & & & $ 0.810 \pm 0.01$ & $ 0.349 \pm 0.04$  \\
      & \xmark & &    $ 0.771 \pm 0.02$ & $ 0.425 \pm 0.05$ \\
      & & \checkmark &    $ 0.802 \pm 0.01$ & $ 0.425 \pm 0.05$ \\
      \checkmark &  \checkmark &  & $ \bm{0.811 \pm 0.01}$ & $ \bm{0.298 \pm 0.03}$ \\\bottomrule

\end{tabular}
\rowcolors{2}{white}{gray!15}
\begin{tabular}{ccccc}
Target & CS & CS+ & Max. DSC($\uparrow$)  & $GED^2$ ($\downarrow$)\\\hline
      \checkmark &  & & $ 0.776 \pm 0.02$ & $ 0.477 \pm 0.04$\\
       & \checkmark & &   $ 0.700 \pm 0.02$ & $ 0.410 \pm 0.05$\\
               & & \checkmark & $ 0.561 \pm 0.02$  & $ 0.867 \pm 0.05$ \\
      \checkmark & \checkmark &  & $\bm{0.813 \pm 0.01}$ &  $ \bm{0.333 \pm 0.04}$\\

        \checkmark & \checkmark & \checkmark & $ 0.808 \pm 0.01$ & $ 0.358 \pm 0.04$\\\bottomrule
    \end{tabular}
    \caption{\textbf{Ablation Study for Tyche variants.} Top: \emph{\tychetrain}, without simulated multi-annotator data, with \emph{SetBlock}, with Standard Deviation in \emph{SetBlock}. Bottom: \emph{Tyche TeS}, with Target, Context and Large Context augmentations.}
    \label{fig:exp:ablation}
\end{table}

%
\subpara{Ablation}
Table \ref{fig:exp:ablation} illustrates several ablations on \emph{Tyche} design choices. For \emph{\tychetrain}, we evaluate the following variants: no simulated multi-annotator images, no \emph{SetBlock} and finally, using the standard deviation of candidate feature representations in addition to the mean in the \emph{SetBlock} (``Std''). We compare the models using best candidate Dice averaged across tasks. 

Table \ref{fig:exp:ablation} shows that the simulated multi-annotator data provides negligible improvement, as does adding the standard deviation. However, \emph{SetBlock} is a crucial part to improve the best candidate Dice score.

 We study performances for three types of TTA in \emph{\tychetest}: on the target, on the context (CS), and on the context while also including the non-augmented context (CS+): $(S, \mathcal{G}(S))$. Table \ref{fig:exp:ablation} shows that adding noise to only one of the target and context yields sub-optimal performance, while augmenting both the target and context improves performances.

\subsection{Inference Runtime}
We compare the inference runtime by predicting 8 segmentation candidates with each method, and repeat the process 300 times. We use an NVIDIA V100 GPU. Table \ref{tab:exp:inftime} shows that \emph{Tyche} is significantly faster and smaller than SegGPT and CIDM, yet, not as fast as some task-specific stochastic models. \emph{\tychetest} has fewer parameters than \emph{\tychetrain}, but needs additional inference time.

\begin{table}
\centering
\rowcolors{2}{white}{gray!15}
    \begin{tabular}{lcc}
         &  Inference Time (ms)  & Parameters\\\hline
    UniverSeg & $96.62 \pm 0.61$ & $1.2$M \\
    SegGPT & $2857.19 \pm 4.38$ &  $370$M \\
    SENet & $14.91 \pm 0.21$ & $0.89$M\\
    FT-SAM & $1036.75\pm4.61$ & $94$M \\
    SAM-Med2D & $188.8 \pm 7.58$ &  $91$M \\
    PhiSeg & $11.35 \pm 0.672 $ & $21.1$M \\
    ProbaUNet & $8.44 \pm 0.46$ & $5$M\\
    CIDM & $1.7\times10^5\pm 2748$ & $85.6$M\\\hline
    \textbf{\tychetest} & $128.57 \pm 2.626 $ & $1.2$M \\\hline
    \textbf{\tychetrain} & $18.09 \pm 0.61$ & $1.7$M \\\hline
    \end{tabular}
    \caption{\textbf{Inference Runtime and Model Parameters} for 8 predictions and a context size of 16. }
    \label{tab:exp:inftime}
\end{table}

\section{Conclusion}
We introduced \emph{Tyche}, the first framework for stochastic in-context segmentation. For any (new) segmentation task, \emph{Tyche} can directly produce diverse segmentation candidates, from which practitioners can select the most suitable one, and draw a better understanding of the underlying uncertainty. \emph{Tyche} can generalize to images from data unseen at training and outperforms in-context and interactive benchmarks. In addition, \emph{Tyche} often matches stochastic models on tasks for which those models have been specifically trained. \emph{Tyche} has two variants, one designed to optimize the best segmentation candidate, with fast inference time, and a test-time augmentation variant that can be used in combination with existing in-context learning methods.
We are excited to further study the different types of uncertainty captured by \emph{\tychetrain} and \emph{\tychetest}. We will also extend the capabilities of \emph{Tyche} with more complex support sets, including variable annotators and multiple image modalities.

\section{Acknowledgement}
Research reported in this paper was supported by the National Institute of Biomedical Imaging and Bioengineering of the National Institutes of Health under award number R01EB033773. This work was also supported in part by funding from the Eric and Wendy Schmidt Center at the Broad Institute of MIT and Harvard as well as Quanta Computer Inc. Finally, some of the computation resources required for this research was performed on computational hardware generously provided by the Massachusetts Life Sciences Center.

{
    \small
    \bibliographystyle{ieeenat_fullname}
    \bibliography{main, megamedical}

@String(BMVC= {Brit. Mach. Vis. Conf.})

@String(BMVC  =	{BMVC})

@article{armato2011lung,
  title={The lung image database consortium (LIDC) and image database resource initiative (IDRI): a completed reference database of lung nodules on CT scans},
  author={Armato III, Samuel G and et al.},
  journal={Medical physics},
  volume={38},
  number={2},
  pages={915--931},
  year={2011},
  publisher={Wiley Online Library}
}

@inproceedings{ayhan2022test,
  title={Test-time data augmentation for estimation of heteroscedastic aleatoric uncertainty in deep neural networks},
  author={Ayhan, Murat Seckin and Berens, Philipp},
  booktitle={Medical Imaging with Deep Learning},
  year={2022}
}

@article{badrinarayanan2017segnet,
  title={Segnet: A deep convolutional encoder-decoder architecture for image segmentation},
  author={Badrinarayanan, Vijay and et al.},
  journal={IEEE transactions on pattern analysis and machine intelligence},
  volume={39},
  number={12},
  pages={2481--2495},
  year={2017},
  publisher={IEEE}
}

@article{balavzevic2023towards,
  title={Towards In-context Scene Understanding},
  author={Bala{\v{z}}evi{\'c}, Ivana and et al.},
  journal={arXiv preprint arXiv:2306.01667},
  year={2023}
}

@inproceedings{baumgartner2019phiseg,
  title={Phiseg: Capturing uncertainty in medical image segmentation},
  author={Baumgartner, Christian F and Tezcan, Kerem C and Chaitanya, Krishna and H{\"o}tker, Andreas M and Muehlematter, Urs J and Schawkat, Khoschy and Becker, Anton S and Donati, Olivio and Konukoglu, Ender},
  booktitle={Medical Image Computing and Computer Assisted Intervention--MICCAI 2019: 22nd International Conference, Shenzhen, China, October 13--17, 2019, Proceedings, Part II 22},
  pages={119--127},
  year={2019},
  organization={Springer}
}

@article{becker2019variability,
  title={Variability of manual segmentation of the prostate in axial T2-weighted MRI: a multi-reader study},
  author={Becker, Anton S and Chaitanya, Krishna and Schawkat, Khoschy and Muehlematter, Urs J and H{\"o}tker, Andreas M and Konukoglu, Ender and Donati, Olivio F},
  journal={European journal of radiology},
  volume={121},
  pages={108716},
  year={2019},
  publisher={Elsevier}
}

@article{bellemare2017cramer,
  title={The cramer distance as a solution to biased wasserstein gradients},
  author={Bellemare, Marc G and Danihelka, Ivo and Dabney, Will and Mohamed, Shakir and Lakshminarayanan, Balaji and Hoyer, Stephan and Munos, R{\'e}mi},
  journal={arXiv preprint arXiv:1705.10743},
  year={2017}
}

@inproceedings{bhat2022generalized,
  title={Generalized probabilistic u-net for medical image segementation},
  author={Bhat, Ishaan and Pluim, Josien PW and Kuijf, Hugo J},
  booktitle={International Workshop on Uncertainty for Safe Utilization of Machine Learning in Medical Imaging},
  pages={113--124},
  year={2022},
  organization={Springer}
}

@article{butoi2023universeg,
  title={UniverSeg: Universal Medical Image Segmentation},
  author={Butoi, Victor Ion and Ortiz, Jose Javier Gonzalez and Ma, Tianyu and Sabuncu, Mert R and Guttag, John and Dalca, Adrian V},
  journal={arXiv preprint arXiv:2304.06131},
  year={2023}
}

@inproceedings{charpiat2008automatic,
  title={Automatic image colorization via multimodal predictions},
  author={Charpiat, Guillaume and Hofmann, Matthias and Sch{\"o}lkopf, Bernhard},
  booktitle={Computer Vision--ECCV 2008: 10th European Conference on Computer Vision, Marseille, France, October 12-18, 2008, Proceedings, Part III 10},
  pages={126--139},
  year={2008},
  organization={Springer}
}

@misc{cheng_sam-med2d_2023,
	title = {{SAM}-{Med2D}},
	url = {http://arxiv.org/abs/2308.16184},
	language = {en},
	urldate = {2023-09-03},
	publisher = {arXiv},
	author = {Cheng, Junlong and Ye, Jin and Deng, Zhongying and Chen, Jianpin and Li, Tianbin and Wang, Haoyu and Su, Yanzhou and Huang, Ziyan and Chen, Jilong and Jiang, Lei and Sun, Hui and He, Junjun and Zhang, Shaoting and Zhu, Min and Qiao, Yu},
	month = aug,
	year = {2023},
	note = {arXiv:2308.16184 [cs]},
}

@inproceedings{czolbe2021segmentation,
  title={Is segmentation uncertainty useful?},
  author={Czolbe, Steffen and Arnavaz, Kasra and Krause, Oswin and Feragen, Aasa},
  booktitle={Information Processing in Medical Imaging: 27th International Conference, IPMI 2021, Virtual Event, June 28--June 30, 2021, Proceedings 27},
  pages={715--726},
  year={2021},
  organization={Springer}
}

@article{der2009aleatory,
  title={Aleatory or epistemic? Does it matter?},
  author={Der Kiureghian, Armen and Ditlevsen, Ove},
  journal={Structural safety},
  volume={31},
  number={2},
  pages={105--112},
  year={2009},
  publisher={Elsevier}
}

@article{dice1945measures,
  title={Measures of the amount of ecologic association between species},
  author={Dice, Lee R},
  journal={Ecology},
  volume={26},
  number={3},
  pages={297--302},
  year={1945},
  publisher={JSTOR}
}

@inproceedings{ding2023few,
  title={Few-shot medical image segmentation with cycle-resemblance attention},
  author={Ding, Hao and Sun, Changchang and Tang, Hao and Cai, Dawen and Yan, Yan},
  booktitle={Proceedings of the IEEE/CVF Winter Conference on Applications of Computer Vision},
  pages={2488--2497},
  year={2023}
}

@inproceedings{shanmugam2021better,
  title={Better aggregation in test-time augmentation},
  author={Shanmugam, Divya and Blalock, Davis and Balakrishnan, Guha and Guttag, John},
  booktitle={Proceedings of the IEEE/CVF international conference on computer vision},
  pages={1214--1223},
  year={2021}
}

@inproceedings{finn2017model,
  title={Model-agnostic meta-learning for fast adaptation of deep networks},
  author={Finn, Chelsea and Abbeel, Pieter and Levine, Sergey},
  booktitle={International conference on machine learning},
  pages={1126--1135},
  year={2017},
  organization={PMLR}
}

@article{guzman2012multiple,
  title={Multiple choice learning: Learning to produce multiple structured outputs},
  author={Guzman-Rivera, Abner and Batra, Dhruv and Kohli, Pushmeet},
  journal={Advances in neural information processing systems},
  volume={25},
  year={2012}
}

@article{hoffmann2021synthmorph,
  title={SynthMorph: learning contrast-invariant registration without acquired images},
  author={Hoffmann, Malte and Billot, Benjamin and Greve, Douglas N and Iglesias, Juan Eugenio and Fischl, Bruce and Dalca, Adrian V},
  journal={IEEE transactions on medical imaging},
  volume={41},
  number={3},
  pages={543--558},
  year={2021},
  publisher={IEEE}
}

@article{hong2021hypernet,
  title={Hypernet-ensemble learning of segmentation probability for medical image segmentation with ambiguous labels},
  author={Hong, Sungmin and Bonkhoff, Anna K and Hoopes, Andrew and Bretzner, Martin and Schirmer, Markus D and Giese, Anne-Katrin and Dalca, Adrian V and Golland, Polina and Rost, Natalia S},
  journal={arXiv preprint arXiv:2112.06693},
  year={2021}
}

@article{hu2023inter,
  title={Inter-Rater Uncertainty Quantification in Medical Image Segmentation via Rater-Specific Bayesian Neural Networks},
  author={Hu, Qingqiao and Wang, Hao and Luo, Jing and Luo, Yunhao and Zhangg, Zhiheng and Kirschke, Jan S and Wiestler, Benedikt and Menze, Bjoern and Zhang, Jianguo and Li, Hongwei Bran},
  journal={arXiv preprint arXiv:2306.16556},
  year={2023}
}

@article{isensee2021nnu,
  title={nnU-Net: a self-configuring method for deep learning-based biomedical image segmentation},
  author={Isensee, Fabian and Jaeger, Paul F and Kohl, Simon AA and Petersen, Jens and Maier-Hein, Klaus H},
  journal={Nature methods},
  volume={18},
  number={2},
  pages={203--211},
  year={2021},
  publisher={Nature Publishing Group US New York}
}

@article{kuhn1955hungarian,
  title={The Hungarian method for the assignment problem},
  author={Kuhn, Harold W},
  journal={Naval research logistics quarterly},
  volume={2},
  number={1-2},
  pages={83--97},
  year={1955},
  publisher={Wiley Online Library}
}

@misc{NerveUS,
    author = {Montoya, Anna and Hasnin and kaggle446 and shirzad and Cukierski, Will and yffud},
    title = {Ultrasound Nerve Segmentation},
    publisher = {Kaggle},
    year = {2016},
    url = {https://kaggle.com/competitions/ultrasound-nerve-segmentation}
}

@article{CT2US,
	title = {{CT2US}: {Cross}-modal transfer learning for kidney segmentation in ultrasound images with synthesized data},
	volume = {122},
	issn = {0041-624X},
	url = {https://www.sciencedirect.com/science/article/pii/S0041624X22000191},
	doi = {https://doi.org/10.1016/j.ultras.2022.106706},
	journal = {Ultrasonics},
	author = {Song, Yuxin and Zheng, Jing and Lei, Long and Ni, Zhipeng and Zhao, Baoliang and Hu, Ying},
	year = {2022},
	pages = {106706},
}

@article{SegThy,
  title={Tracked 3D ultrasound and deep neural network-based thyroid segmentation reduce interobserver variability in thyroid volumetry},
  author={Kr{\"o}nke, Markus and Eilers, Christine and Dimova, Desislava and K{\"o}hler, Melanie and Buschner, Gabriel and Schweiger, Lilit and Konstantinidou, Lemonia and Makowski, Marcus and Nagarajah, James and Navab, Nassir and others},
  journal={Plos one},
  volume={17},
  number={7},
  pages={e0268550},
  year={2022},
  publisher={Public Library of Science San Francisco, CA USA}
}

@misc{DukeLiver,
	title = {Duke liver dataset ({MRI}) v2},
	url = {https://doi.org/10.5281/zenodo.7774566},
	doi = {10.5281/zenodo.7774566},
	publisher = {Zenodo},
	author = {Macdonald, Jacob A. and Zhu, Zhe and Konkel, Brandon and Mazurowski, Maciej and Wiggins, Walter and Bashir, Mustafa},
	month = apr,
	year = {2023},
}

@article{PanNuke,
  title={PanNuke dataset extension, insights and baselines. arXiv. 2020 doi: 10.48550},
  author={Gamper, J and Koohbanani, NA and Benes, K and Graham, S and Jahanifar, M and Khurram, SA and Azam, A and Hewitt, K and Rajpoot, N},
  journal={ARXIV},
  year={2003}
}

@inproceedings{islam2021spatially,
  title={Spatially varying label smoothing: Capturing uncertainty from expert annotations},
  author={Islam, Mobarakol and Glocker, Ben},
  booktitle={Information Processing in Medical Imaging: 27th International Conference, IPMI 2021, Virtual Event, June 28--June 30, 2021, Proceedings 27},
  pages={677--688},
  year={2021},
  organization={Springer}
}

@article{joskowicz2019inter,
  title={Inter-observer variability of manual contour delineation of structures in CT},
  author={Joskowicz, Leo and Cohen, D and Caplan, N and Sosna, Jacob},
  journal={European radiology},
  volume={29},
  pages={1391--1399},
  year={2019},
  publisher={Springer}
}

@article{kendall2015bayesian,
  title={Bayesian segnet: Model uncertainty in deep convolutional encoder-decoder architectures for scene understanding},
  author={Kendall, Alex and Badrinarayanan, Vijay and Cipolla, Roberto},
  journal={arXiv preprint arXiv:1511.02680},
  year={2015}
}

@article{kendall2017uncertainties,
  title={What uncertainties do we need in bayesian deep learning for computer vision?},
  author={Kendall, Alex and Gal, Yarin},
  journal={Advances in neural information processing systems},
  volume={30},
  year={2017}
}

@article{kim2023universal,
  title={Universal few-shot learning of dense prediction tasks with visual token matching},
  author={Kim, Donggyun and Kim, Jinwoo and Cho, Seongwoong and Luo, Chong and Hong, Seunghoon},
  journal={arXiv preprint arXiv:2303.14969},
  year={2023}
}

@article{kirillov2023segment,
  title={Segment anything},
  author={Kirillov, Alexander and et al.},
  journal={arXiv preprint arXiv:2304.02643},
  year={2023}
}

@article{kohl2018probabilistic,
  title={A probabilistic u-net for segmentation of ambiguous images},
  author={Kohl, Simon and Romera-Paredes, Bernardino and Meyer, Clemens and De Fauw, Jeffrey and Ledsam, Joseph R and Maier-Hein, Klaus and Eslami, SM and Jimenez Rezende, Danilo and Ronneberger, Olaf},
  journal={Advances in neural information processing systems},
  volume={31},
  year={2018}
}

@article{kohl2019hierarchical,
  title={A hierarchical probabilistic u-net for modeling multi-scale ambiguities},
  author={Kohl, Simon AA and Romera-Paredes, Bernardino and Maier-Hein, Klaus H and Rezende, Danilo Jimenez and Eslami, SM and Kohli, Pushmeet and Zisserman, Andrew and Ronneberger, Olaf},
  journal={arXiv preprint arXiv:1905.13077},
  year={2019}
}

@inproceedings{lambert2023triadnet,
	title = {TriadNet: Sampling-Free Predictive Intervals for Lesional Volume in 3D Brain MR Images},
	shorttitle = {TriadNet},
	booktitle = {Uncertainty for Safe Utilization of Machine Learning in Medical Imaging – {MICCAI} 2023},
	publisher = {Springer Nature Switzerland},
	author = {Lambert, Benjamin and Forbes, Florence and Doyle, Senan and Dojat, Michel},
	year = {2023},
}

@article{larrazabal2021maximum,
  title={Maximum Entropy on Erroneous Predictions (MEEP): Improving model calibration for medical image segmentation},
  author={Larrazabal, Agostina and Martinez, Cesar and Dolz, Jose and Ferrante, Enzo},
  journal={arXiv preprint arXiv:2112.12218},
  year={2021}
}

@inproceedings{li2018interactive,
  title={Interactive image segmentation with latent diversity},
  author={Li, Zhuwen and Chen, Qifeng and Koltun, Vladlen},
  booktitle={Proceedings of the IEEE Conference on Computer Vision and Pattern Recognition},
  pages={577--585},
  year={2018}
}

@article{li2022prototypical,
  title={Prototypical few-shot segmentation for cross-institution male pelvic structures with spatial registration},
  author={Li, Yiwen and Fu, Yunguan and Gayo, Iani and Yang, Qianye and Min, Zhe and Saeed, Shaheer and Yan, Wen and Wang, Yipei and Noble, J Alison and Emberton, Mark and others},
  journal={arXiv preprint arXiv:2209.05160},
  year={2022}
}

@article{matsunaga2017image,
  title={Image classification of melanoma, nevus and seborrheic keratosis by deep neural network ensemble},
  author={Matsunaga, Kazuhisa and Hamada, Akira and Minagawa, Akane and Koga, Hiroshi},
  journal={arXiv preprint arXiv:1703.03108},
  year={2017}
}

@inproceedings{milletari2016v,
  title={V-net: Fully convolutional neural networks for volumetric medical image segmentation},
  author={Milletari, Fausto and Navab, Nassir and Ahmadi, Seyed-Ahmad},
  booktitle={2016 fourth international conference on 3D vision (3DV)},
  pages={565--571},
  year={2016},
  organization={Ieee}
}

@article{monteiro2020stochastic,
  title={Stochastic segmentation networks: Modelling spatially correlated aleatoric uncertainty},
  author={Monteiro, Miguel and Le Folgoc, Lo{\"\i}c and Coelho de Castro, Daniel and Pawlowski, Nick and Marques, Bernardo and Kamnitsas, Konstantinos and van der Wilk, Mark and Glocker, Ben},
  journal={Advances in Neural Information Processing Systems},
  volume={33},
  pages={12756--12767},
  year={2020}
}

@inproceedings{nguyen2019feature,
  title={Feature weighting and boosting for few-shot segmentation},
  author={Nguyen, Khoi and Todorovic, Sinisa},
  booktitle={Proceedings of the IEEE/CVF International Conference on Computer Vision},
  pages={622--631},
  year={2019}
}

@article{nichol2018first,
  title={On first-order meta-learning algorithms},
  author={Nichol, Alex and Achiam, Joshua and Schulman, John},
  journal={arXiv preprint arXiv:1803.02999},
  year={2018}
}

@article{nichyporuk2022rethinking,
  title={Rethinking Generalization: The Impact of Annotation Style on Medical Image Segmentation},
  author={Nichyporuk, Brennan and Cardinell, Jillian and Szeto, Justin and Mehta, Raghav and Falet, Jean-Pierre R and Arnold, Douglas L and Tsaftaris, Sotirios A and Arbel, Tal},
  journal={arXiv preprint arXiv:2210.17398},
  year={2022}
}

@article{pandey2023robust,
  title={Robust Prototypical Few-Shot Organ Segmentation with Regularized Neural-ODEs},
  author={Pandey, Prashant and Chasmai, Mustafa and Sur, Tanuj and Lall, Brejesh},
  journal={IEEE Transactions on Medical Imaging},
  year={2023},
  publisher={IEEE}
}

@inproceedings{wolleb_diffusion_2021,
	title = {Diffusion {Models} for {Implicit} {Image} {Segmentation} {Ensembles}},
	author = {Wolleb, Julia and Sandkühler, Robin and Bieder, 
                  Florentin and Valmaggia, Philippe and Cattin, Philippe C.},
	year = {2021},
        booktitle = {Medical Imaging with Deep Learning}
}

@inproceedings{rahman2023ambiguous,
  title={Ambiguous medical image segmentation using diffusion models},
  author={Rahman, Aimon and Valanarasu, Jeya Maria Jose and Hacihaliloglu, Ilker and Patel, Vishal M},
  booktitle={Proceedings of the IEEE/CVF Conference on Computer Vision and Pattern Recognition},
  pages={11536--11546},
  year={2023}
}

@inproceedings{ronneberger_u-net_2015,
	title = {U-{Net}: {Convolutional} {Networks} for {Biomedical} {Image} {Segmentation}},
	shorttitle = {U-{Net}},
	booktitle = {Medical {Image} {Computing} and {Computer} {Assisted} {Intervention} – {MICCAI} 2015},
	publisher = {Springer International Publishing},
	author = {Ronneberger, Olaf and Fischer, Philipp and Brox, Thomas},
	year = {2015},
}

@article{salimans2018improving,
  title={Improving GANs using optimal transport},
  author={Salimans, Tim and Zhang, Han and Radford, Alec and Metaxas, Dimitris},
  journal={arXiv preprint arXiv:1803.05573},
  year={2018}
}

@article{seo2022task,
  title={Task-Adaptive Feature Transformer with Semantic Enrichment for Few-Shot Segmentation},
  author={Seo, Jun and Park, Young-Hyun and Yoon, Sung Whan and Moon, Jaekyun},
  journal={arXiv preprint arXiv:2202.06498},
  year={2022}
}

@article{shen2022q,
  title={Q-net: Query-informed few-shot medical image segmentation},
  author={Shen, Qianqian and Li, Yanan and Jin, Jiyong and Liu, Bin},
  journal={arXiv preprint arXiv:2208.11451},
  year={2022}
}

@inproceedings{schmidt2023probabilistic,
  title={Probabilistic Modeling of Inter-and Intra-observer Variability in Medical Image Segmentation},
  author={Schmidt, Arne and Morales-{\'A}lvarez, Pablo and Molina, Rafael},
  booktitle={Proceedings of the IEEE/CVF International Conference on Computer Vision},
  pages={21097--21106},
  year={2023}
}

@article{snell2017prototypical,
  title={Prototypical networks for few-shot learning},
  author={Snell, Jake and Swersky, Kevin and Zemel, Richard},
  journal={Advances in neural information processing systems},
  volume={30},
  year={2017}
}

@article{szekely2013energy,
  title={Energy statistics: A class of statistics based on distances},
  author={Sz{\'e}kely, G{\'a}bor J and Rizzo, Maria L},
  journal={Journal of statistical planning and inference},
  volume={143},
  number={8},
  pages={1249--1272},
  year={2013},
  publisher={Elsevier}
}

@article{vinyals2016matching,
  title={Matching networks for one shot learning},
  author={Vinyals, Oriol and Blundell, Charles and Lillicrap, Timothy and Wierstra, Daan and others},
  journal={Advances in neural information processing systems},
  volume={29},
  year={2016}
}

@inproceedings{tanno2019learning,
  title={Learning from noisy labels by regularized estimation of annotator confusion},
  author={Tanno, Ryutaro and Saeedi, Ardavan and Sankaranarayanan, Swami and Alexander, Daniel C and Silberman, Nathan},
  booktitle={Proceedings of the IEEE/CVF conference on computer vision and pattern recognition},
  pages={11244--11253},
  year={2019}
}

@inproceedings{wang2023images,
  title={Images speak in images: A generalist painter for in-context visual learning},
  author={Wang, Xinlong and Wang, Wen and Cao, Yue and Shen, Chunhua and Huang, Tiejun},
  booktitle={Proceedings of the IEEE/CVF Conference on Computer Vision and Pattern Recognition},
  pages={6830--6839},
  year={2023}
}

@inproceedings{wang2023seggpt,
  title={SegGPT: Towards Segmenting Everything in Context},
  author={Wang, Xinlong and Zhang, Xiaosong and Cao, Yue and Wang, Wen and Shen, Chunhua and Huang, Tiejun},
  booktitle={Proceedings of the IEEE/CVF International Conference on Computer Vision},
  pages={1130--1140},
  year={2023}
}

@article{wong2023scribbleprompt,
  title={ScribblePrompt: Fast and Flexible Interactive Segmentation for Any Medical Image},
  author={Wong, Hallee E and Rakic, Marianne and Guttag, John and Dalca, Adrian V},
  journal={arXiv preprint arXiv:2312.07381},
  year={2023}
}

@article{ye2023confidence,
  title={Confidence Contours: Uncertainty-Aware Annotation for Medical Semantic Segmentation},
  author={Ye, Andre and Chen, Quan Ze and Zhang, Amy},
  journal={arXiv preprint arXiv:2308.07528},
  year={2023}
}

@inproceedings{zbinden2023stochastic,
  title={Stochastic segmentation with conditional categorical diffusion models},
  author={Zbinden, Lukas and Doorenbos, Lars and Pissas, Theodoros and Huber, Adrian Thomas and Sznitman, Raphael and M{\'a}rquez-Neila, Pablo},
  booktitle={Proceedings of the IEEE/CVF International Conference on Computer Vision},
  pages={1119--1129},
  year={2023}
}

@inproceedings{zhang2019canet,
  title={Canet: Class-agnostic segmentation networks with iterative refinement and attentive few-shot learning},
  author={Zhang, Chi and Lin, Guosheng and Liu, Fayao and Yao, Rui and Shen, Chunhua},
  booktitle={Proceedings of the IEEE/CVF conference on computer vision and pattern recognition},
  pages={5217--5226},
  year={2019}
}

@article{ResNetTTAColon,
  title={A comprehensive study on colorectal polyp segmentation with ResUNet++, conditional random field and test-time augmentation},
  author={Jha, Debesh and Smedsrud, Pia H and Johansen, Dag and de Lange, Thomas and Johansen, H{\aa}vard D and Halvorsen, P{\aa}l and Riegler, Michael A},
  journal={IEEE journal of biomedical and health informatics},
  volume={25},
  number={6},
  pages={2029--2040},
  year={2021},
  publisher={IEEE}
}

@article{TTA2019aleatoric,
  title={Aleatoric uncertainty estimation with test-time augmentation for medical image segmentation with convolutional neural networks},
  author={Wang, Guotai and Li, Wenqi and Aertsen, Michael and Deprest, Jan and Ourselin, S{\'e}bastien and Vercauteren, Tom},
  journal={Neurocomputing},
  volume={338},
  pages={34--45},
  year={2019},
  publisher={Elsevier}
}

@inproceedings{wang2019automatic,
  title={Automatic brain tumor segmentation using convolutional neural networks with test-time augmentation},
  author={Wang, Guotai and Li, Wenqi and Ourselin, S{\'e}bastien and Vercauteren, Tom},
  booktitle={Brainlesion: Glioma, Multiple Sclerosis, Stroke and Traumatic Brain Injuries: 4th International Workshop, BrainLes 2018, Held in Conjunction with MICCAI 2018, Granada, Spain, September 16, 2018, Revised Selected Papers, Part II 4},
  pages={61--72},
  year={2019},
  organization={Springer}
}

@article{cohen2023boosting,
  title={Boosting anomaly detection using unsupervised diverse test-time augmentation},
  author={Cohen, Seffi and Goldshlager, Niv and Rokach, Lior and Shapira, Bracha},
  journal={Information Sciences},
  volume={626},
  pages={821--836},
  year={2023},
  publisher={Elsevier}
}

@article{amiri2020two,
  title={Two-stage ultrasound image segmentation using U-Net and test time augmentation},
  author={Amiri, Mina and Brooks, Rupert and Behboodi, Bahareh and Rivaz, Hassan},
  journal={International journal of computer assisted radiology and surgery},
  volume={15},
  pages={981--988},
  year={2020},
  publisher={Springer}
}

@article{moshkov2020test,
  title={Test-time augmentation for deep learning-based cell segmentation on microscopy images},
  author={Moshkov, Nikita and Mathe, Botond and Kertesz-Farkas, Attila and Hollandi, Reka and Horvath, Peter},
  journal={Scientific reports},
  volume={10},
  number={1},
  pages={5068},
  year={2020},
  publisher={Nature Publishing Group UK London}
}

@inproceedings{huang2021style,
  title={Style-invariant cardiac image segmentation with test-time augmentation},
  author={Huang, Xiaoqiong and Chen, Zejian and Yang, Xin and Liu, Zhendong and Zou, Yuxin and Luo, Mingyuan and Xue, Wufeng and Ni, Dong},
  booktitle={Statistical Atlases and Computational Models of the Heart. M\&Ms and EMIDEC Challenges: 11th International Workshop, STACOM 2020, Held in Conjunction with MICCAI 2020, Lima, Peru, October 4, 2020, Revised Selected Papers 11},
  pages={305--315},
  year={2021},
  organization={Springer}
}

@inproceedings{melanieTAAL,
    author = {Gaillochet, Mélanie and Desrosiers, Christian and Lombaert, Hervé},
    title = {TAAL: Test-Time Augmentation for Active Learning in Medical Image Segmentation},
    booktitle = {Data Augmentation, Labelling, and Imperfections},
    year = {2022},
    publisher={Springer Nature Switzerland},
}

@inproceedings{sinha2019variational,
  title={Variational adversarial active learning},
  author={Sinha, Samarth and Ebrahimi, Sayna and Darrell, Trevor},
  booktitle={Proceedings of the IEEE/CVF International Conference on Computer Vision},
  pages={5972--5981},
  year={2019}
}

@inproceedings{kim2021task,
  title={Task-aware variational adversarial active learning},
  author={Kim, Kwanyoung and Park, Dongwon and Kim, Kwang In and Chun, Se Young},
  booktitle={Proceedings of the IEEE/CVF Conference on Computer Vision and Pattern Recognition},
  pages={8166--8175},
  year={2021}
}

@article{HipXRay,
	title = {X-ray images of the hip joints},
	volume = {1},
	url = {https://data.mendeley.com/datasets/zm6bxzhmfz/1},
	doi = {10.17632/zm6bxzhmfz.1},
	language = {en},
	urldate = {2023-09-03},
	author = {Gut, Daniel},
	month = jul,
	year = {2021},
	note = {Publisher: Mendeley Data},
}

@inproceedings{PAXRay,
    author    = {Seibold,Constantin and Reiß,Simon and Sarfraz,Saquib and Fink,Matthias A. and Mayer,Victoria and Sellner,Jan and Kim,Moon Sung and Maier-Hein, Klaus H.  and Kleesiek, Jens  and Stiefelhagen,Rainer}, 
    title     = {Detailed Annotations of Chest X-Rays via CT Projection for Report Understanding}, 
    booktitle = {Proceedings of the 33th British Machine Vision Conference (BMVC)},
    year  = {2022}
}

@article{CT_ORG,
    title = {{CT}-{ORG}, a new dataset for multiple organ segmentation in computed tomography},
    volume = {7},
    issn = {2052-4463},
    url = {https://www.nature.com/articles/s41597-020-00715-8},
    doi = {10.1038/s41597-020-00715-8},
    number = {1},
    journal = {Scientific Data},
    author = {Rister, Blaine and Yi, Darvin and Shivakumar, Kaushik and Nobashi, Tomomi and Rubin, Daniel L.},
    month = nov,
    year = {2020},
    pages = {381},
}

@article{ACDC,
  title={Deep learning techniques for automatic MRI cardiac multi-structures segmentation and diagnosis: is the problem solved?},
  author={Bernard, Olivier and Lalande, Alain and Zotti, Clement and Cervenansky, Frederick and Yang, Xin and Heng, Pheng-Ann and Cetin, Irem and Lekadir, Karim and Camara, Oscar and Ballester, Miguel Angel Gonzalez and others},
  journal={IEEE transactions on medical imaging},
  volume={37},
  number={11},
  pages={2514--2525},
  year={2018},
  publisher={ieee}
}

@article{DRIVE,
  title={Ridge-based vessel segmentation in color images of the retina},
  author={Staal, Joes and Abr{\`a}moff, Michael D and Niemeijer, Meindert and Viergever, Max A and Van Ginneken, Bram},
  journal={IEEE transactions on medical imaging},
  volume={23},
  number={4},
  pages={501--509},
  year={2004},
  publisher={IEEE}
}

@article{FeTA,
  title={An automatic multi-tissue human fetal brain segmentation benchmark using the fetal tissue annotation dataset},
  author={Payette, Kelly and de Dumast, Priscille and Kebiri, Hamza and Ezhov, Ivan and Paetzold, Johannes C and Shit, Suprosanna and Iqbal, Asim and Khan, Romesa and Kottke, Raimund and Grehten, Patrice and others},
  journal={Scientific Data},
  volume={8},
  number={1},
  pages={1--14},
  year={2021},
  publisher={Nature Publishing Group}
}

@article{LiTS,
  title={The liver tumor segmentation benchmark (lits)},
  author={Bilic, Patrick and Christ, Patrick Ferdinand and Vorontsov, Eugene and Chlebus, Grzegorz and Chen, Hao and Dou, Qi and Fu, Chi-Wing and Han, Xiao and Heng, Pheng-Ann and Hesser, J{\"u}rgen and others},
  journal={arXiv preprint arXiv:1901.04056},
  year={2019}
}

@article{MSD,
  title={A large annotated medical image dataset for the development and evaluation of segmentation algorithms},
  author={Simpson, Amber L and Antonelli, Michela and Bakas, Spyridon and Bilello, Michel and Farahani, Keyvan and Van Ginneken, Bram and Kopp-Schneider, Annette and Landman, Bennett A and Litjens, Geert and Menze, Bjoern and others},
  journal={arXiv preprint arXiv:1902.09063},
  year={2019}
}

@article{OASIS-data,
  title={Open Access Series of Imaging Studies (OASIS): cross-sectional MRI data in young, middle aged, nondemented, and demented older adults},
  author={Marcus, Daniel S and Wang, Tracy H and Parker, Jamie and Csernansky, John G and Morris, John C and Buckner, Randy L},
  journal={Journal of cognitive neuroscience},
  volume={19},
  number={9},
  pages={1498--1507},
  year={2007},
  publisher={MIT Press One Rogers Street, Cambridge, MA 02142-1209, USA journals-info}
}

@inproceedings{OASIS-proc,
    title = "Learning the Effect of Registration Hyperparameters with HyperMorph",
    author = "Hoopes, Andrew and Hoffmann, Malte and Greve, Douglas N. and Fischl, Bruce and Guttag, John and Dalca, Adrian V.",
    journal = "Machine Learning for Biomedical Imaging",
    volume = "1",
    issue = "IPMI 2021 special issue",
    year = "2022",
    pages = "1--30",
    issn = "2766-905X",
    url = "https://melba-journal.org/2022:003"
}

@inproceedings{SegTHOR,
  title={SegTHOR: segmentation of thoracic organs at risk in CT images},
  author={Lambert, Zo{\'e} and Petitjean, Caroline and Dubray, Bernard and Kuan, Su},
  booktitle={2020 Tenth International Conference on Image Processing Theory, Tools and Applications (IPTA)},
  pages={1--6},
  year={2020},
  organization={IEEE}
}

@article{WBC,
  title={Fast and Robust Segmentation of White Blood Cell Images by Self-supervised Learning},
  author={Xin Zheng and Yong Wang and Guoyou Wang and Jianguo Liu},
  journal={Micron},
  volume={107},
  pages={55--71},
  year={2018},
  publisher={Elsevier},
  doi={https://doi.org/10.1016/j.micron.2018.01.010},
  url={https://www.sciencedirect.com/science/article/pii/S0968432817303037}
}

@article{qubiq,
    author ={Bjoern Menze and Leo Joskowicz and Spyridon Bakas and Andras Jakab and Ender Konukoglu and Anton Becker and Amber Simpson and Richard D},
    title = {Quantification of Uncertainties in Biomedical Image Quantification 2021},
    journal = {4th International Conference on Medical Image Computing and Computer Assisted Intervention (MICCAI 2021)},
    year = {2021},
    doi={https://doi.org/10.5281/zenodo.4575204},
    publisher={zenodo}
}

@article{gousias2012magnetic,
  title={Magnetic resonance imaging of the newborn brain: manual segmentation of labelled atlases in term-born and preterm infants},
  author={Gousias, Ioannis S and Edwards, A David and Rutherford, Mary A and Counsell, Serena J and Hajnal, Jo V and Rueckert, Daniel and Hammers, Alexander},
  journal={Neuroimage},
  volume={62},
  number={3},
  pages={1499--1509},
  year={2012},
  publisher={Elsevier}
}

@article{serag2012construction,
  title={Construction of a consistent high-definition spatio-temporal atlas of the developing brain using adaptive kernel regression},
  author={Serag, Ahmed and Aljabar, Paul and Ball, Gareth and Counsell, Serena J and Boardman, James P and Rutherford, Mary A and Edwards, A David and Hajnal, Joseph V and Rueckert, Daniel},
  journal={Neuroimage},
  volume={59},
  number={3},
  pages={2255--2265},
  year={2012},
  publisher={Elsevier}
}

@article{BrainDevelopment,
  title={A dynamic 4D probabilistic atlas of the developing brain},
  author={Kuklisova-Murgasova, Maria and Aljabar, Paul and Srinivasan, Latha and Counsell, Serena J and Doria, Valentina and Serag, Ahmed and Gousias, Ioannis S and Boardman, James P and Rutherford, Mary A and Edwards, A David and others},
  journal={NeuroImage},
  volume={54},
  number={4},
  pages={2750--2763},
  year={2011},
  publisher={Elsevier}
}

@article{BrainDevFetal,
  title={Automatic segmentation of brain MRIs of 2-year-olds into 83 regions of interest},
  author={Gousias, Ioannis S and Rueckert, Daniel and Heckemann, Rolf A and Dyet, Leigh E and Boardman, James P and Edwards, A David and Hammers, Alexander},
  journal={Neuroimage},
  volume={40},
  number={2},
  pages={672--684},
  year={2008},
  publisher={Elsevier}
}

@article{STARE,
  title={Locating blood vessels in retinal images by piecewise threshold probing of a matched filter response},
  author={Hoover, AD and Kouznetsova, Valentina and Goldbaum, Michael},
  journal={IEEE Transactions on Medical imaging},
  volume={19},
  number={3},
  pages={203--210},
  year={2000},
  publisher={IEEE}
}

@article{AMOS,
  title={AMOS: A Large-Scale Abdominal Multi-Organ Benchmark for Versatile Medical Image Segmentation},
  author={Ji, Yuanfeng and Bai, Haotian and Yang, Jie and Ge, Chongjian and Zhu, Ye and Zhang, Ruimao and Li, Zhen and Zhang, Lingyan and Ma, Wanling and Wan, Xiang and others},
  journal={arXiv preprint arXiv:2206.08023},
  year={2022}
}

@article{LUNA,
  title={Validation, comparison, and combination of algorithms for automatic detection of pulmonary nodules in computed tomography images: the LUNA16 challenge},
  author={Setio, Arnaud Arindra Adiyoso and Traverso, Alberto and De Bel, Thomas and Berens, Moira SN and Van Den Bogaard, Cas and Cerello, Piergiorgio and Chen, Hao and Dou, Qi and Fantacci, Maria Evelina and Geurts, Bram and others},
  journal={Medical image analysis},
  volume={42},
  pages={1--13},
  year={2017},
  publisher={Elsevier}
}

@article{OCTA500,
  title={Ipn-v2 and octa-500: Methodology and dataset for retinal image segmentation},
  author={Li, Mingchao and Zhang, Yuhan and Ji, Zexuan and Xie, Keren and Yuan, Songtao and Liu, Qinghuai and Chen, Qiang},
  journal={arXiv preprint arXiv:2012.07261},
  year={2020}
}

@article{SCD,
  title={Evaluation framework for algorithms segmenting short axis cardiac MRI},
  author={Radau, Perry and Lu, Yingli and Connelly, Kim and Paul, Gideon and Dick, AJWG and Wright, Graham},
  journal={The MIDAS Journal-Cardiac MR Left Ventricle Segmentation Challenge},
  volume={49},
  year={2009}
}

@article{I2CVB,
  title={Computer-aided detection and diagnosis for prostate cancer based on mono and multi-parametric MRI: a review},
  author={Lema{\^\i}tre, Guillaume and Mart{\'\i}, Robert and Freixenet, Jordi and Vilanova, Joan C and Walker, Paul M and Meriaudeau, Fabrice},
  journal={Computers in biology and medicine},
  volume={60},
  pages={8--31},
  year={2015},
  publisher={Elsevier}
}

@article{Promise12,
  title={Evaluation of prostate segmentation algorithms for MRI: the PROMISE12 challenge},
  author={Litjens, Geert and Toth, Robert and van de Ven, Wendy and Hoeks, Caroline and Kerkstra, Sjoerd and van Ginneken, Bram and Vincent, Graham and Guillard, Gwenael and Birbeck, Neil and Zhang, Jindang and others},
  journal={Medical image analysis},
  volume={18},
  number={2},
  pages={359--373},
  year={2014},
  publisher={Elsevier}
}

@article{BBBC038,
	title = {Nucleus segmentation across imaging experiments: the 2018 {Data} {Science} {Bowl}},
	volume = {16},
	issn = {1548-7105},
	url = {https://doi.org/10.1038/s41592-019-0612-7},
	doi = {10.1038/s41592-019-0612-7},
	number = {12},
	journal = {Nature Methods},
	author = {Caicedo, Juan C. and Goodman, Allen and Karhohs, Kyle W. and Cimini, Beth A. and Ackerman, Jeanelle and Haghighi, Marzieh and Heng, CherKeng and Becker, Tim and Doan, Minh and McQuin, Claire and Rohban, Mohammad and Singh, Shantanu and Carpenter, Anne E.},
	month = dec,
	year = {2019},
	pages = {1247--1253},
}

@article{BUID,
	title = {Dataset of breast ultrasound images},
	volume = {28},
	issn = {2352-3409},
	url = {https://www.sciencedirect.com/science/article/pii/S2352340919312181},
	doi = {https://doi.org/10.1016/j.dib.2019.104863},
	journal = {Data in Brief},
	author = {Al-Dhabyani, Walid and Gomaa, Mohammed and Khaled, Hussien and Fahmy, Aly},
	year={2020},
	pages = {104863},
}

@misc{ToothSeg,
    title = "Teeth Segmentation Dataset",
    url = {https://humansintheloop.org/resources/datasets/teeth-segmentation-dataset/},
    author = {Humans in the Loop}
}

@article{CHAOS_1,
	title = {{CHAOS} {Challenge} - combined ({CT}-{MR}) healthy abdominal organ segmentation},
	volume = {69},
	issn = {1361-8415},
	url = {https://www.sciencedirect.com/science/article/pii/S1361841520303145},
	doi = {https://doi.org/10.1016/j.media.2020.101950},
	journal = {Medical Image Analysis},
	author = {Kavur, A. Emre and Gezer, N. Sinem and Barış, Mustafa and Aslan, Sinem and Conze, Pierre-Henri and Groza, Vladimir and Pham, Duc Duy and Chatterjee, Soumick and Ernst, Philipp and Özkan, Savaş and Baydar, Bora and Lachinov, Dmitry and Han, Shuo and Pauli, Josef and Isensee, Fabian and Perkonigg, Matthias and Sathish, Rachana and Rajan, Ronnie and Sheet, Debdoot and Dovletov, Gurbandurdy and Speck, Oliver and Nürnberger, Andreas and Maier-Hein, Klaus H. and Bozdağı Akar, Gözde and Ünal, Gözde and Dicle, Oğuz and Selver, M. Alper},
	year = {2021},
	pages = {101950},
}

@misc{CHAOS_2,
  author       = {Ali Emre Kavur and M. Alper Selver and Oğuz Dicle and Mustafa Barış and  N. Sinem Gezer},
  title        = {{CHAOS - Combined (CT-MR) Healthy Abdominal Organ Segmentation Challenge Data}},
  month        = Apr,
  year         = 2019,
  publisher    = {Zenodo},
  version      = {v1.03},
  doi          = {10.5281/zenodo.3362844},
  url          = {https://doi.org/10.5281/zenodo.3362844}
}

@article{BBBC003,
  title={Annotated high-throughput microscopy image sets for validation.},
  author={Ljosa, Vebjorn and Sokolnicki, Katherine L and Carpenter, Anne E},
  journal={Nature methods},
  volume={9},
  number={7},
  pages={637--637},
  year={2012}
}

@article{Rose,
  title={ROSE: a retinal OCT-angiography vessel segmentation dataset and new model},
  author={Ma, Yuhui and Hao, Huaying and Xie, Jianyang and Fu, Huazhu and Zhang, Jiong and Yang, Jianlong and Wang, Zhen and Liu, Jiang and Zheng, Yalin and Zhao, Yitian},
  journal={IEEE Transactions on Medical Imaging},
  year={2021},
  volume={40},
  number={3},
  pages={928--939},
  doi={10.1109/TMI.2020.3042802},
  publisher={IEEE}
}

@data{Idrid,
doi = {10.21227/H25W98},
url = {https://dx.doi.org/10.21227/H25W98},
author = {Porwal, Prasanna and Pachade, Samiksha and Kamble, Ravi and Kokare, Manesh and Deshmukh, Girish and Sahasrabuddhe, Vivek and Meriaudeau, Fabrice},
publisher = {IEEE Dataport},
title = {Indian Diabetic Retinopathy Image Dataset (IDRiD)},
year = {2018} }

@article{NCI-ISBI,
  title={NCI-ISBI 2013 challenge: automated segmentation of prostate structures},
  author={Bloch, Nicholas and Madabhushi, Anant and Huisman, Henkjan and Freymann, John and Kirby, Justin and Grauer, Michael and Enquobahrie, Andinet and Jaffe, Carl and Clarke, Larry and Farahani, Keyvan},
  journal={The Cancer Imaging Archive},
  volume={370},
  number={6},
  pages={5},
  year={2015}
}

@article{ISBI_EM,
title= {ISBI Challenge: Segmentation of neuronal structures in EM stacks},
journal= {},
author= {Albert Cardon and Stephan Saalfeld and Stephan Preibisch and Benjamin Schmid and Anchi Cheng and Jim Pulokas and Pavel Tomancak and Volker Hartenstein},
year= {},
url= {},
license= {}
}

@article{Word,
  title={WORD: Revisiting Organs Segmentation in the Whole Abdominal Region},
  author={Luo, Xiangde and Liao, Wenjun and Xiao, Jianghong and Song, Tao and Zhang, Xiaofan and Li, Kang and Wang, Guotai and Zhang, Shaoting},
  journal={arXiv preprint arXiv:2111.02403},
  year={2021}
}

@article{PanDental,
  title={Automatic segmentation of mandible in panoramic x-ray},
  author={Abdi, Amir Hossein and Kasaei, Shohreh and Mehdizadeh, Mojdeh},
  journal={Journal of Medical Imaging},
  volume={2},
  number={4},
  pages={044003},
  year={2015},
  publisher={SPIE}
}

@article{SpineWeb,
  title={Evaluation and comparison of 3D intervertebral disc localization and segmentation methods for 3D T2 MR data: A grand challenge},
  author={Zheng, Guoyan and Chu, Chengwen and Belav{\`y}, Daniel L and Ibragimov, Bulat and Korez, Robert and Vrtovec, Toma{\v{z}} and Hutt, Hugo and Everson, Richard and Meakin, Judith and Andrade, Isabel L{\u{o}}pez and others},
  journal={Medical image analysis},
  volume={35},
  pages={327--344},
  year={2017},
  publisher={Elsevier}
}

@inproceedings{BUS,
  title={BUSIS: A Benchmark for Breast Ultrasound Image Segmentation},
  author={Zhang, Yingtao and Xian, Min and Cheng, Heng-Da and Shareef, Bryar and Ding, Jianrui and Xu, Fei and Huang, Kuan and Zhang, Boyu and Ning, Chunping and Wang, Ying},
  booktitle={Healthcare},
  volume={10},
  number={4},
  pages={729},
  year={2022},
  organization={MDPI}
}

@inproceedings{BTCV,
  title={Miccai multi-atlas labeling beyond the cranial vault--workshop and challenge},
  author={Landman, Bennett and Xu, Zhoubing and Igelsias, J and Styner, Martin and Langerak, T and Klein, Arno},
  booktitle={Proc. MICCAI Multi-Atlas Labeling Beyond Cranial Vault—Workshop Challenge},
  volume={5},
  pages={12},
  year={2015}
}

@article{roy2020SEnet,
  title={Squeeze \& excite guided few-shot segmentation of volumetric images},
  author={Roy, Abhijit Guha and Siddiqui, Shayan and P{\"o}lsterl, Sebastian and Navab, Nassir and Wachinger, Christian},
  journal={Medical image analysis},
  volume={59},
  pages={101587},
  year={2020},
  publisher={Elsevier}
}

@article{MMOTU,
  author    = {Qi Zhao and
               Shuchang Lyu and
               Wenpei Bai and
               Linghan Cai and
               Binghao Liu and
               Meijing Wu and
               Xiubo Sang and
               Min Yang and
               Lijiang Chen},
  title     = {A Multi-Modality Ovarian Tumor Ultrasound Image Dataset for Unsupervised
               Cross-Domain Semantic Segmentation},
  journal   = {CoRR},
  volume    = {abs/2207.06799},
  year      = {2022},
}

@inproceedings{BUSIS, 
  title={BUSIS: A Benchmark for Breast Ultrasound Image Segmentation},
  author={Zhang, Yingtao and Xian, Min and Cheng, Heng-Da and Shareef, Bryar and Ding, Jianrui and Xu, Fei and Huang, Kuan and Zhang, Boyu and Ning, Chunping and Wang, Ying},
  booktitle={Healthcare},
  volume={10},
  number={4},
  pages={729},
  year={2022},
  organization={MDPI}
}

@article{ISIC,
  author       = {Noel C. F. Codella and
                  David A. Gutman and
                  M. Emre Celebi and
                  Brian Helba and
                  Michael A. Marchetti and
                  Stephen W. Dusza and
                  Aadi Kalloo and
                  Konstantinos Liopyris and
                  Nabin K. Mishra and
                  Harald Kittler and
                  Allan Halpern},
  title        = {Skin Lesion Analysis Toward Melanoma Detection: {A} Challenge at the
                  2017 International Symposium on Biomedical Imaging (ISBI), Hosted
                  by the International Skin Imaging Collaboration {(ISIC)}},
  journal      = {CoRR},
  volume       = {abs/1710.05006},
  year         = {2017},
  url          = {http://arxiv.org/abs/1710.05006},
  eprinttype    = {arXiv},
  eprint       = {1710.05006},
  bibsource    = {dblp computer science bibliography, https://dblp.org}
}

@article{ssTEM,
    author = "Stephan Gerhard and Jan Funke and Julien Martel and Albert Cardona and Richard Fetter",
    title = "{Segmented anisotropic ssTEM dataset of neural tissue}",
journal={figshare},
  pages={0--0},
  publisher={figshare},
    year = "2013",
    month = "11",
    url = "https://figshare.com/articles/dataset/Segmented_anisotropic_ssTEM_dataset_of_neural_tissue/856713",
    doi = "10.6084/m9.figshare.856713.v1"
}

@article{CheXplanation,
  title={Deep learning saliency maps do not accurately highlight diagnostically relevant regions for medical image interpretation},
  author={Saporta, Adriel and Gui, Xiaotong and Agrawal, Ashwin and Pareek, Anuj and Truong, SQ and Nguyen, CD and Ngo, Van-Doan and Seekins, Jayne and Blankenberg, Francis G and Ng, AY and others},
  journal={MedRxiv},
  year={2021}
}

@article{CHAOS2021Challenge,
  title = {{CHAOS Challenge - combined (CT-MR) healthy abdominal organ segmentation}},
  journal = {Medical Image Analysis},
  publisher= {Elsevier BV},
  volume = {69},
  pages = {101950},
  year = {2021},
  month = Apr,
  issn = {1361-8415},
  doi = {https://doi.org/10.1016/j.media.2020.101950},
  url = {http://www.sciencedirect.com/science/article/pii/S1361841520303145},
  author = {A. Emre Kavur and N. Sinem Gezer and Mustafa Barış and Sinem Aslan and Pierre-Henri Conze and Vladimir Groza and Duc Duy Pham and Soumick Chatterjee and Philipp Ernst and Savaş Özkan and Bora Baydar and Dmitry Lachinov and Shuo Han and Josef Pauli and Fabian Isensee and Matthias Perkonigg and Rachana Sathish and Ronnie Rajan and Debdoot Sheet and Gurbandurdy Dovletov and Oliver Speck and Andreas Nürnberger and Klaus H. Maier-Hein and Gözde {Bozdağı Akar} and Gözde Ünal and Oğuz Dicle and M. Alper Selver},
}

@article{CHAOSdata2019,
  author       = {Ali Emre Kavur and M. Alper Selver and Oğuz Dicle and Mustafa Barış and  N. Sinem Gezer},
  title        = {{CHAOS - Combined (CT-MR) Healthy Abdominal Organ Segmentation Challenge Data}},
  month        = Apr,
  year         = 2019,
  publisher    = {Zenodo},
  version      = {v1.03},
  doi          = {10.5281/zenodo.3362844},
  url          = {https://doi.org/10.5281/zenodo.3362844}
}

@article{HMC-QU,
  title={Early detection of myocardial infarction in low-quality echocardiography},
  author={Degerli, Aysen and Zabihi, Morteza and Kiranyaz, Serkan and Hamid, Tahir and Mazhar, Rashid and Hamila, Ridha and Gabbouj, Moncef},
  journal={IEEE Access},
  volume={9},
  pages={34442--34453},
  year={2021},
  publisher={IEEE}
}

@article{kiranyaz2020left,
  title={Left ventricular wall motion estimation by active polynomials for acute myocardial infarction detection},
  author={Kiranyaz, Serkan and Degerli, Aysen and Hamid, Tahir and Mazhar, Rashid and Ahmed, Rayyan El Fadil and Abouhasera, Rayaan and Zabihi, Morteza and Malik, Junaid and Hamila, Ridha and Gabbouj, Moncef},
  journal={IEEE Access},
  volume={8},
  pages={210301--210317},
  year={2020},
  publisher={IEEE}
}

@inproceedings{FetoPlac,
  title={Deep placental vessel segmentation for fetoscopic mosaicking},
  author={Bano, Sophia and Vasconcelos, Francisco and Shepherd, Luke M and Vander Poorten, Emmanuel and Vercauteren, Tom and Ourselin, Sebastien and David, Anna L and Deprest, Jan and Stoyanov, Danail},
  booktitle={Medical Image Computing and Computer Assisted Intervention--MICCAI 2020: 23rd International Conference, Lima, Peru, October 4--8, 2020, Proceedings, Part III 23},
  pages={763--773},
  year={2020},
  organization={Springer}
}

@article{LGGFlair,
  title={Radiogenomics of lower-grade glioma: algorithmically-assessed tumor shape is associated with tumor genomic subtypes and patient outcomes in a multi-institutional study with The Cancer Genome Atlas data},
  author={Mazurowski, Maciej A and Clark, Kal and Czarnek, Nicholas M and Shamsesfandabadi, Parisa and Peters, Katherine B and Saha, Ashirbani},
  journal={Journal of neuro-oncology},
  volume={133},
  pages={27--35},
  year={2017},
  publisher={Springer}
}

@article{KiTS,
  title={The state of the art in kidney and kidney tumor segmentation in contrast-enhanced CT imaging: Results of the KiTS19 Challenge},
  author={Heller, Nicholas and Isensee, Fabian and Maier-Hein, Klaus H and Hou, Xiaoshuai and Xie, Chunmei and Li, Fengyi and Nan, Yang and Mu, Guangrui and Lin, Zhiyong and Han, Miofei and others},
  journal={Medical Image Analysis},
  pages={101821},
  year={2020},
  publisher={Elsevier}
}

@article{ISLES,
  title={ISLES 2022: A multi-center magnetic resonance imaging stroke lesion segmentation dataset},
  author={Hernandez Petzsche, Moritz R and de la Rosa, Ezequiel and Hanning, Uta and Wiest, Roland and Valenzuela, Waldo and Reyes, Mauricio and Meyer, Maria and Liew, Sook-Lei and Kofler, Florian and Ezhov, Ivan and others},
  journal={Scientific data},
  volume={9},
  number={1},
  pages={762},
  year={2022},
  publisher={Nature Publishing Group UK London}
}

@article{EOphtha,
  title={TeleOphta: Machine learning and image processing methods for teleophthalmology},
  author={Decenciere, Etienne and Cazuguel, Guy and Zhang, Xiwei and Thibault, Guillaume and Klein, J-C and Meyer, Fernand and Marcotegui, Beatriz and Quellec, Gw{\'e}nol{\'e} and Lamard, Mathieu and Danno, Ronan and others},
  journal={Irbm},
  volume={34},
  number={2},
  pages={196--203},
  year={2013},
  publisher={Elsevier}
}

@article{MCIC,
  title={The MCIC collection: a shared repository of multi-modal, multi-site brain image data from a clinical investigation of schizophrenia},
  author={Gollub, Randy L and Shoemaker, Jody M and King, Margaret D and White, Tonya and Ehrlich, Stefan and Sponheim, Scott R and Clark, Vincent P and Turner, Jessica A and Mueller, Bryon A and Magnotta, Vince and others},
  journal={Neuroinformatics},
  volume={11},
  pages={367--388},
  year={2013},
  publisher={Springer}
}

@article{PPMI,
  title={The Parkinson progression marker initiative (PPMI)},
  author={Marek, Kenneth and Jennings, Danna and Lasch, Shirley and Siderowf, Andrew and Tanner, Caroline and Simuni, Tanya and Coffey, Chris and Kieburtz, Karl and Flagg, Emily and Chowdhury, Sohini and others},
  journal={Progress in neurobiology},
  volume={95},
  number={4},
  pages={629--635},
  year={2011},
  publisher={Elsevier}
}

@article{buda2019association,
  title={Association of genomic subtypes of lower-grade gliomas with shape features automatically extracted by a deep learning algorithm},
  author={Buda, Mateusz and Saha, Ashirbani and Mazurowski, Maciej A},
  journal={Computers in biology and medicine},
  volume={109},
  pages={218--225},
  year={2019},
  publisher={Elsevier}
}

@article{BRATS,
  title={The RSNA-ASNR-MICCAI BraTS 2021 benchmark on brain tumor segmentation and radiogenomic classification},
  author={Baid, Ujjwal and Ghodasara, Satyam and Mohan, Suyash and Bilello, Michel and Calabrese, Evan and Colak, Errol and Farahani, Keyvan and Kalpathy-Cramer, Jayashree and Kitamura, Felipe C and Pati, Sarthak and others},
  journal={arXiv preprint arXiv:2107.02314},
  year={2021}
}

@article{menze2014multimodal,
  title={The multimodal brain tumor image segmentation benchmark (BRATS)},
  author={Menze, Bjoern H and Jakab, Andras and Bauer, Stefan and Kalpathy-Cramer, Jayashree and Farahani, Keyvan and Kirby, Justin and Burren, Yuliya and Porz, Nicole and Slotboom, Johannes and Wiest, Roland and others},
  journal={IEEE transactions on medical imaging},
  volume={34},
  number={10},
  pages={1993--2024},
  year={2014},
  publisher={IEEE}
}

@article{bakas2017advancing,
  title={Advancing the cancer genome atlas glioma MRI collections with expert segmentation labels and radiomic features},
  author={Bakas, Spyridon and Akbari, Hamed and Sotiras, Aristeidis and Bilello, Michel and Rozycki, Martin and Kirby, Justin S and Freymann, John B and Farahani, Keyvan and Davatzikos, Christos},
  journal={Scientific data},
  volume={4},
  number={1},
  pages={1--13},
  year={2017},
  publisher={Nature Publishing Group}
}

@article{CAMUS,
  title={Deep learning for segmentation using an open large-scale dataset in 2D echocardiography},
  author={Leclerc, Sarah and Smistad, Erik and Pedrosa, Joao and {\O}stvik, Andreas and Cervenansky, Frederic and Espinosa, Florian and Espeland, Torvald and Berg, Erik Andreas Rye and Jodoin, Pierre-Marc and Grenier, Thomas and others},
  journal={IEEE transactions on medical imaging},
  volume={38},
  number={9},
  pages={2198--2210},
  year={2019},
  publisher={IEEE}
}

@article{CDemris,
  title={Evaluation of current algorithms for segmentation of scar tissue from late gadolinium enhancement cardiovascular magnetic resonance of the left atrium: an open-access grand challenge},
  author={Karim, Rashed and Housden, R James and Balasubramaniam, Mayuragoban and Chen, Zhong and Perry, Daniel and Uddin, Ayesha and Al-Beyatti, Yosra and Palkhi, Ebrahim and Acheampong, Prince and Obom, Samantha and others},
  journal={Journal of Cardiovascular Magnetic Resonance},
  volume={15},
  number={1},
  pages={1--17},
  year={2013},
  publisher={BioMed Central}
}

@article{CoNSeP,
  title={Hover-net: Simultaneous segmentation and classification of nuclei in multi-tissue histology images},
  author={Graham, Simon and Vu, Quoc Dang and Raza, Shan E Ahmed and Azam, Ayesha and Tsang, Yee Wah and Kwak, Jin Tae and Rajpoot, Nasir},
  journal={Medical Image Analysis},
  volume={58},
  pages={101563},
  year={2019},
  publisher={Elsevier}
}

@article{WMH,
  title={Standardized assessment of automatic segmentation of white matter hyperintensities and results of the WMH segmentation challenge},
  author={Kuijf, Hugo J and Biesbroek, J Matthijs and De Bresser, Jeroen and Heinen, Rutger and Andermatt, Simon and Bento, Mariana and Berseth, Matt and Belyaev, Mikhail and Cardoso, M Jorge and Casamitjana, Adria and others},
  journal={IEEE transactions on medical imaging},
  volume={38},
  number={11},
  pages={2556--2568},
  year={2019},
  publisher={IEEE}
}

@article{VerSe,
  title={A vertebral segmentation dataset with fracture grading},
  author={L{\"o}ffler, Maximilian T and Sekuboyina, Anjany and Jacob, Alina and Grau, Anna-Lena and Scharr, Andreas and El Husseini, Malek and Kallweit, Mareike and Zimmer, Claus and Baum, Thomas and Kirschke, Jan S},
  journal={Radiology: Artificial Intelligence},
  volume={2},
  number={4},
  pages={e190138},
  year={2020},
  publisher={Radiological Society of North America}
}
}

\clearpage
\setcounter{page}{1}
\maketitlesupplementary
\appendix

\section{Overview}
\label{overview}
We first present a brief overview of the analysis and information in this Supplemental Material.

\subpara{Tyche Model Choices}

We present additional data details, including for MegaMedical, multi-annotator data, and simulated data. We give additional details on the \emph{\tychetrain} training strategy, as well as the augmentations used at inference for \emph{\tychetest}. We also detail how we trained each benchmark, and their respective limitations. 

We provide an intuition behind the best candidate Dice loss, and show that with \emph{Tyche}, it is possible to optimize for objectives that apply to the candidate predictions as a group, and show results when optimizing GED. 

\subpara{Tyche Analysis}

For \emph{\tychetrain}, we investigate four aspects: noise, context set, number of predictions, and \textit{SetBlock}. We show how different noise levels impact the predictions. Moreover, we give examples of how prediction changes when the context or the noise changes. We also show that the number of predictions at inference impacts the diversity of the segmentation candidates predicted. %

For \emph{\tychetest}, we investigate how different augmentations affect performance.

We also analyze a scenario where only few of annotated examples are available, using the LIDC-IDRI dataset. We compare \emph{Tyche} with these samples in the context to PhiSeg, trained on these few annotated samples.

\subpara{Further Evaluation}

Given the nature of stochastic segmentation tasks, no single metric fits all purposes, and the optimal metric depends on the downstream goals. We provide \textit{sample diversity} and \textit{Hungarian Matching}, showing that \emph{Tyche} performs well on these as well. 

We also present performance per dataset and show that the relative performance of each model stays generally unchanged, even though the difficulty of each dataset is widely different. %

We provide additional visualizations on how \emph{Tyche} and the baselines perform for each datasets, both on single and multi-annotator data. 

\subpara{Data split} All analysis results in this supplemental material use the validation set of the out-of-distribution datasets, to avoid making modeling decisions on the test sets.

\section{Tyche Model Choices}
We present the implementation details for the data, our \emph{Tyche} models, and the benchmarks. 

\subsection{Medical Image Data}
\subpara{Megamedical}
 \label{sup:megamedicaldata}
 We build on the dataset collection proposed in~\cite{butoi2023universeg,wong2023scribbleprompt}, using the similar preprocessing methods. The complete list of datasets is presented Table \ref{tab:sup:megamedical2}.

\subpara{Processing}
Each image is normalized between 0 and 1 and is resized to $128\times128$. When the full 3D volume is available, we take two slices for each task: the slice in the middle of the volume and the slice with the largest count of pixels labelled for that task.  

\subpara{Task Definition} We consider a task as labelling a certain structure from a certain modality for a certain dataset. If a given medical image has different structures labeled, we consider each structure a different task. For 3D volumes, we consider each axis a different task. %

\begin{figure}
    \centering
    \includegraphics[width=0.45\textwidth]{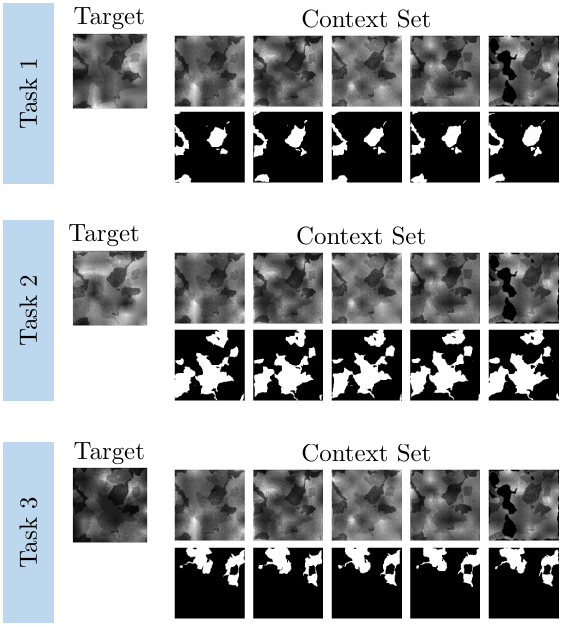}
    \caption{\textbf{Synthetic Data Examples.} Example of training images on the single annotator synthetic data. Each row is an example of a target-context pair corresponding to a different task.}
    \label{fig:sup:synth}
\end{figure}

 \subpara{Synthetic Data} We use a set of synthetic tasks to enhance the performance of our networks, similar to~\cite{butoi2023universeg}. Some examples are shown in Figure \ref{fig:sup:synth}.

\begin{figure}
    \centering
    \includegraphics[width=0.45\textwidth]{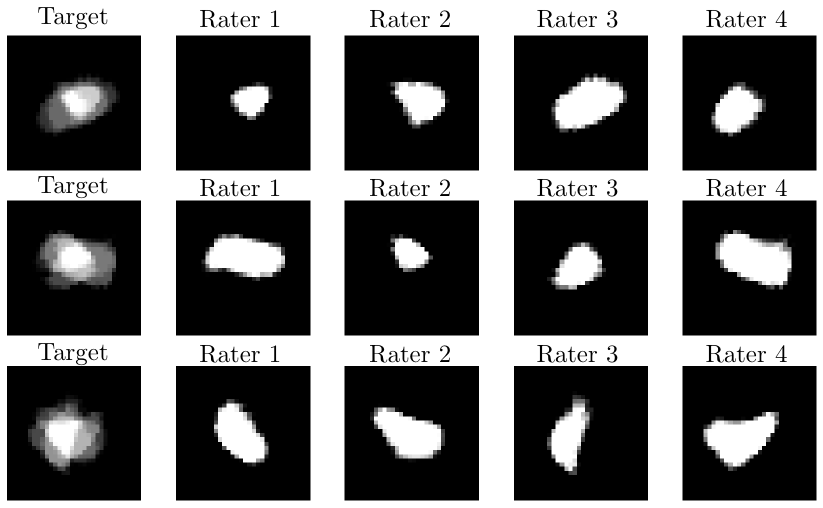}
    \caption{\textbf{Synthetic Multi-Annotated Examples.} Each blob is generated by deforming a white disk using a random smooth deformation field, each representing a different rater. The blobs are then averaged to form the target image to segment.}
    \label{fig:sup:blob}
\end{figure}

\subpara{Synthetic Multi-Annotator Data}
We use synthetic data to encourage diversity in our predictions. Figure \ref{fig:sup:blob} shows examples of blob targets with the corresponding simulated annotations. 

 \subsection{Training of \tychetrain}
 \label{sup:aug:training}
 We train \emph{\tychetrain} with a context size of 16 and a batch size of 4. We use a learning rate of 0.0001, Adam as an optimizer, and a kernel size of 3.  We apply the augmentations shown Table \ref{tab:sup:tychetrsaug}. To obtain $K$ segmentation candidates, we generate $K$ noise samples $z_{k}\sim \mathcal{N}(0, \mathbb{I})$. We duplicate the input $K$ times and concatenate each noise sample with a duplicated target to form $K$ stochastic inputs to the network. 

\begin{table}[h]
\begin{subtable}[h]{0.45\textwidth}
\centering
\rowcolors{2}{white}{gray!15}
    \begin{tabular}{lcc}
   \textbf{Augmentations} & $p$ & Parameters\\
    \hline
        \cellcolor{gray!15}    &\cellcolor{gray!15} & \cellcolor{gray!15} degrees $\in[-25, 25]$ \\
        \cellcolor{gray!15}    & \cellcolor{gray!15} & \cellcolor{gray!15} translate$\in[0, 0.1]$ \\
    \multirow{-3}{*}{\cellcolor{gray!15} Random Affine}     & \multirow{-3}{*}{\cellcolor{gray!15} 0.25} & \cellcolor{gray!15} scale$\in[0.9, 1.1]$\\
       \cellcolor{white}  & \cellcolor{white} & \cellcolor{white}brightness$\in[-0.1, 0.1]$, \\
     \multirow{-2}{*}{\cellcolor{white}{Brightness Contrast}} &  \multirow{-2}{*}{\cellcolor{white}{0.5}} & \cellcolor{white} contrast$\in[0.5, 1.5]$ \\
     \cellcolor{gray!15}    & \cellcolor{gray!15} & \cellcolor{gray!15}$\alpha\in[1, 2.5]$\\
    \multirow{-2}{*}{\cellcolor{gray!15} Elastic Transform} & \multirow{-2}{*}{\cellcolor{gray!15} 0.8} & \cellcolor{gray!15} $\sigma\in[7,9]$\\
    Sharpness & 0.25 & sharpness$=5$\\
    Flip Intensities & 0.5 & None\\
        &  &  $\sigma\in[0.1, 1.0]$ \\
    \multirow{-2}{*}{\cellcolor{white} Gaussian Blur}     & \multirow{-2}{*}{\cellcolor{white} 0.25} & \cellcolor{white} k=5 \\
    \cellcolor{gray!15} & \cellcolor{gray!15} & \cellcolor{gray!15} $\mu\in[0, 0.05]$\\
    \multirow{-2}{*}{\cellcolor{gray!15} Gaussian Noise} & \multirow{-2}{*}{\cellcolor{gray!15} 0.25} &  \cellcolor{gray!15} $ \sigma \in [0, 0.05]$\\\hline
    \end{tabular}
    \caption{In-Task Augmentation}
\end{subtable}
\begin{subtable}[h]{0.45\textwidth}
\centering
\rowcolors{2}{white}{gray!15}
    \begin{tabular}{lcc}
    \textbf{Augmentations} & $p$ & Parameters\\
    \hline
           &  &  degrees $\in[0, 360]$ \\
        \cellcolor{gray!15}   & \cellcolor{gray!15} & \cellcolor{gray!15} translate$\in[0, 0.2]$ \\
    \multirow{-3}{*}{ Random Affine}     & \multirow{-3}{*}{ 0.5} & scale$\in[0.8, 1.1]$\\
         &  & brightness$\in[-0.1, 0.1]$, \\
     \multirow{-2}{*}{\cellcolor{white}{Brightness Contrast}} &  \multirow{-2}{*}{\cellcolor{white}{0.5}} & \cellcolor{white}contrast$\in[0.8, 1.2]$ \\
     \cellcolor{gray!15}   &  \cellcolor{gray!15} & \cellcolor{gray!15}$\sigma\in[0.1, 1.1]$ \\
    \multirow{-2}{*}{\cellcolor{gray!15} Gaussian Blur}     & \multirow{-2}{*}{\cellcolor{gray!15} 0.5} & \cellcolor{gray!15} $k=5$ \\
    \cellcolor{white} & \cellcolor{white} & \cellcolor{white} $\mu\in[0, 0.05]$\\
    \multirow{-2}{*}{\cellcolor{white} Gaussian Noise} & \multirow{-2}{*}{\cellcolor{white} 0.5} & \cellcolor{white} $ \sigma \in [0, 0.05]$\\
      \cellcolor{gray!15}   & \cellcolor{gray!15} & \cellcolor{gray!15} $\alpha\in[1, 2]$\\
    \multirow{-2}{*}{\cellcolor{gray!15} Elastic Transform} & \multirow{-2}{*}{\cellcolor{gray!15} 0.5} & \cellcolor{gray!15} $\sigma\in[6,8]$\\
    Sharpness & 0.5 & sharpness$=5$\\
    \cellcolor{white} Horizontal Flip & \cellcolor{white} 0.5 & \cellcolor{white} None \\
     \cellcolor{gray!15}Vertical Flip & \cellcolor{gray!15} 0.5 & \cellcolor{gray!15} None \\
     \cellcolor{white}Sobel Edges Label & \cellcolor{white} 0.5 & \cellcolor{white} None \\\hline
    \end{tabular}
    \caption{Task Augmentation}
    \end{subtable}
\caption{\textbf{Set of augmentations used to train \tychetrain.} We distinguish between augmentations aimed at increasing the diversity inside a task (Top) and the augmentations aimed at increasing the diversity of tasks (Bottom). An augmentation is applied with probability $p$ (second column).}
    \label{tab:sup:tychetrsaug}
\end{table}

\begin{figure*}
    \centering
    \includegraphics[width=\textwidth]{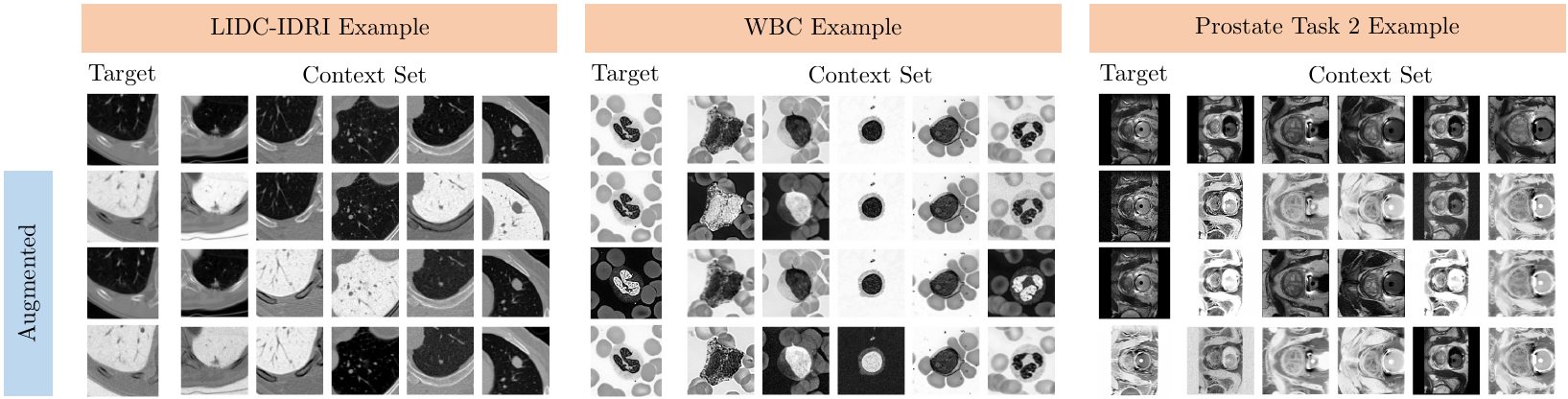}
    \caption{\textbf{Example Augmentations for \tychetest.} We show for samples from different datasets how three new segmentation candidates are generated. Starting from the top row, augmentations are applied to both the target and context set to obtain new prediction. Each row represents a new target-context pair. We do not show the corresponding labels as for \emph{\tychetest}, we only augment with intensity-based transforms.}
    \label{fig:sup:aug:tycheteS}
\end{figure*}

 \subsection{Network for \tychetest}
 \label{sup:tes:info}
For the \emph{\tychetest} network, we use the baseline UniverSeg~\cite{butoi2023universeg}, trained with the same data as \emph{\tychetrain}, same batch size, and same context size. At test time, we use the augmentations in Table \ref{tab:sup:aug:tycheteS}. Figure \ref{fig:sup:aug:tycheteS} shows example augmentation on different samples.

\begin{table}[]
    \centering
   \rowcolors{2}{white}{gray!15}
    \begin{tabular}{lcc}
   \textbf{\emph{\tychetest} Augment.} & $p$ & Parameters\\
    \hline
        &  & $\sigma\in[0.1, 1.0]$ \\
    \multirow{-2}{*}{\cellcolor{gray!15} Gaussian Blur}     & \multirow{-2}{*}{\cellcolor{gray!15} 0.25} & \cellcolor{gray!15} $k=5$ \\
    \cellcolor{white} & \cellcolor{white} & \cellcolor{white} $\mu\in[0, 0.05]$\\
    \multirow{-2}{*}{\cellcolor{white} Gaussian Noise} & \multirow{-2}{*}{\cellcolor{white} 0.25} &  $ \sigma \in [0, 0.05]$\\
    Flip Intensities & 0.5 & None\\
    Sharpness & 0.25 & sharpness=5\\
     &  & brightness$\in[-0.1, 0.1]$, \\
     \multirow{-2}{*}{\cellcolor{gray!15}{Brightness Contrast}} &  \multirow{-2}{*}{\cellcolor{gray!15}{0.25}} & \cellcolor{gray!15} contrast$\in[0.5, 1.5]$ \\\hline
    \end{tabular}
    \caption{\textbf{Augmentations used for \emph{\tychetest}.} We focus on intensity transforms, to avoid inverting the prediction. For each image, an augmentation is sampled with probability $p$.}
    \label{tab:sup:aug:tycheteS}
\end{table}

\begin{figure}
    \centering
    \includegraphics[width=0.47\textwidth]{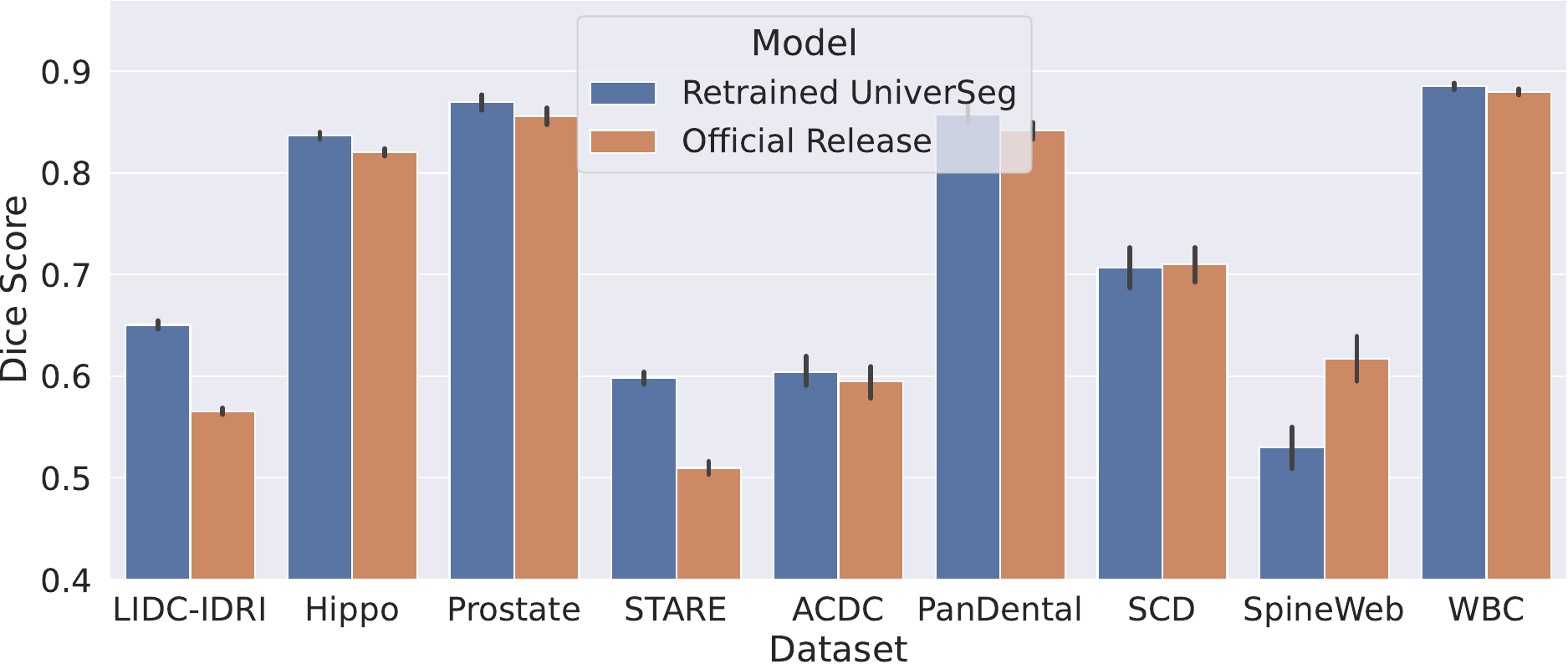}
    \caption{\textbf{Improvement of UniverSeg by training with more data.} By training UniverSeg on more data, we obtain a better average Dice score than the official UniverSeg release.}
    \label{fig:sup:OfficialvsRetraineduniverseg}
\end{figure}

\subsection{Benchmarks}
We distinguish three types of benchmarks: in-context methods, that can take as input a context set, the interactive frameworks, and the task-trained specialized upper bounds.

\subpara{In-context baselines}
Because they were trained on medical data, we use SegGPT and SAM-Med2D as provided in their official release. We train from scratch UniverSeg and SENet. We use a batch size of 4 and the same set of data augmentation transforms as the ones used for \emph{\tychetrain} to encourage within task and across task diversity. Training UniverSeg on the same datasets as \emph{Tyche} improves performances compared to the official UniverSeg release, as shown in Figure \ref{fig:sup:OfficialvsRetraineduniverseg}.

\subpara{Specialized benchmarks}
For the specialized models (CIDM, PhiSeg, and Probabilistic UNet), we train a model for each task, for a total of 20 tasks. For each task, we train three model variants, one for each different data augmentation scheme, shown Table \ref{tab:sup:benchmarksaug}. We select the model with the data augmentation strategy that does best on the \textit{out-of-distribution} validation set. We use a batch size of 4. For CIDM and Probabilistic UNet, we use the official PyTorch release. We found Probabilistic UNet particularly unstable to train on some datasets when applying augmentations. To avoid diverging loss values, we had to initialize the Probabilistic UNet models from versions trained on datasets without augmentation. For PhiSeg, the official TensorFLow release was unstable, and we worked with the authors to run a more recent PyTorch version of their code. Per the authors' request, we smooth the predicted segmentations of CIDM to remove irregularities in the final segmentations.  
\begin{table}[]
\begin{subtable}[ht]{0.45\textwidth}
        \centering
   \rowcolors{2}{white}{gray!15}
       \begin{tabular}{lcc}
   \textbf{Augmentation Set 1} & $p$ & Parameters\\
    \hline
     None   &  1 &  None \\\hline
    \end{tabular}
    \caption{No Augmentation. The images are given to the specialized model as is. }

\end{subtable}
\hfill
\begin{subtable}[h]{0.45\textwidth}
\centering
    \rowcolors{2}{white}{gray!15}
       \begin{tabular}{lcc}
   \textbf{Light Augmentation} & $p$ & Parameters\\
    \hline
        &  & $\sigma\in[0.1, 1.5]$ \\
    \multirow{-2}{*}{\cellcolor{gray!15} Gaussian Blur}     & \multirow{-2}{*}{\cellcolor{gray!15} 0.5} & \cellcolor{gray!15} $k=7$ \\
    \cellcolor{white} & \cellcolor{white} & \cellcolor{white} $\mu\in[0, 0.1]$\\
    \multirow{-2}{*}{\cellcolor{white} Gaussian Noise} & \multirow{-2}{*}{\cellcolor{white} 0.5} &  $ \sigma \in [0, 0.1]$\\
     &  & $\alpha\in[1, 2]$\\
    \multirow{-2}{*}{\cellcolor{gray!15} Variable Elastic Transform} & \multirow{-2}{*}{\cellcolor{gray!15} 0.25} & \cellcolor{gray!15}$\sigma\in[6,8]$\\\hline
    \end{tabular}
    \caption{Light Augmentation. Gaussian bur, Gaussian noise, elastic transforms are each applied to the target with probability $p$. }
\end{subtable}
\hfill
\vfill
\begin{subtable}[h]{0.45\textwidth}
\centering
\rowcolors{2}{white}{gray!15}
    \begin{tabular}{lcc}
   \textbf{Heavy Augmentation} & $p$ & Parameters\\
    \hline
        &  & $\sigma\in[0.1, 1.5]$ \\
    \multirow{-2}{*}{\cellcolor{gray!15} Gaussian Blur}     & \multirow{-2}{*}{\cellcolor{gray!15} 0.5} & \cellcolor{gray!15} $k=7$ \\
    \cellcolor{white} & \cellcolor{white} & \cellcolor{white} $\mu\in[0, 0.1]$\\
    \multirow{-2}{*}{\cellcolor{white} Gaussian Noise} & \multirow{-2}{*}{\cellcolor{white} 0.25} &  $ \sigma \in [0, 0.1]$\\
    
         &  & $\alpha\in[1, 2]$\\
    \multirow{-2}{*}{\cellcolor{gray!15} Elastic Transform} & \multirow{-2}{*}{\cellcolor{gray!15} 0.25} & \cellcolor{gray!15}$\sigma\in[6,8]$\\
    
        \cellcolor{white}    & \cellcolor{white} & \cellcolor{white} degrees $\in[0, 360]$ \\
        \cellcolor{white}    & \cellcolor{white} & \cellcolor{white} translate$\in[0, 0.2]$ \\
    \multirow{-3}{*}{\cellcolor{white} Random Affine}     & \multirow{-3}{*}{\cellcolor{white} 0.5} & \cellcolor{white} scale$\in[0.8, 1.1]$\\
   \cellcolor{gray!15}  & \cellcolor{gray!15} & \cellcolor{gray!15}brightness$\in[-0.1, 0.1]$, \\
     \multirow{-2}{*}{\cellcolor{gray!15}{Brightness Contrast}} &  \multirow{-2}{*}{\cellcolor{gray!15}{0.5}} & \cellcolor{gray!15}contrast$\in[0.5, 1.5]$ \\

     Horizontal Flip & 0.5 & None\\
     \cellcolor{gray!15}Vertical Flip & \cellcolor{gray!15} 0.5 & \cellcolor{gray!15}None \\
    Sharpness & 0.5 & sharpness$=5$\\\hline

    \end{tabular}
    \caption{Full Set of Augmentations. To form the last augmentation set, we add the following transforms: Affine transform, Sharpness, Brightness and Flips. }
    \end{subtable}
    
    \caption{\textbf{Set of Augmentations used to train the specialized benchmarks.} For evaluation, we select the model with the training augmentations that does best on the \textit{out-of-distribution} validation set.}
    \label{tab:sup:benchmarksaug}
\end{table}

\subpara{Interactive Methods} 
We use the SAM-Med2D and SAM as interactive segmentation baselines. In our setting, user interaction is not available. We simulate it by averaging the ground truth labels from the context set and sample clicks and bounding boxes from averages. 

We use SAM-Med2D's official release as it is an adaptation of SAM exclusively meant for 2D medical image segmentation tasks. We found empirically that SAM-Med2D performs best when 3 positive and 2 negative clicks are sampled. For the negative clicks, we sample areas inside the bounding box that are not covered by the context set. 

We also fine-tune SAM on our data. We sample 5 positive clicks and 5 negative clicks uniformly. We also provide a bounding box and the average of the context label maps as input.

\section{Advantages of candidates loss} 
\subsection{Intuition behind best candidate loss}
\begin{figure}
    \centering
    \includegraphics[width=0.46\textwidth]{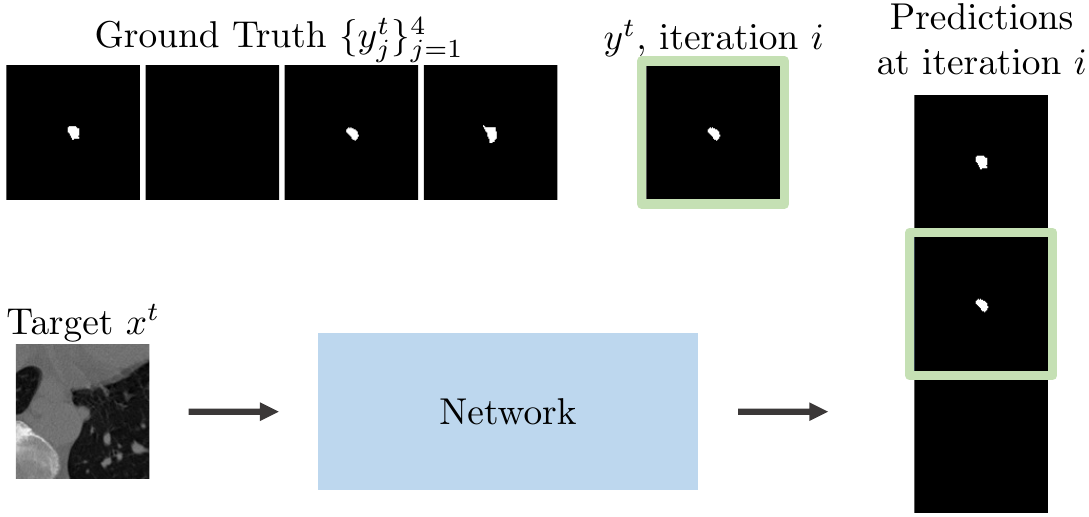}
    \caption{\textbf{Best candidate Loss Explanation.} Because of ambiguity in the target, there are multiple plausible ground truths. With a standard Dice loss, all the candidates segmentation converge to the mean segmentation. The best candidate loss enables the model to explore different plausible segmentation, since even one proposal matching the rater segmentation used at that iteration leads to a good loss value.}
    \label{fig:method:minlossexplanation}
\end{figure}

The best candidate loss only evaluates the candidate that yields the lowest loss. We provide intuition for this loss function (Figure \ref{fig:method:minlossexplanation}). 

Given as input a target $x^t$ and a context $\mathcal{S}^t$, the model outputs $K$ segmentation candidates $\hat{y}_j$. Despite potentially high target ambiguity, for each training iteration we use only one annotation $y^{t}_r$, .

The loss element $l(\hat{y}_k, y^t_ r)$ captures volume overlap between a candidate prediction~$\hat{y}_k$ and the rater annotation~$y^t_r$. If a regular loss is used across all predictions, then \textit{each} prediction tends towards the lowest expect cost over the space of possible segmentations for that image, leading to a {mean} segmentation among raters. This is particularly harmful when target ambiguity is high (high rater disagreement). Then, the mean segmentation may be very different than any one rater and may not be representative of the ambiguity. For the best candidate loss, the model is encouraged to make very different guesses for different segmentation candidates. This increases the chance that one prediction matches the rater being used as ground truth at that iteration.

 \begin{figure}
    \centering
    \includegraphics[width=0.45\textwidth]{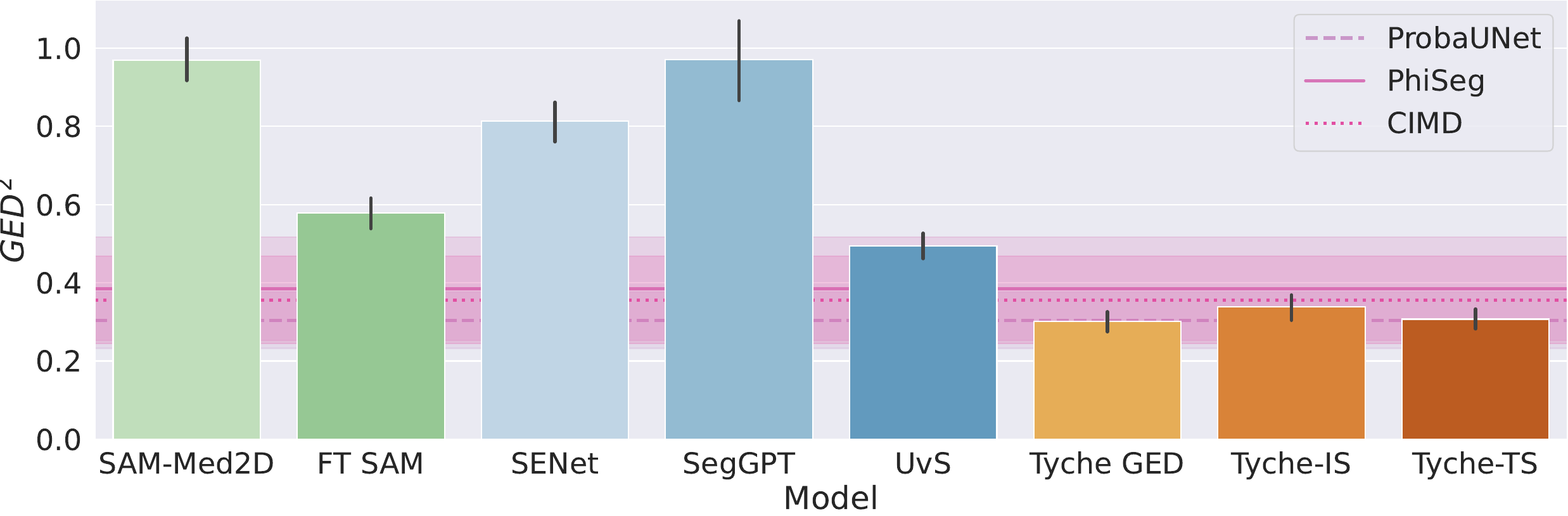}
    \caption{\textbf{\emph{\tychetrain} with GED loss}  (yellow) improved test-time GED performance (lower is better).}
    \label{fig:sup:ged}
\end{figure}

\subsection{Distribution loss function}
Since \emph{\tychetrain} enables multiple candidates that can coordinate with one another, we can employ loss functions that apply to a collection of raters and predictions.
We demonstrate this concept by fine-tuning \emph{\tychetrain}, using a $GED^2$ loss function. 

Figure \ref{fig:sup:ged} shows that the corresponding model produces samples with lower $GED^2$, as expected.

\begin{figure*}[h]
    \centering
    \includegraphics[width=\textwidth]{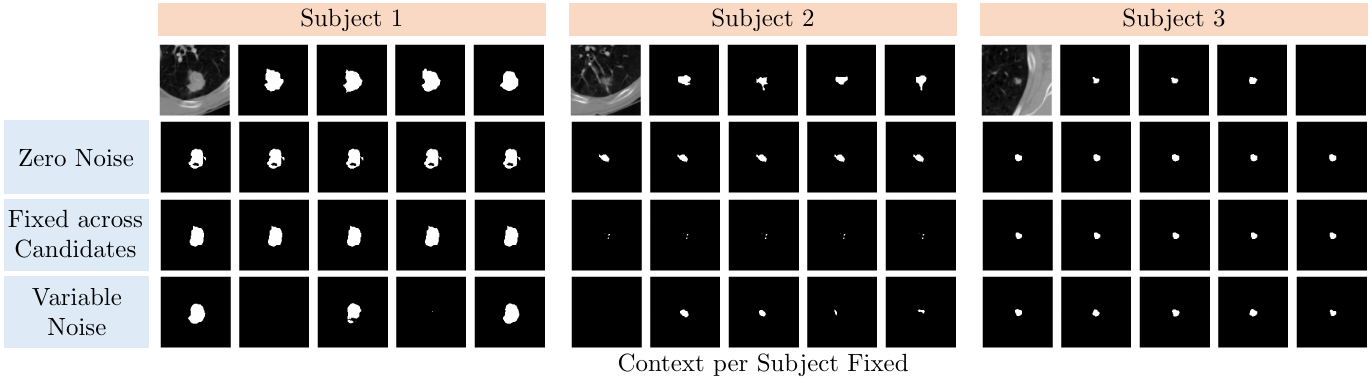}
    \caption{\textbf{Prediction variability as a function of injected noise.} Examples of predictions for three subjects with three different types of input noise. Top to Bottom: zero noise, Gaussian noise constant across segmentation candidates, and randomly sampled Gaussian noise. The context set is fixed. With random noise as an input \emph{Tyche} can output diverse segmentation candidates.}
    \label{fig:sup:noise}
\end{figure*}

\begin{figure*}[h]
    \centering
    \includegraphics[width=\textwidth]{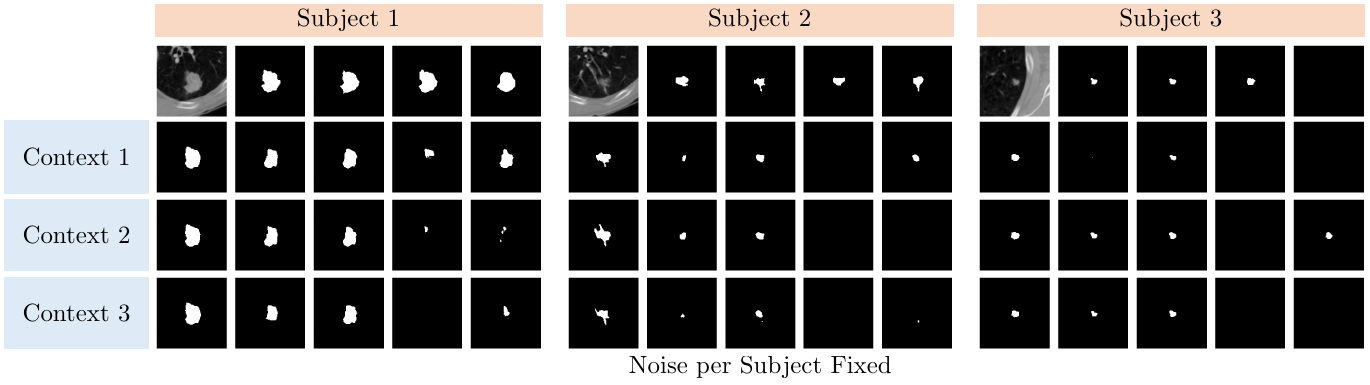}
    \caption{\textbf{Prediction variability as a function of the context set.} The context set significantly contributes to the variability of the output. For three different subjects, we show how the corresponding segmentation candidates are affected by different context sets. The random Gaussian noise given as input is fixed for each set of candidates.}
    \label{fig:sup:support}
\end{figure*}

\begin{figure}
    \centering
    \includegraphics[width=0.45\textwidth]{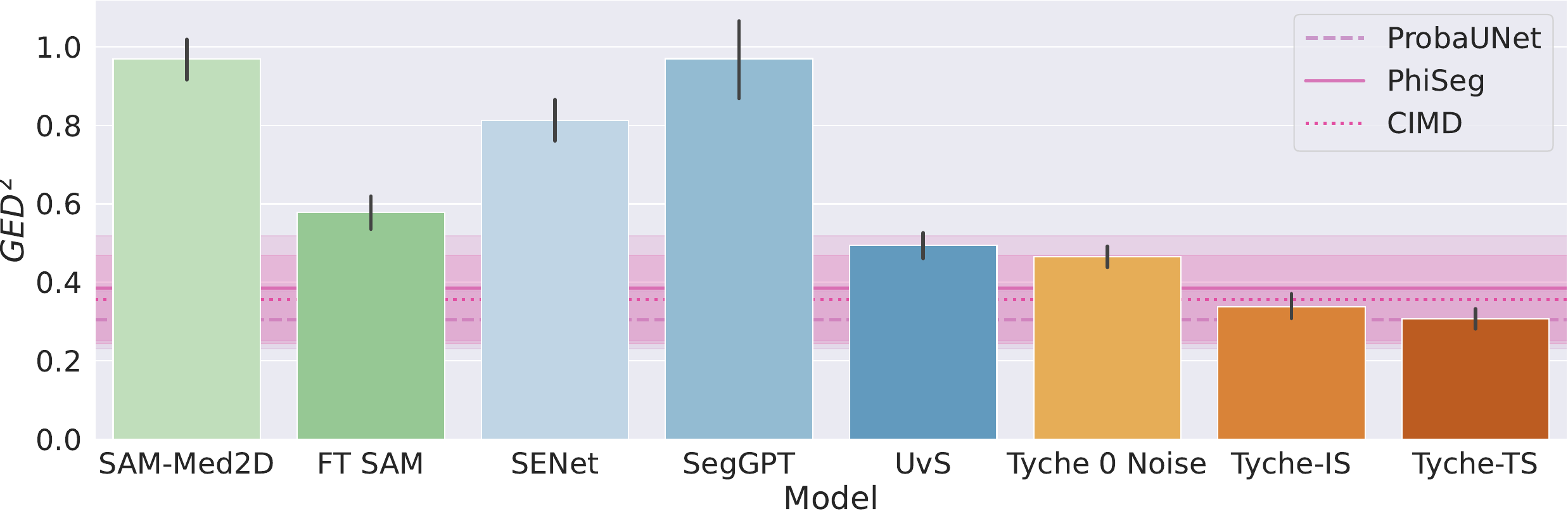}
    \includegraphics[width=0.45\textwidth]{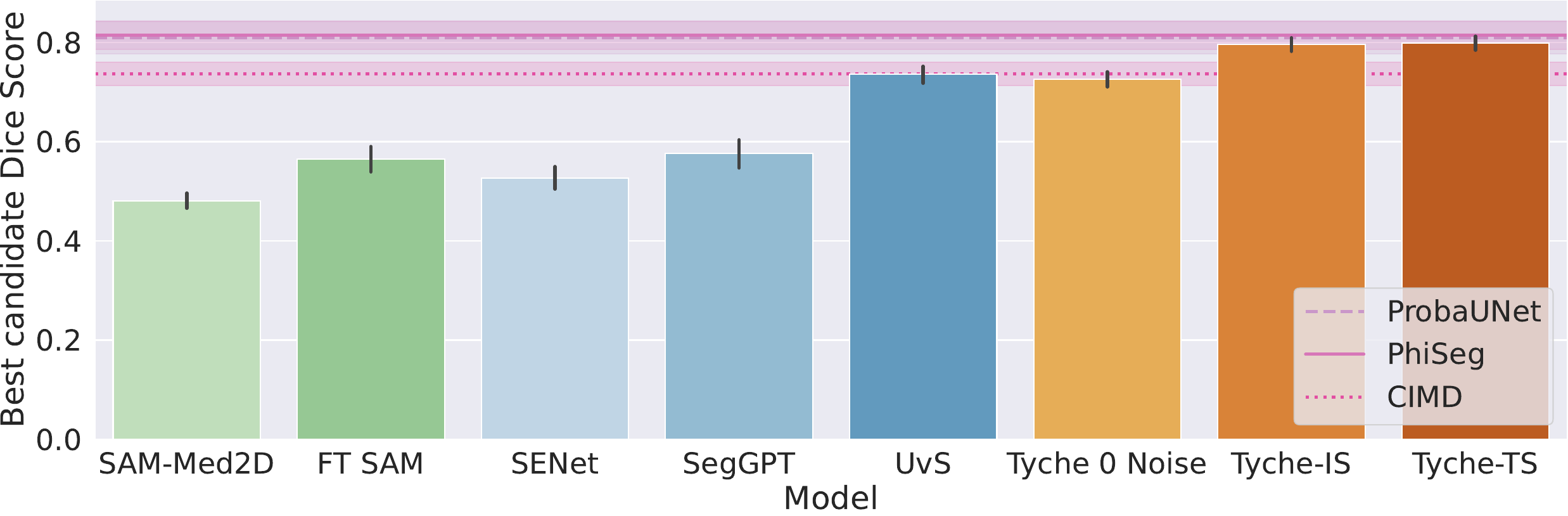}
    \caption{\textbf{\emph{\tychetrain} with no noise gives similar results to UniverSeg.} Setting the noise to 0 for \emph{\tychetrain} gives similar best candidate Dice score and similar $GED^2$ to UniverSeg. Top: Generalized Energy Distance. Bottom: Best candidate Dice Score.}
    \label{fig:sup:nonoise}
\end{figure}

\begin{figure*}[h]
    \centering
    \includegraphics[width=\textwidth]{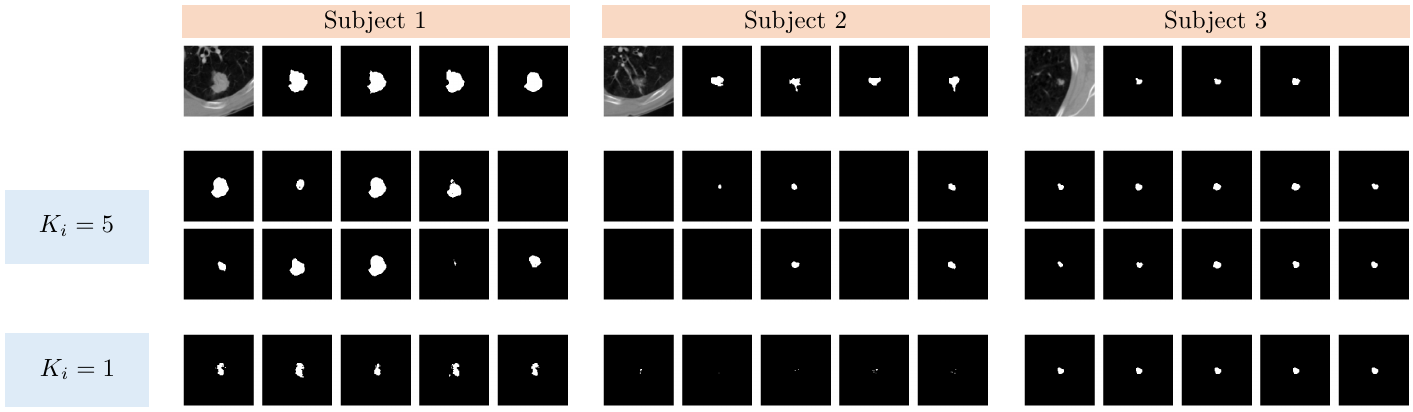}
    \caption{\textbf{Setting a number of predictions larger than 1 allows for more diversity.} The outputs of \emph{\tychetrain} are a lot more uniform when the number of prediction is set to 1 than when it is set to 5.}
    \label{fig:sup:1v5}
\end{figure*}
\begin{figure*}[h]
    \centering
    \includegraphics[width=\textwidth]{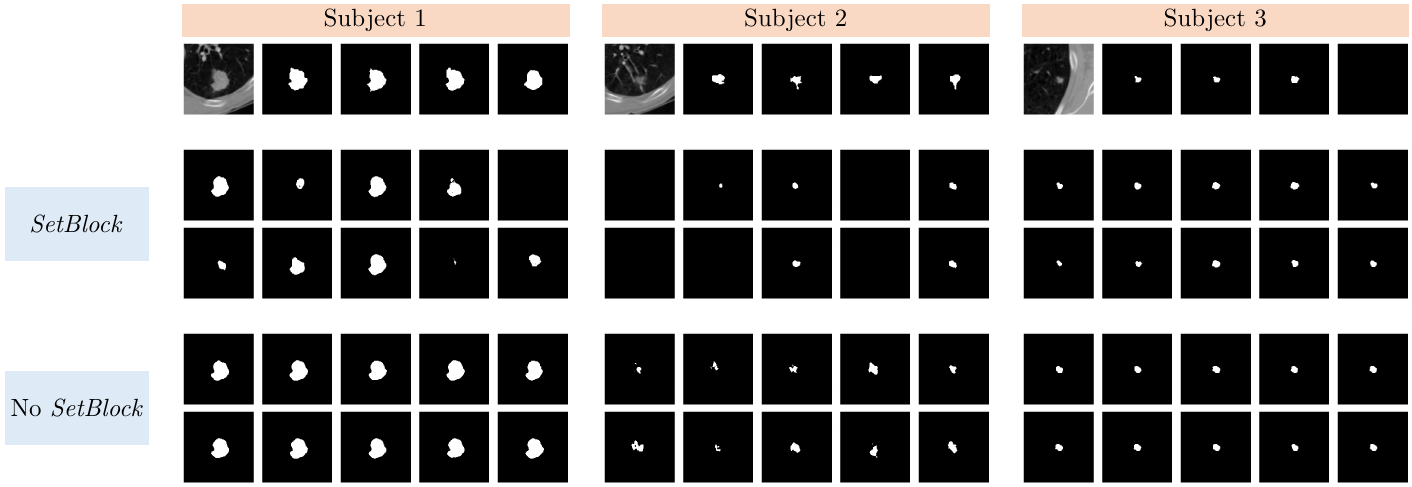}
    \caption{\textbf{Removing the SetBlock leads to less diversity in the output.} The segmentation candidates output by \emph{\tychetrain} are more diverse than the candidates of a \emph{\tychetrain} model trained  without the \textit{SetBlock} interaction.}
    \label{fig:sup:setblock}
\end{figure*}
\newpage
\section{Tyche Analysis}
We first analyse the different components that lead to variability in \emph{\tychetrain}: the noise, the context set, the number of prediction in on forward pass, and the \textit{SetBlock} mechanism. We then investigate how the size of the context set and the number of predictions impact performances. For \emph{\tychetest}, we investigate different plausible sets of augmentations to apply. Finally, we show that \emph{Tyche} can produce good results even in data scarcity settings when a stochastic methods trained solely on a few samples would have limited performance.

\subsection{Diversity from noise}
Figure \ref{fig:sup:noise} shows how three predictions vary depending on three noise variants: zero-noise, constant noise across candidates, and variable noise. Both zero noise and constant noise lead systematically to identical segmentation candidates. For each example of segmentation candidates, the context set is fixed.

Figure \ref{fig:sup:nonoise} shows that setting the noise to 0 for \emph{\tychetrain} produces similar performances to UniverSeg, both for GED and best candidate Dice score. 

\subsection{Diversity from context}
Figure \ref{fig:sup:support} shows example predictions for three subjects and three context sets. Different contexts lead to different predictions, even if the noise is the same, sometimes drastically. For instance, the predictions for subject 1 and context 3 contain a candidate with no annotations.

\subsection{Diversity in the number of predictions}
While \emph{\tychetest} predicts each segmentation sample sequentially, \emph{\tychetrain} has a one-shot mechanism that predicts all the candidates at once. One may wonder how the number of predictions requested at inference impacts the output of \emph{\tychetrain}. Figure \ref{fig:sup:1v5} shows that with the number of prediction set to 1, \emph{\tychetrain} loses a lot of its diversity compared to 5 predictions. 

\begin{figure}
    \centering
    \includegraphics[width=0.47\textwidth]{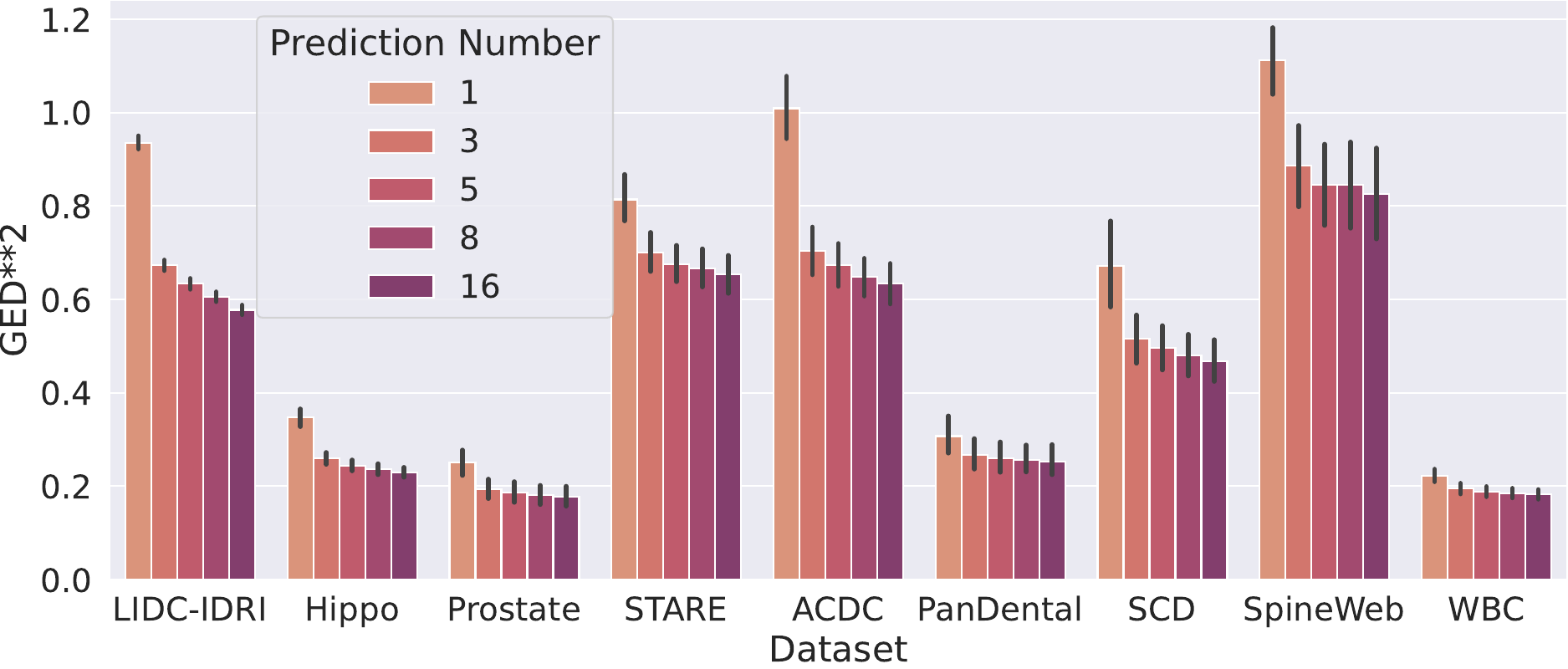}
    \caption{\textbf{$GED^2$ for Tyche as a function of the number of predictions, $K_i$.} Performances improve with the number of prediction but with diminishing returns.}
    \label{fig:exp:kistudy_ged}
\end{figure}

This is also shown quantitatively in Figure \ref{fig:exp:kistudy_ged}, where Generalized Energy Distance decreases as the number of predictions increases.

\subsection{Diversity from the SetBlock}
\emph{\tychetrain} has an intrinsic mechanism to encourage the segmentation candidates to be diverse: \textit{SetBlock}. Figure \ref{fig:sup:setblock} visually compares the segmentations predicted with the mechanisms and the segmentations predicted by a \emph{Tyche} model trained without the \textit{SetBlock}. Without the \textit{SetBlock}, the predictions output by \emph{Tyche} are a lot less diverse.

\subsection{Size of the Context Set}
\begin{figure}
    \centering
    \includegraphics[width=0.47\textwidth]{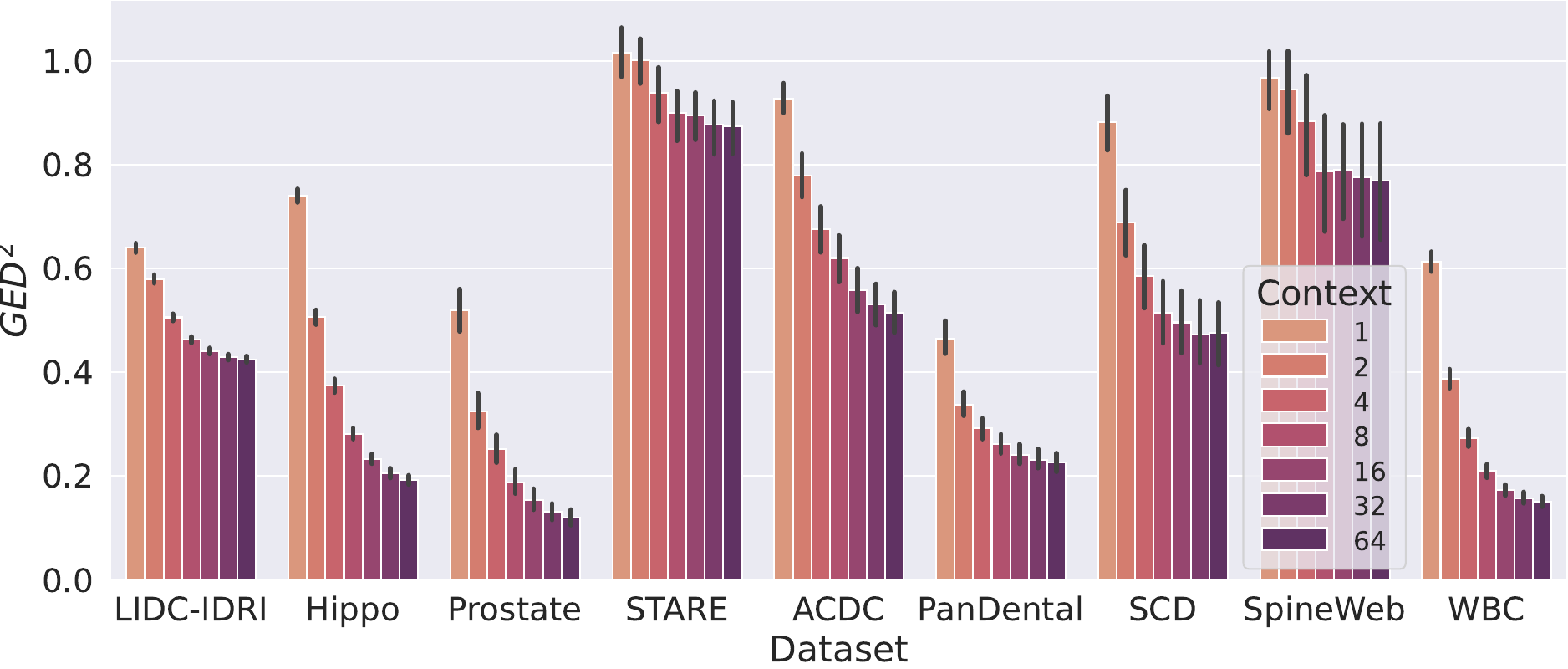}
    \caption{\textbf{$GED^2$ for Tyche as a function of the inference context size.} Performances improve as the context size increases but with diminishing returns.}
    \label{fig:exp:ssstudy_ged}
\end{figure}

Generally, a larger context set improves performances. Figure \ref{fig:exp:ssstudy_ged} shows that Generalized Energy Distance improves as the size of the context set increases. However, the improvement decreases beyond 16 samples.

\subsection{Augmentations}
\begin{figure}
    \centering
    \includegraphics[width=0.45\textwidth]{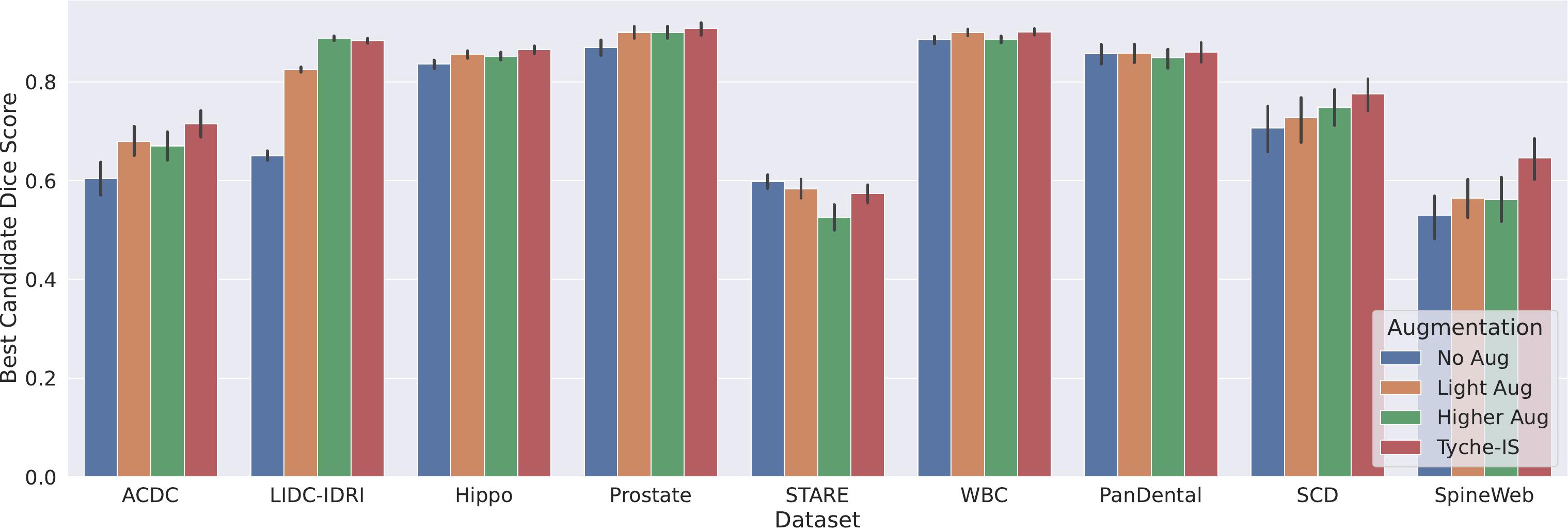}
    \includegraphics[width=0.45\textwidth]{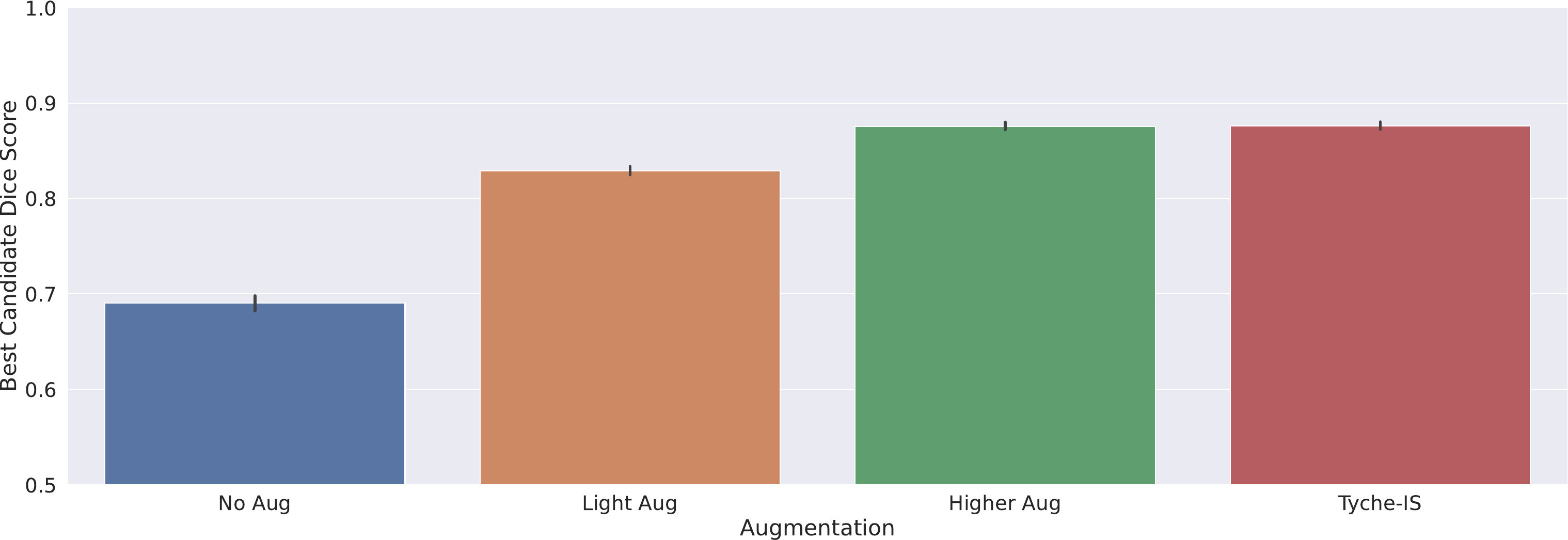} 
    \caption{\textbf{Performance of different augmentation schemes for \emph{\tychetest}}. We consider four augmentations: none, light, the one used in \emph{\tychetest} and a stronger version of the later. Top: Per Dataset. Bottom: Overall. The one that we selected for \emph{\tychetest} is the most promising.}
    \label{fig:augexp:TycheTeS}
\end{figure}

We study different augmentation for \emph{\tychetest} and how the influence the quality of predictions. We consider four settings: no augmentation, only light augmentation such as Gaussian Noise and Gaussian Blur, described Table \ref{tab:sup:aug:tycheteS}, the \emph{\tychetest} augmentations shown Table \ref{tab:sup:aug:tycheteS} and the \emph{\tychetest} with slightly stronger parameters. shown Table \ref{tab:sup:aug:tycheteS_aggr}. The results are shown in Figure \ref{fig:augexp:TycheTeS} both aggregated across datasets and for each dataset individually. Overall, adding augmentations improves the best candidate Dice score. However, it can degrades the quality of the predictions, for instance for STARE. The augmentation we selected for \emph{\tychetest} is the most promising so far, without requiring inversion of the transform applied.

\begin{table}[]
    \centering
   \rowcolors{2}{white}{gray!15}
    \begin{tabular}{lcc}
   \textbf{\emph{\tychetest} High Aug.} & $p$ & Parameters\\
    \hline
        &  & $\sigma\in[0.5, 1.0]$ \\
    \multirow{-2}{*}{\cellcolor{gray!15} Gaussian Blur}     & \multirow{-2}{*}{\cellcolor{gray!15} 0.25} & \cellcolor{gray!15} $k=5$ \\
    \cellcolor{white} & \cellcolor{white} & \cellcolor{white} $\mu\in[0.4, 0.5]$\\
    \multirow{-2}{*}{\cellcolor{white} Gaussian Noise} & \multirow{-2}{*}{\cellcolor{white} 0.5} &  $ \sigma \in [0.1, 0.2]$\\
    Flip Intensities & 0.5 & None\\
    Sharpness & 0.25 & sharpness=5\\
     &  & brightness$\in[-0.1, 0.1]$, \\
     \multirow{-2}{*}{\cellcolor{gray!15}{Brightness Contrast}} &  \multirow{-2}{*}{\cellcolor{gray!15}{0.25}} & \cellcolor{gray!15} contrast$\in[0.5, 1.5]$ \\\hline
    \end{tabular}
    \caption{\textbf{High augmentations used to validate \emph{\tychetest}.} We focus on intensity transforms, to avoid inverting the prediction. For each image, an augmentation is sampled with probability $p$.}
    \label{tab:sup:aug:tycheteS_aggr}
\end{table}

\subsection{Few Shot Regime}
\begin{figure}
    \centering
    \includegraphics[width=0.5\textwidth]{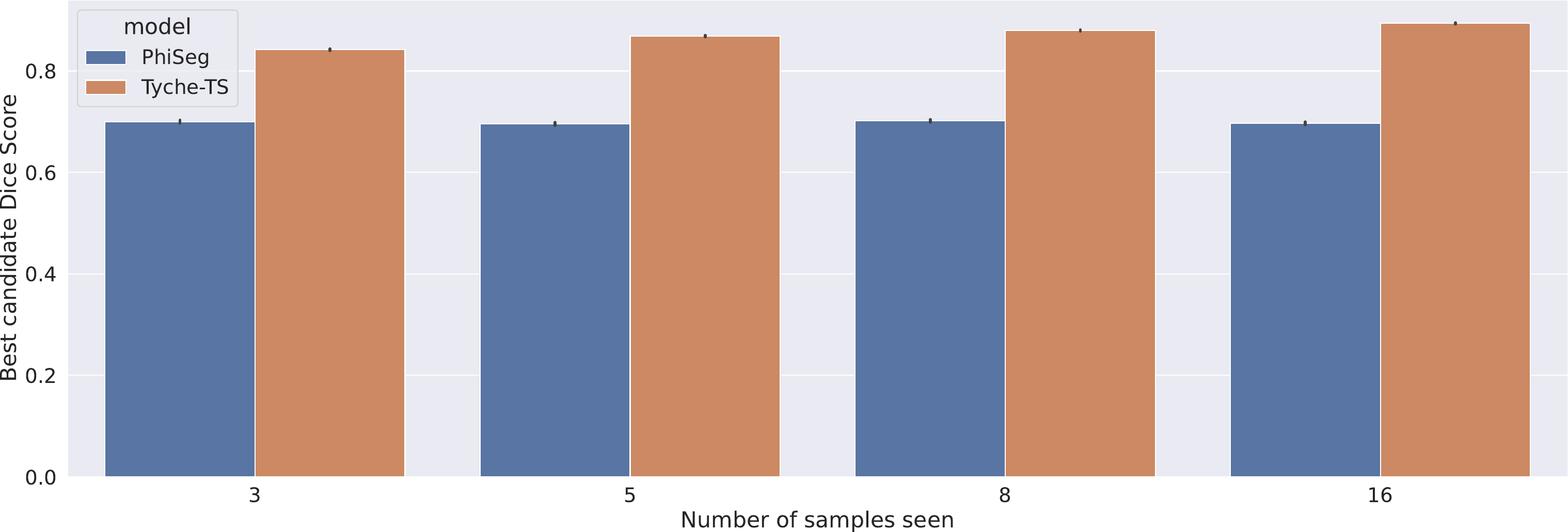}
    \includegraphics[width=0.5\textwidth]{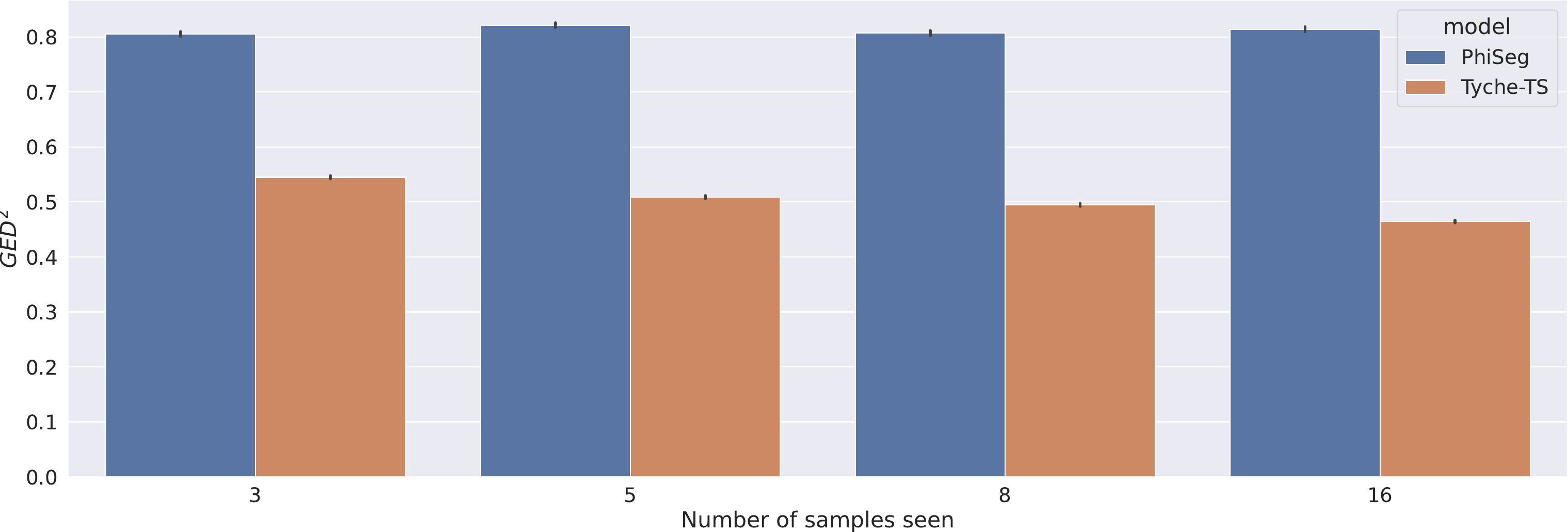}
    \caption{\textbf{Few-Shot Regime.} Comparison between \emph{\tychetrain} and PhiSeg trained on the few-shot examples. PhiSeg fails to learn from very few samples both on max. Dice score and GED, compared to \emph{Tyche}.}
    \label{fig:fsbdged}
\end{figure}

We train PhiSeg on a subset of LIDC-IDRI, $\tilde{\mathcal{S}}$, and examine how this network generalizes compared to Tyche, where the context set is $\tilde{\mathcal{S}}$. We investigate four few-shot settings: 3, 5, 8 and 16. For each, we train 30 PhiSegs: 10 seeds to account for variability in our samples and three data augmentation regimes (none, light, normal). For each seed and each few-shot setting, we select the PhiSeg that does best on the validation set. Figure \ref{fig:fsbdged} shows that \emph{Tyche} can leverage the data available much more effectively than PhiSeg, which fails to learn with so little samples.

\newpage


\section{Further Evaluation}

\begin{figure}
    \centering
    \includegraphics[width=0.45\textwidth]{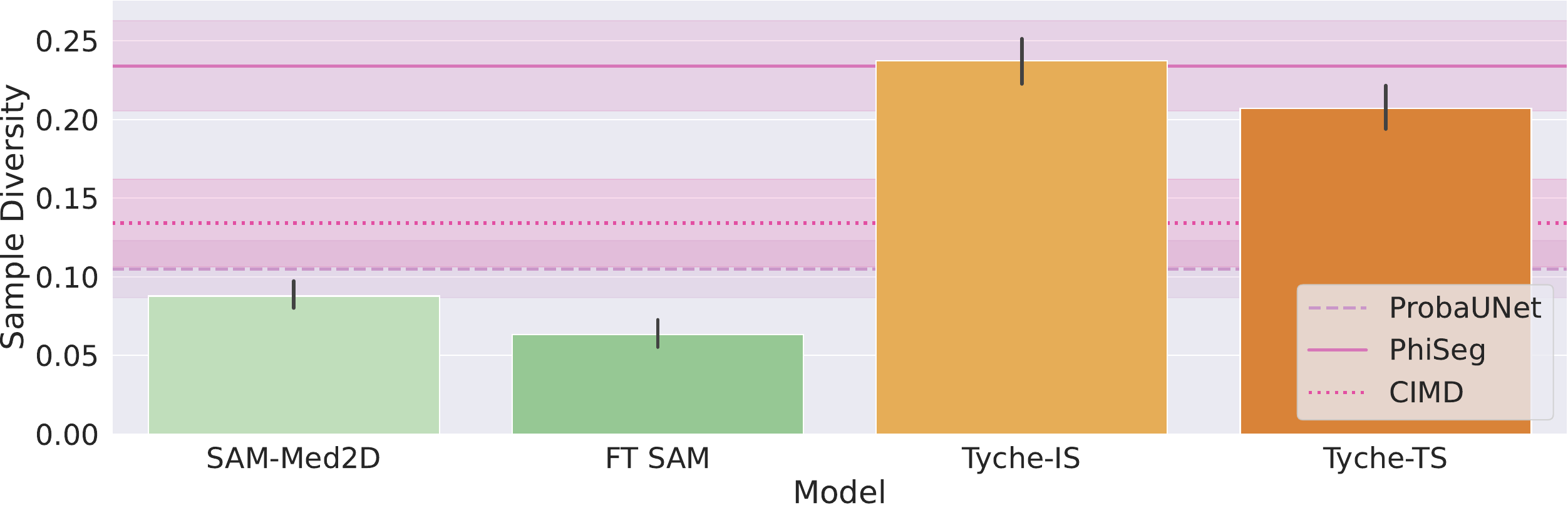}

    \caption{\textbf{Sample Diversity on Multi-Annotator Data.} Since the sample diversity of the deterministic methods is 0, we do not show them here. \emph{\tychetest} produces the most diverse samples, while Fine-Tuned SAM has very low diversity, despite varying clicks and bounding box locations. Higher is better.}
    \label{fig:sup:div}
\end{figure}

\begin{figure}
    \centering
\includegraphics[width=0.45\textwidth]{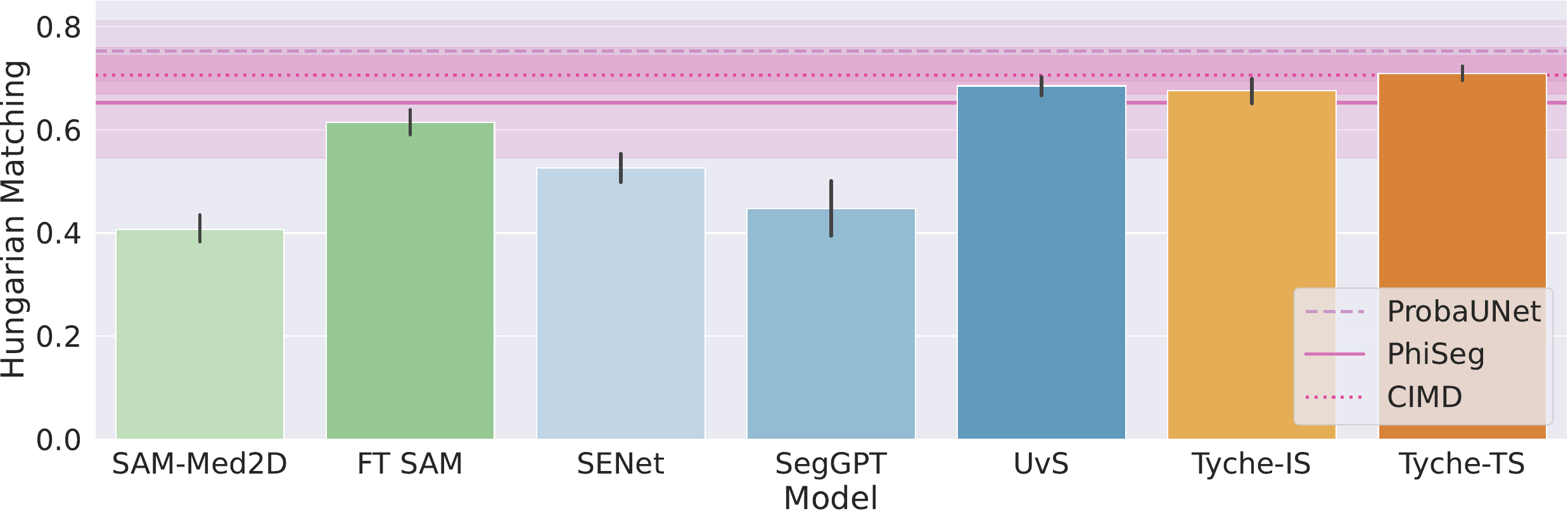}
    \caption{\textbf{Hungarian Matching on Multi-Annotator Data.} Both \emph{\tychetest} and UniverSeg perform well. \emph{\tychetrain} performs best. Higher is better.}
    \label{fig:sup:hm}
\end{figure}

\subsection{Performances on other Metrics}
Evaluating the quality of different predictions can be challenging especially when different annotators are available. We proposed best candidate Dice score and Generalized Energy Distance. Some also analysed sample diversity~\cite{monteiro2020stochastic}, and Hungarian Matching~\cite{kohl2019hierarchical,zbinden2023stochastic}. Sample diversity consists in measuring the agreement between candidate predictions $\hat{\mathcal{Y}}$, rewarding most diverse sets of candidates: 
\begin{equation}
    D_{SD}(\hat{\mathcal{Y}}) = \mathbb{E} \left[d(\hat{p}, \hat{p}')\right],
\end{equation}
where $\hat{p}, \hat{p}' \sim \hat{\mathcal{Y}}$ and $d(\cdot, \cdot)$ is Dice score. One limitation of this metric is that it blindly rewards diversity without taking into account the natural ambiguity in the target. Ideally, when there is high ambiguity in the target, the samples are very diverse, and inversely, when there is low ambiguity, the segmentation candidates are not diverse. 

In the context of stochastic predictions, Hungarian Matching~\cite{kuhn1955hungarian} consists in matching the set of predictions with the set of annotations, so that an overall metric is minimized. We use negative Dice score. One limitation of this method is that it has to be adapted when the number of annotators does not match the number of prediction. The most widely used fix is to artificially inflate the number of predictions and the number of annotations to reach the least common multiple~\cite{kohl2019hierarchical}. We use this strategy here as well. 

Figures \ref{fig:sup:div} and \ref{fig:sup:hm} show the performances for sample diversity and Hungarian Matching respectively. 

\begin{figure}
    \centering
    \includegraphics[width=0.45\textwidth]{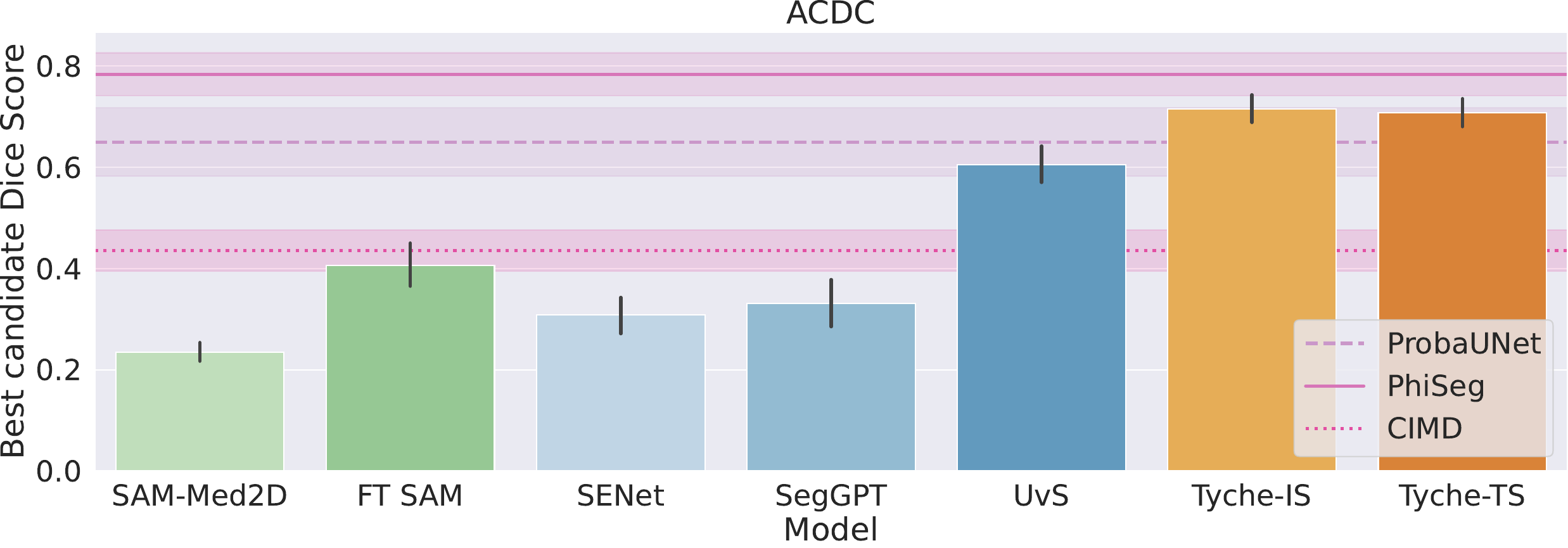}
    \includegraphics[width=0.45\textwidth]{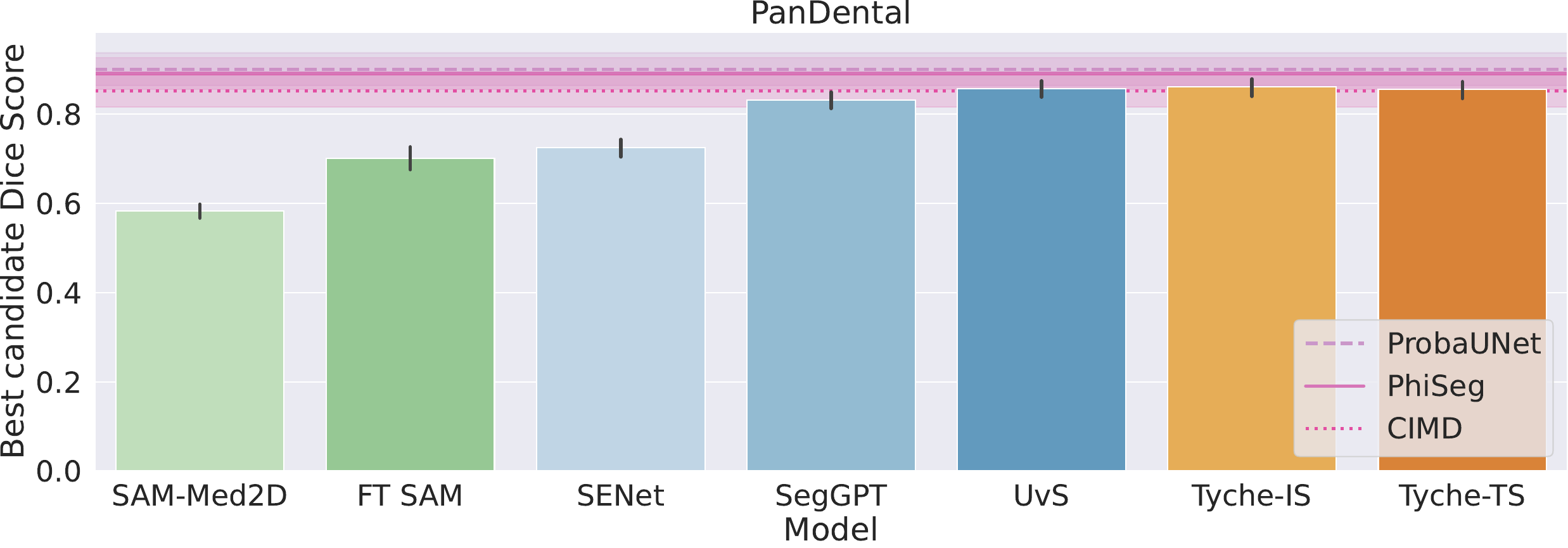}
    \includegraphics[width=0.45\textwidth]{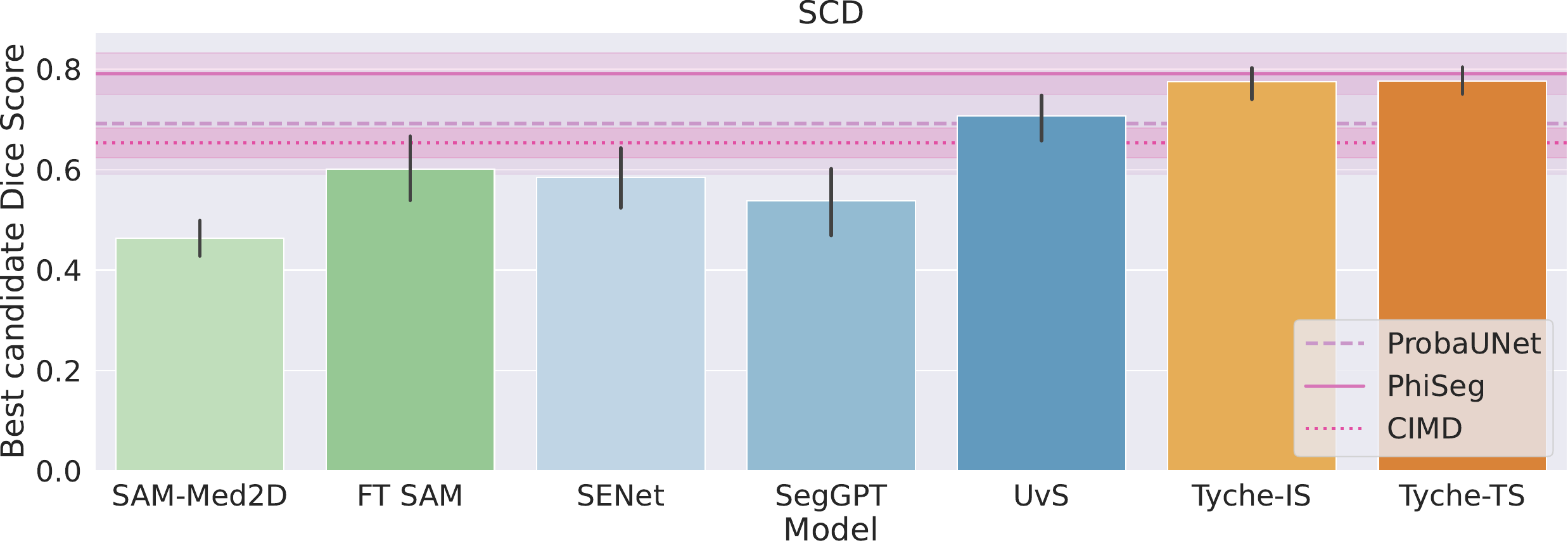}
    \includegraphics[width=0.45\textwidth]{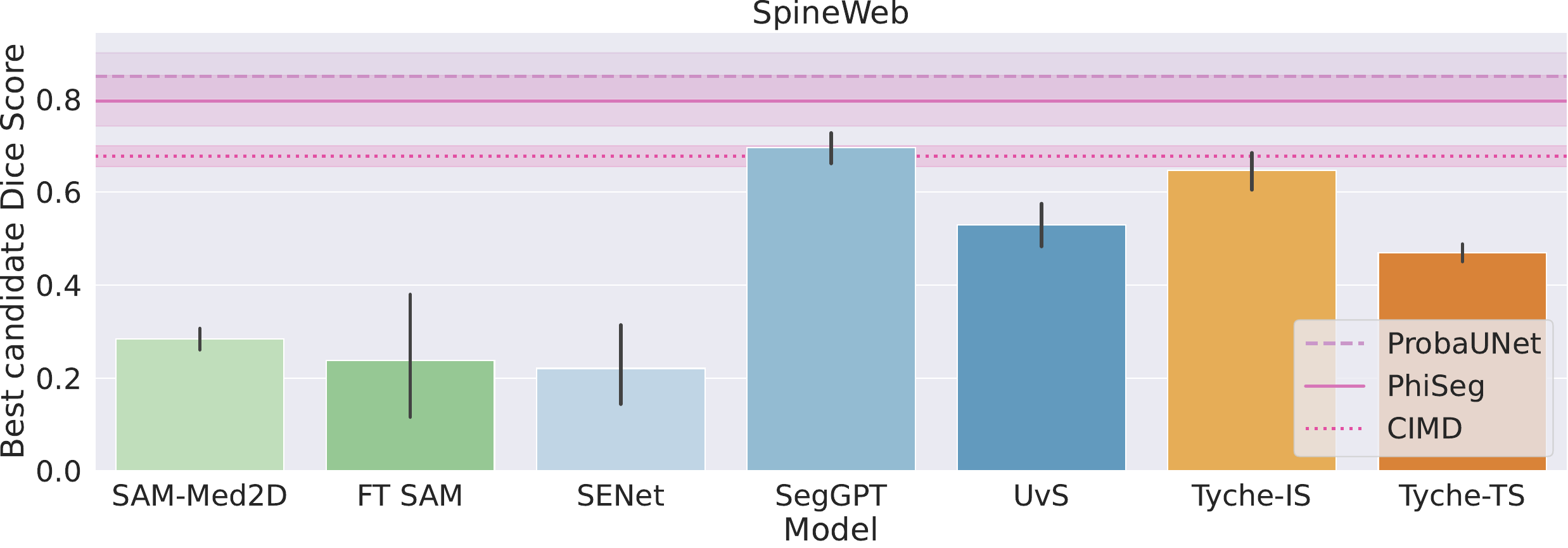}
    \includegraphics[width=0.45\textwidth]{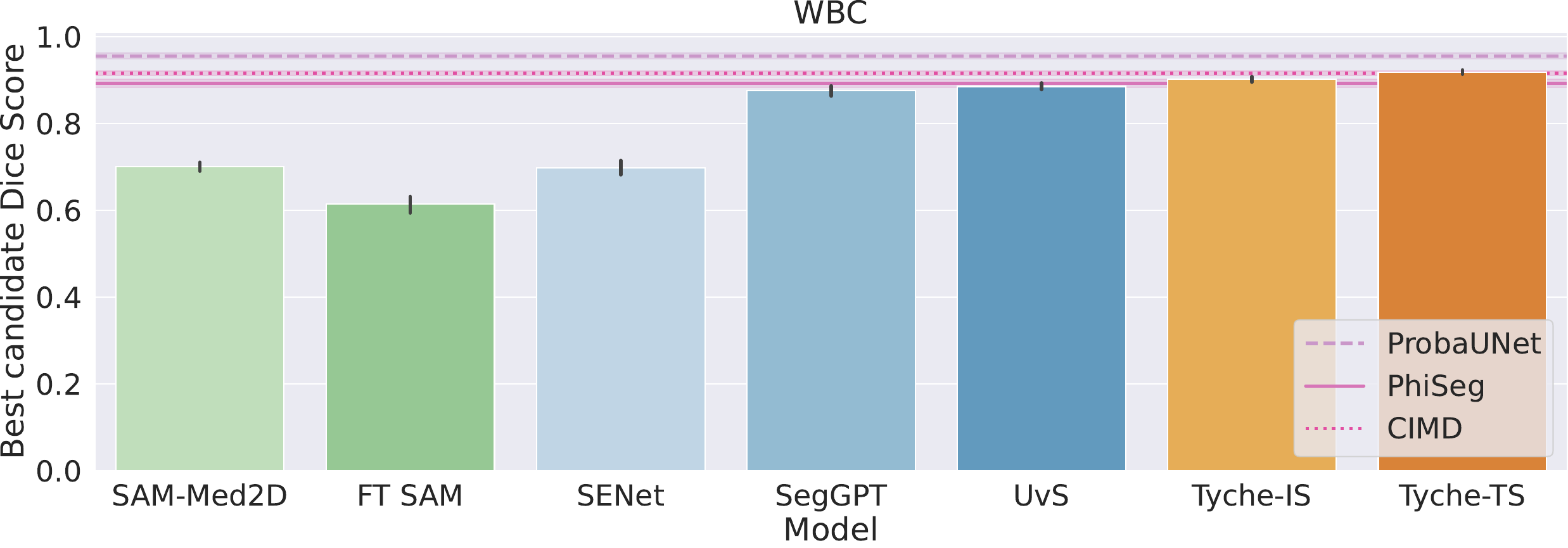}
    \caption{\textbf{Best candidate Dice score for the Single-Annotator Datasets.} Top to Bottom:  ACDC, PanDental, SCD, SpineWeb and WBC. \emph{Tyche} performs well in general except for SpineWeb.}
    \label{fig:SR_BD}
\end{figure}

\subsection{Per Dataset Results}
We show for each dataset and each method the best candidate Dice score. We also show for \emph{Tyche} and the different benchmarks Generalized Energy Distance, Hungarian Matching and Sample Diversity for the datasets with multiple annotations. 

\subpara{Best candidate Dice Score}
\label{sup:sr:BD}
Figure \ref{fig:SR_BD} shows the best candidate Dice score for the single-annotator datasets. \emph{Tyche} performs well across datasets. Both versions of \emph{Tyche} seem to dominate both types of benchmarks and be comparable to the upper bounds, except for one dataset: SpineWeb. We hypothesize that part of the performance drop is due to the nature of the structure to segment in SpineWeb: individual vertebra. Most of our training data contains single structures to segment. 

\begin{figure}
    \centering
    \includegraphics[width=0.45\textwidth]{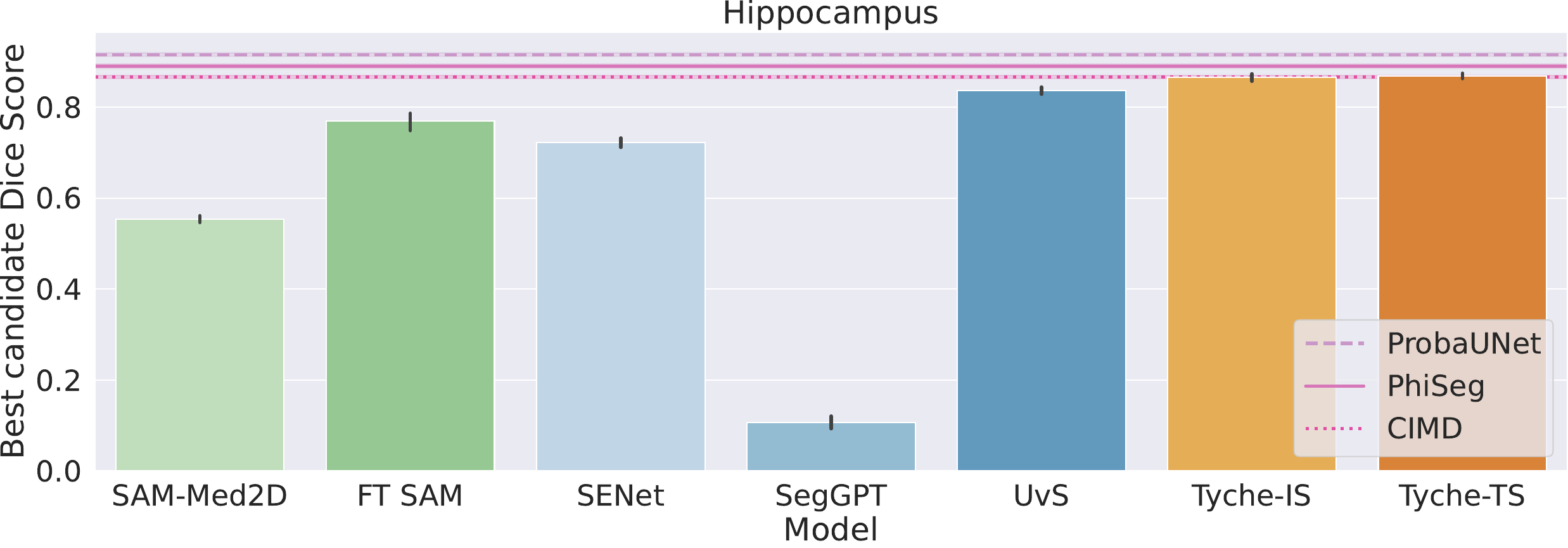}    

    \includegraphics[width=0.45\textwidth]{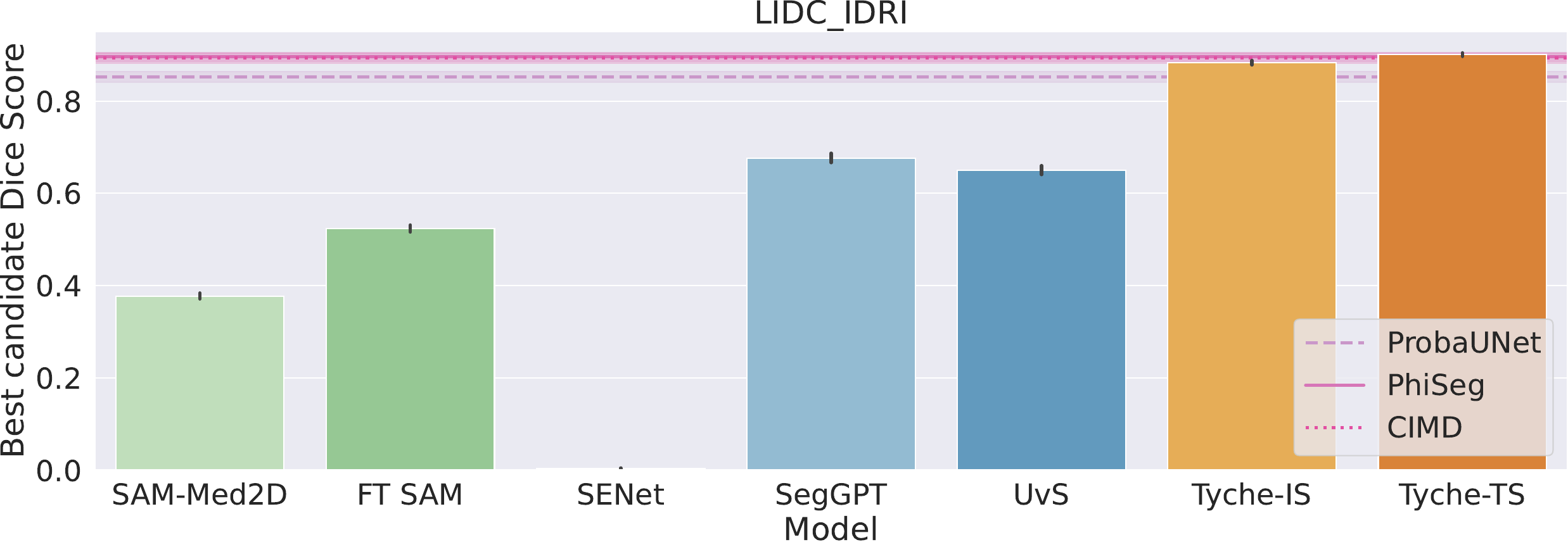}
    \includegraphics[width=0.45\textwidth]{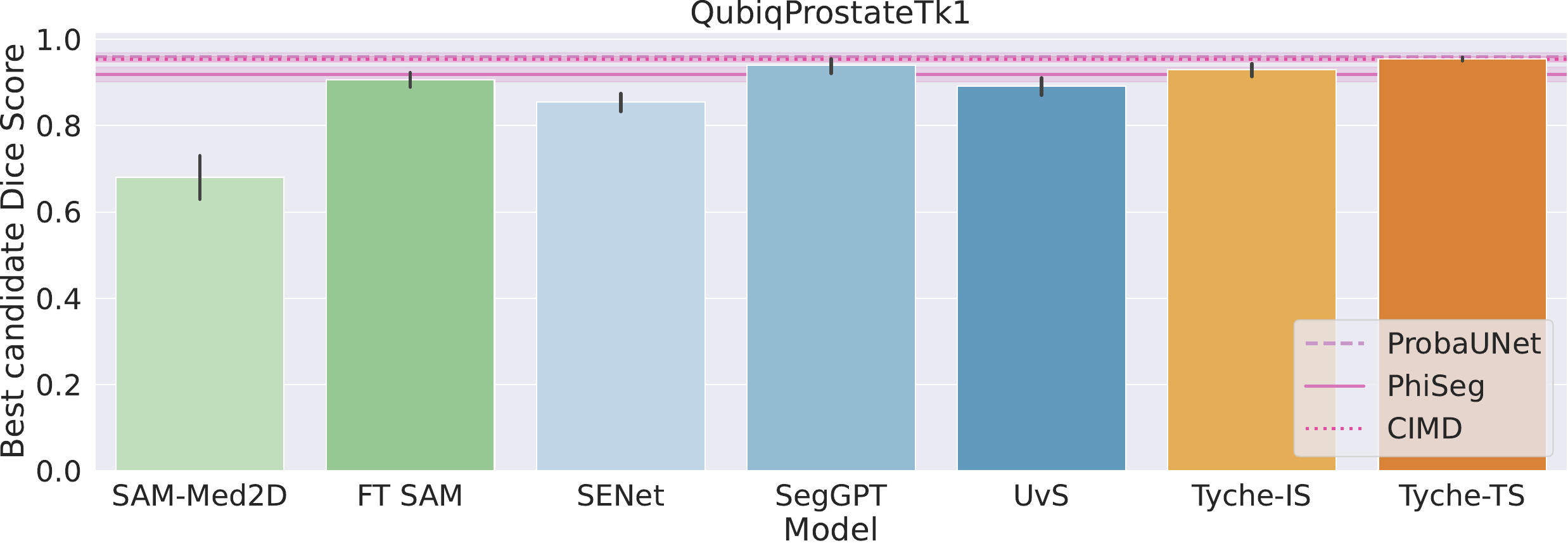}
    \includegraphics[width=0.45\textwidth]{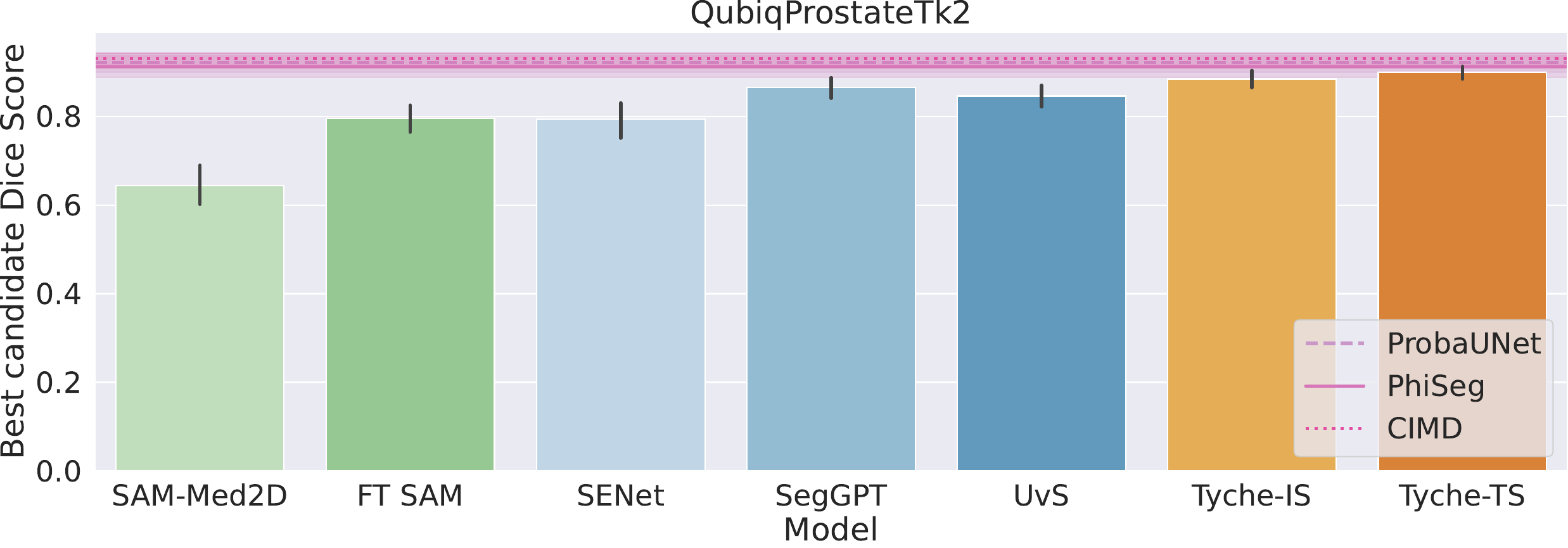}
    \includegraphics[width=0.45\textwidth]{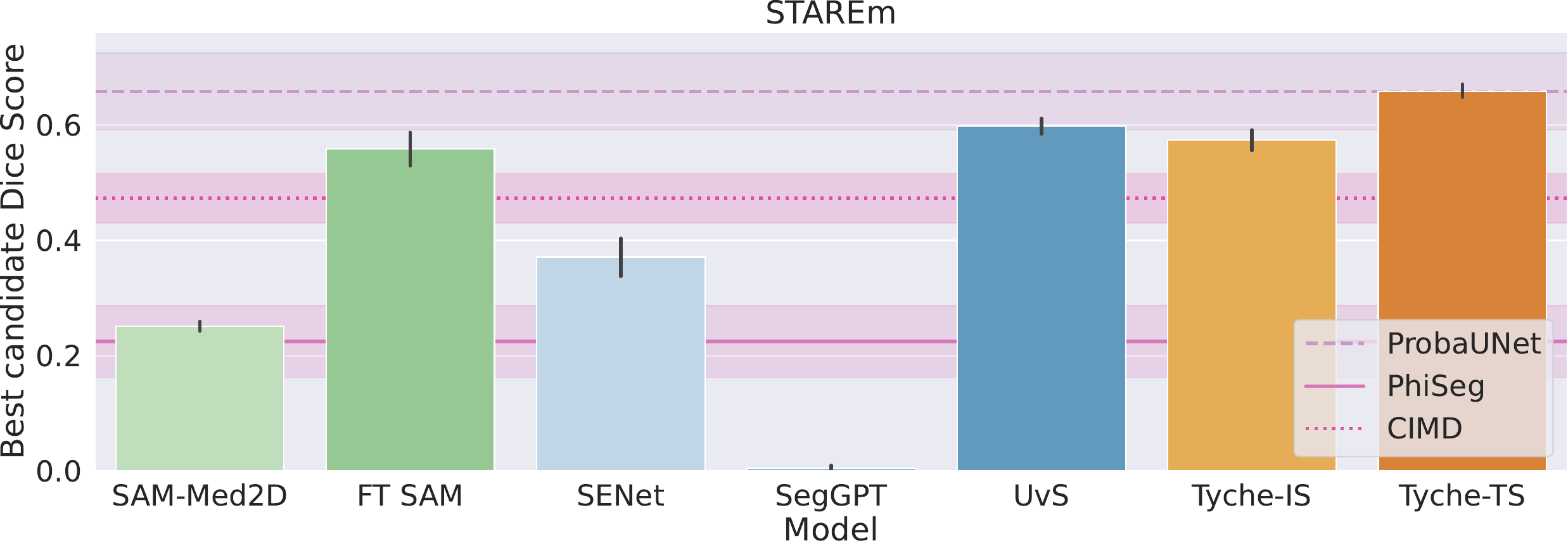}
    \caption{\textbf{Best candidate Dice Score for Multi-Annotator Datasets} Top to bottom: Hippocampus, LIDC-IDRI, Prostate Task 1, Prostate Task 2 and STARE. \emph{Tyche} performs well across datasets. (Higher is better.)}
    \label{fig:MR_BD}
\end{figure}

Figure \ref{fig:MR_BD} shows the best candidate Dice score for the multi-annotator datasets. Similar conclusions can be drawn as for the single-annotator data. Some benchmarks are particularly sensitive to the data they are evaluated on, for instance SegGPT. This methods performs really well on the Prostate data but quite poorly on the STARE and the Hippocampus data. We assume that because SegGPT was designed for images of 448x448, our images of dimension 128x128 might affect performances. 

\begin{figure}
    \centering
\includegraphics[width=0.45\textwidth]{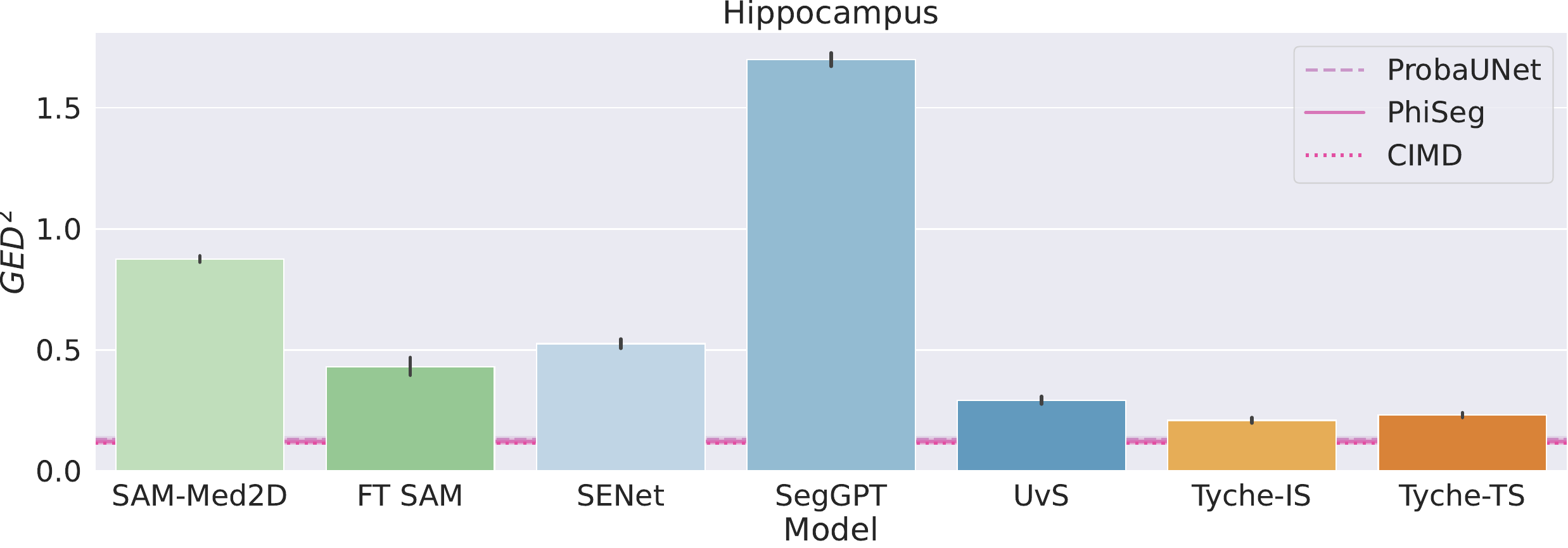}
    \includegraphics[width=0.45\textwidth]{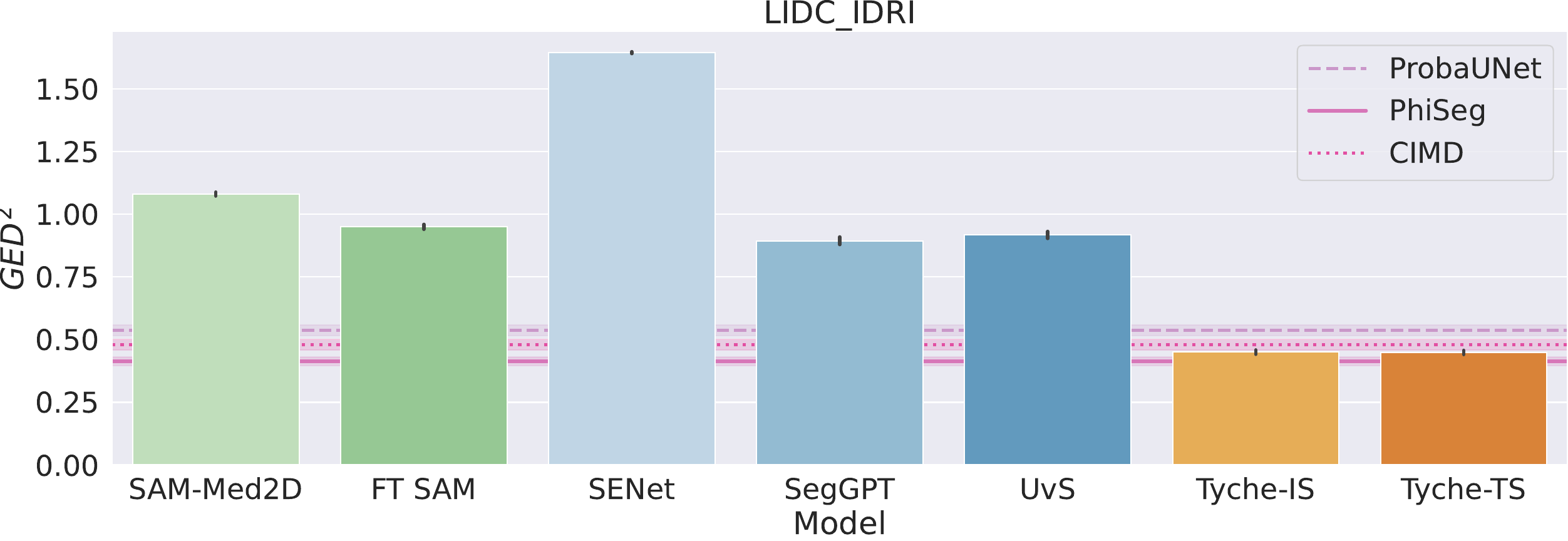}
\includegraphics[width=0.45\textwidth]{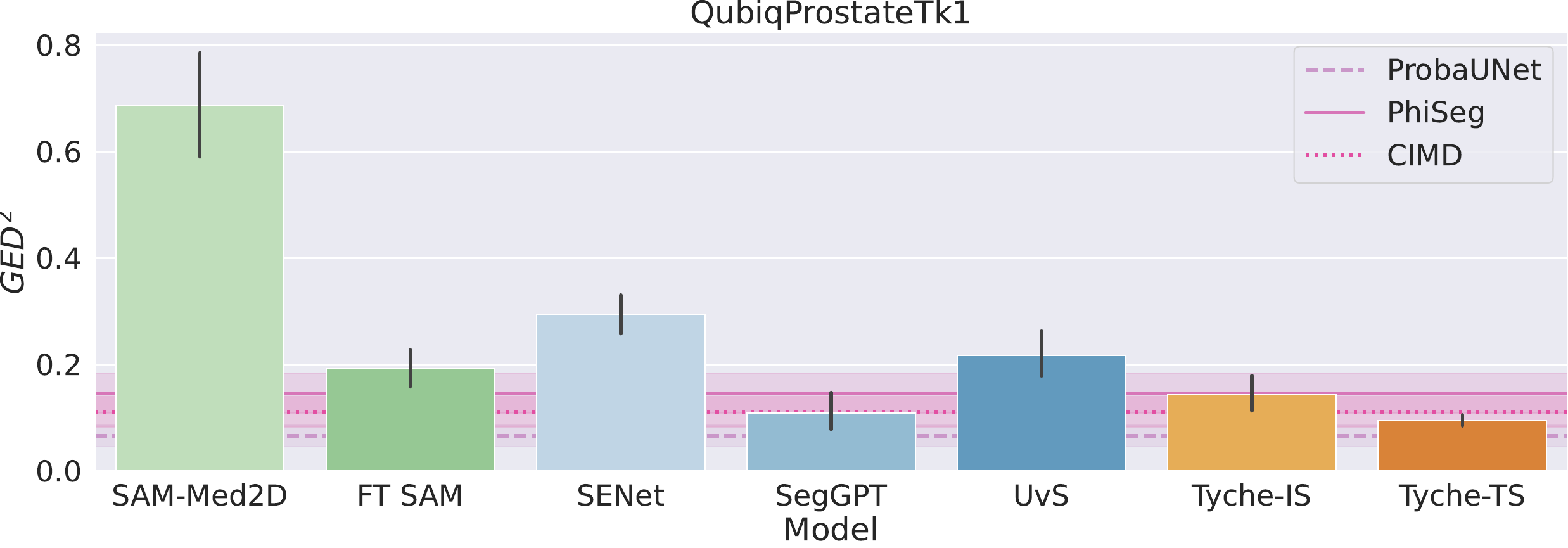}
\includegraphics[width=0.45\textwidth]{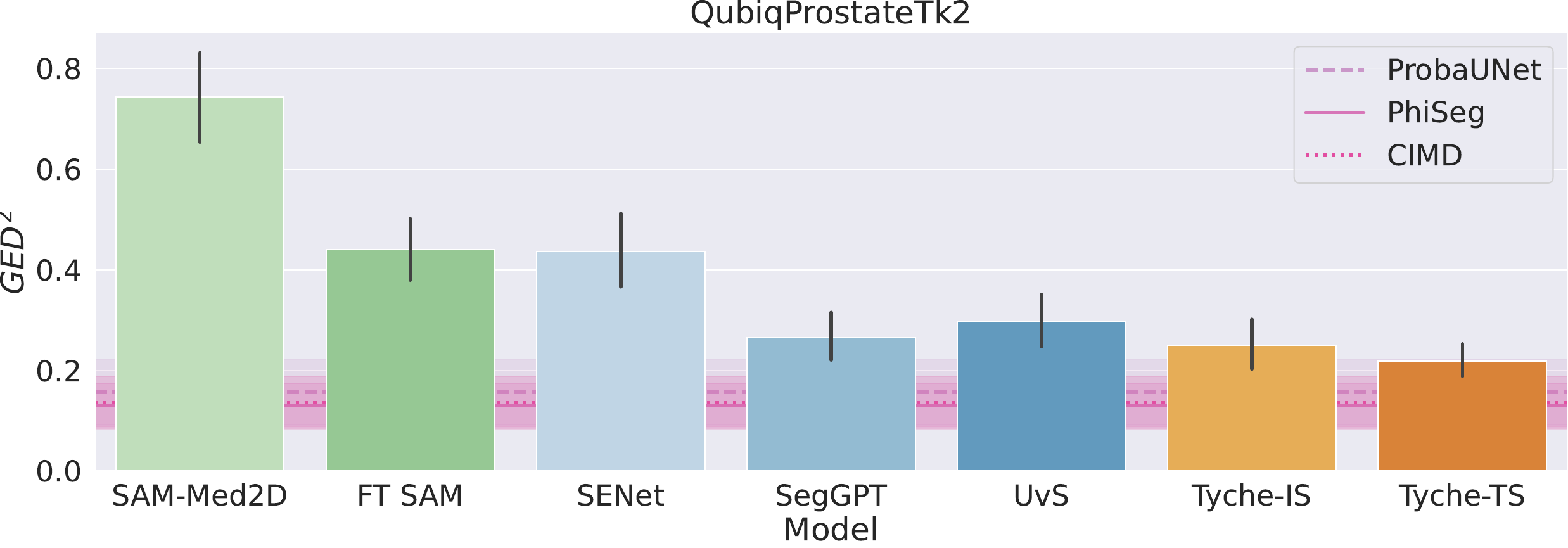}
    \includegraphics[width=0.45\textwidth]{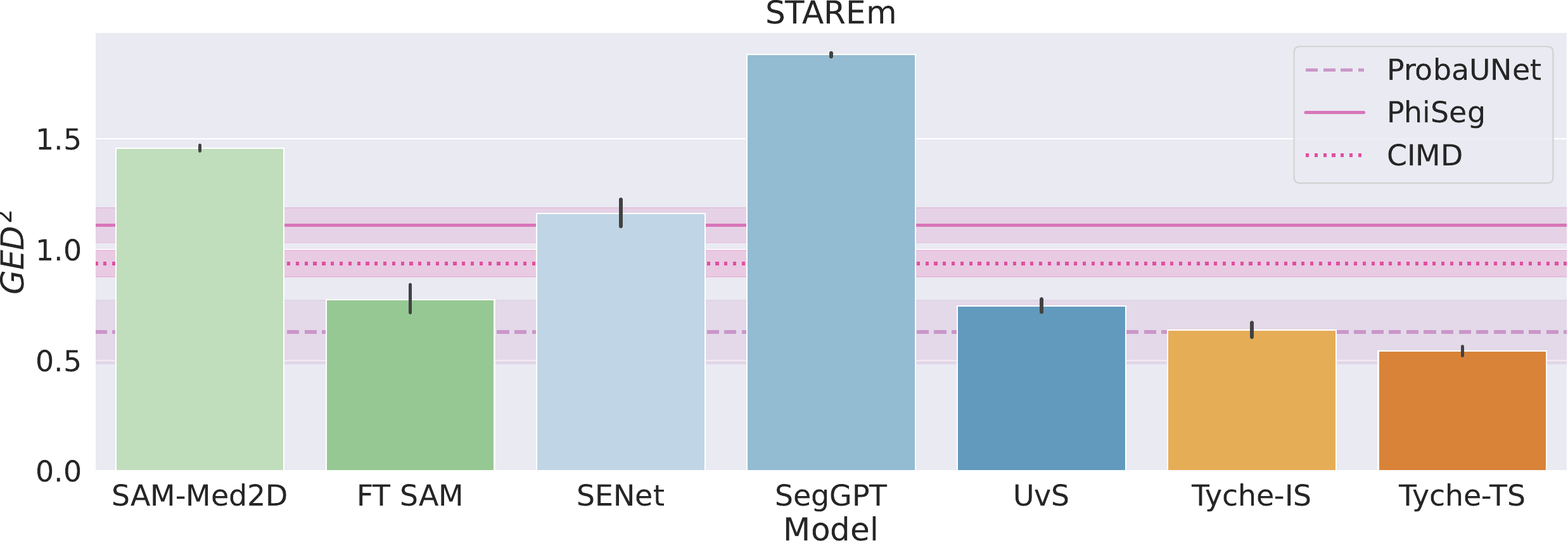}
    \caption{\textbf{Generalized Energy Distance for Multi-Annotator Datasets.} Top to bottom: Hippocampus, LIDC-IDRI, Prostate Task 1, Prostate Task 2 and STARE. \emph{Tyche} performs well across datasets. (Lower is better.)}
    \label{fig:MR_GED}
\end{figure}

\subpara{Generalized Energy Distance}
Figure \ref{fig:MR_GED} shows Generalized Energy distance for the multi-rater datasets. 

\begin{figure}
    \centering
    \includegraphics[width=0.45\textwidth]{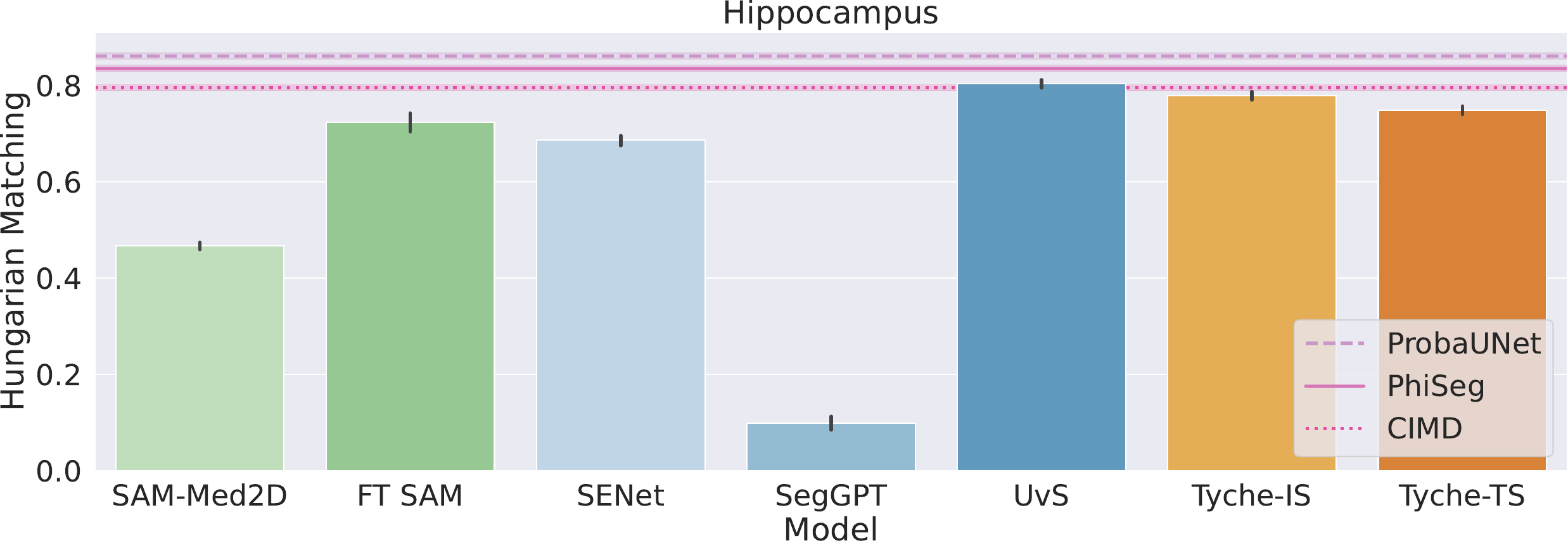}

    \includegraphics[width=0.45\textwidth]{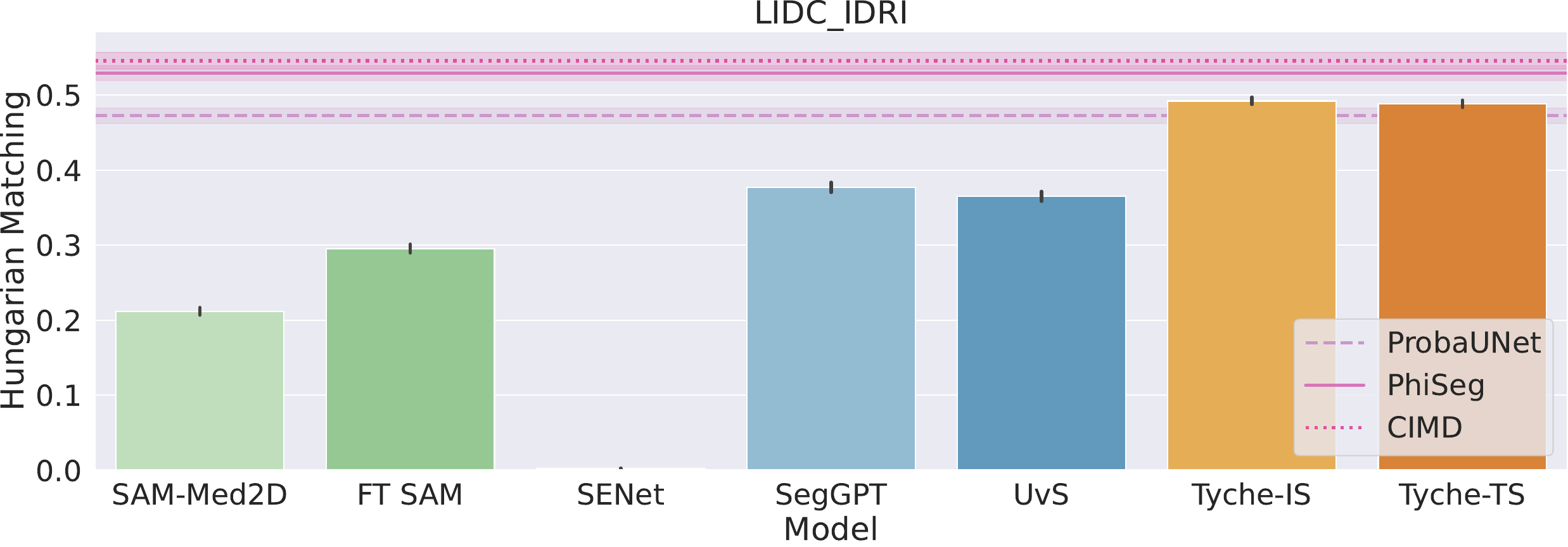}

    \includegraphics[width=0.45\textwidth]{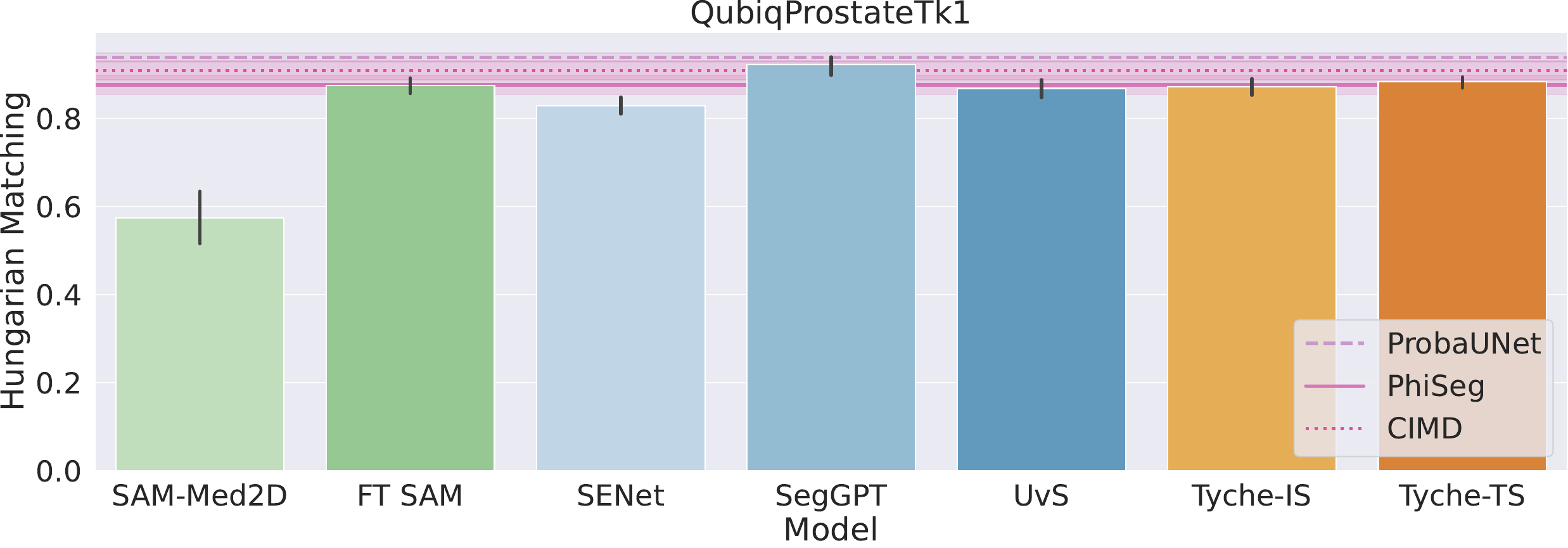}
    \includegraphics[width=0.45\textwidth]{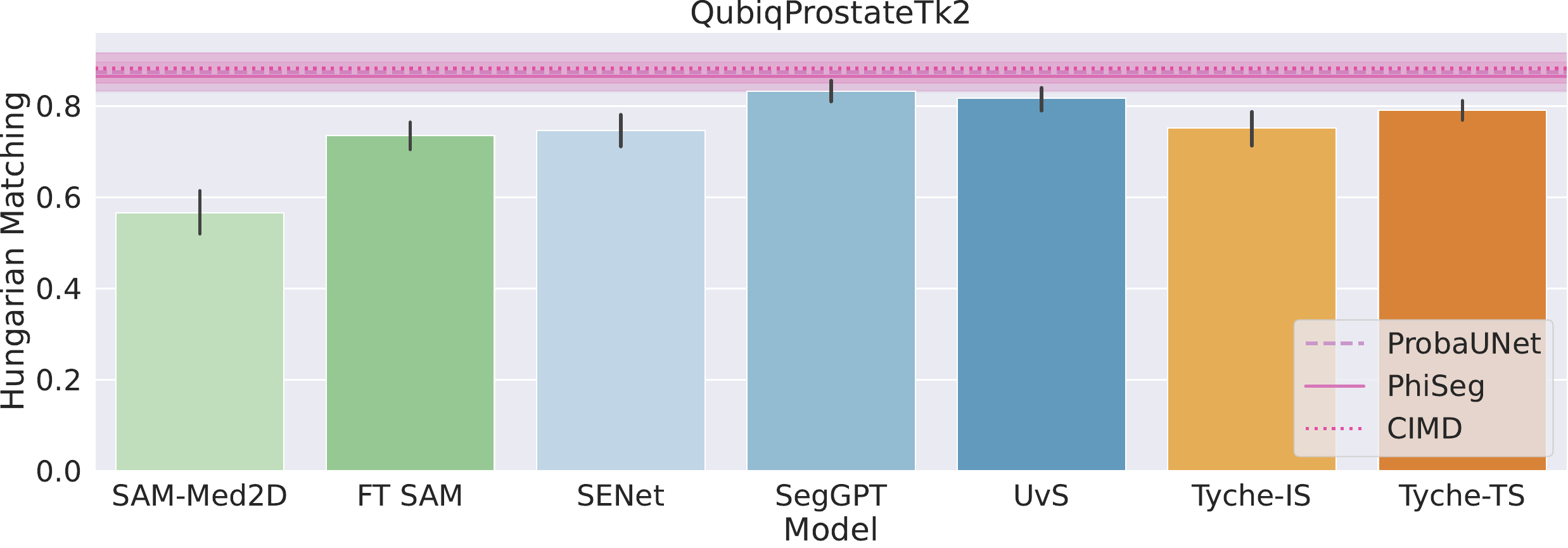}
    \includegraphics[width=0.45\textwidth]{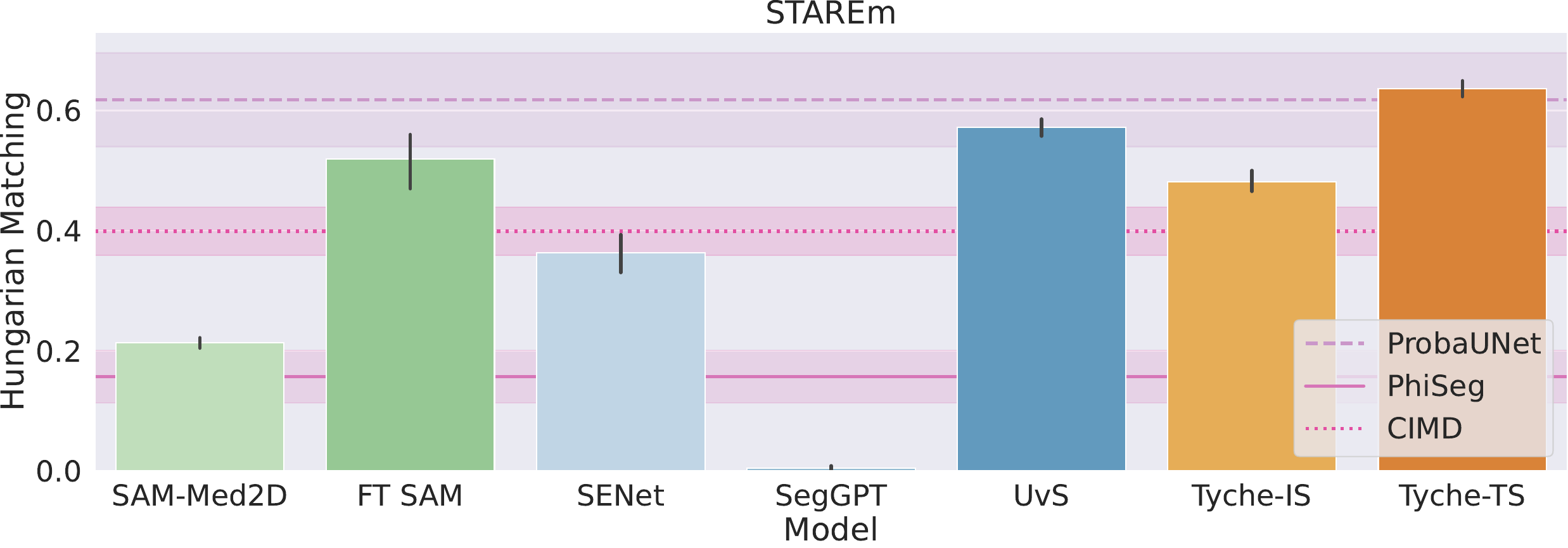}
    \caption{\textbf{Hungarian Matching Dice for Multi-Annotator Datasets.} Top to bottom: Hippocampus, LIDC-IDRI, Prostate Task 1, Prostate Task 2 and STARE. \emph{Tyche} performs well across datasets. (Higher is better.)}
    \label{fig:MR_HM}
\end{figure}

\subpara{Hungarian Matching}
Figure \ref{fig:MR_HM} shows the Hungarian Matching metric for the multi-rater datasets. We find that UniverSeg performs particularly well. We suspect that because we have to artificially duplicate our samples to compute this metric, the resulting scenario favors methods that are closer to the mean. 

\begin{figure}
    \centering
    \includegraphics[width=0.45\textwidth]{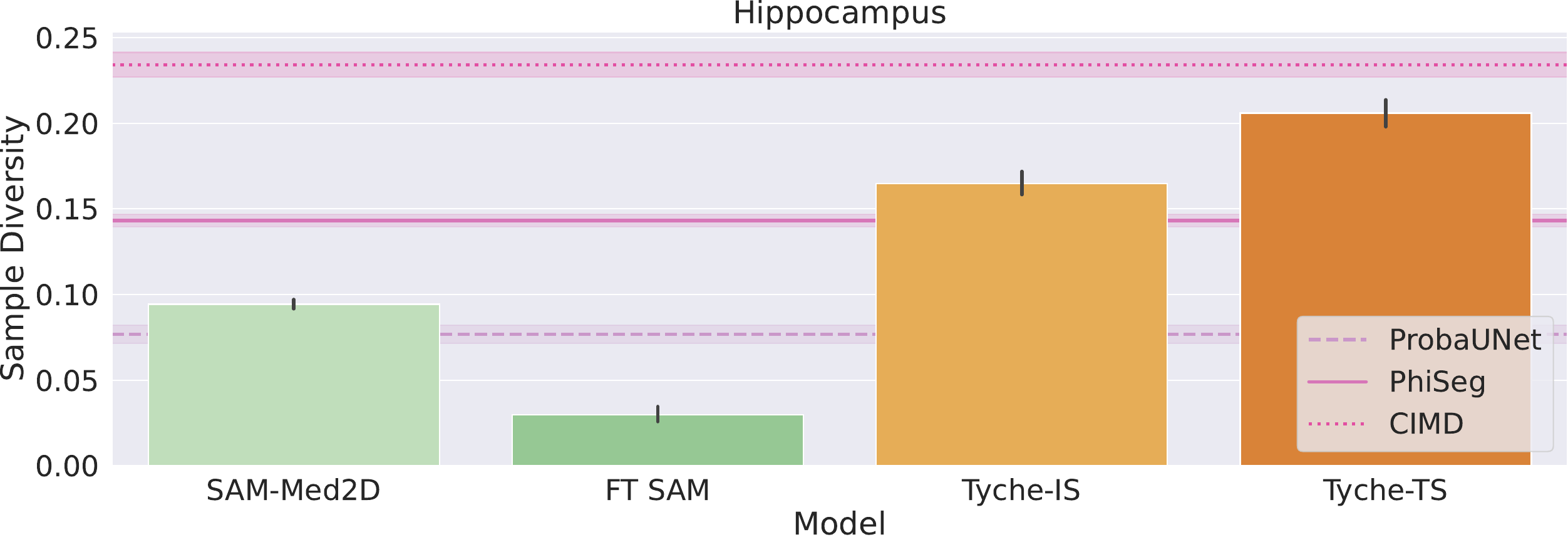}
    \includegraphics[width=0.45\textwidth]{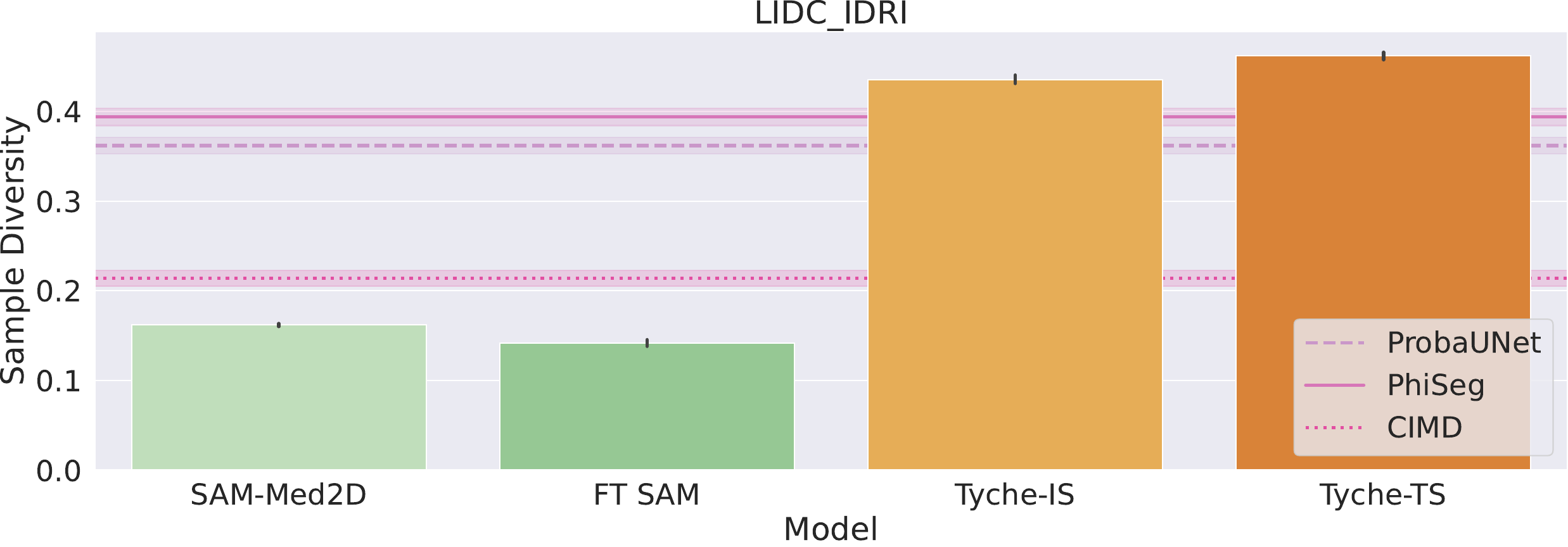}
    \includegraphics[width=0.45\textwidth]{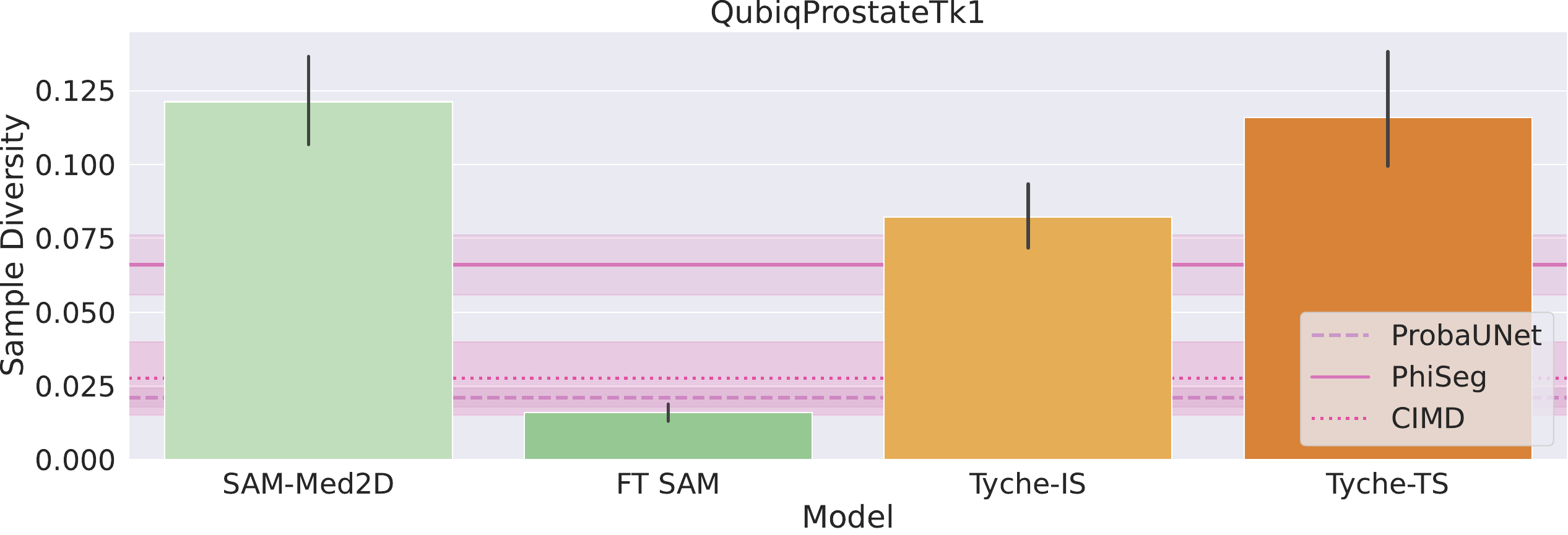}
    \includegraphics[width=0.45\textwidth]{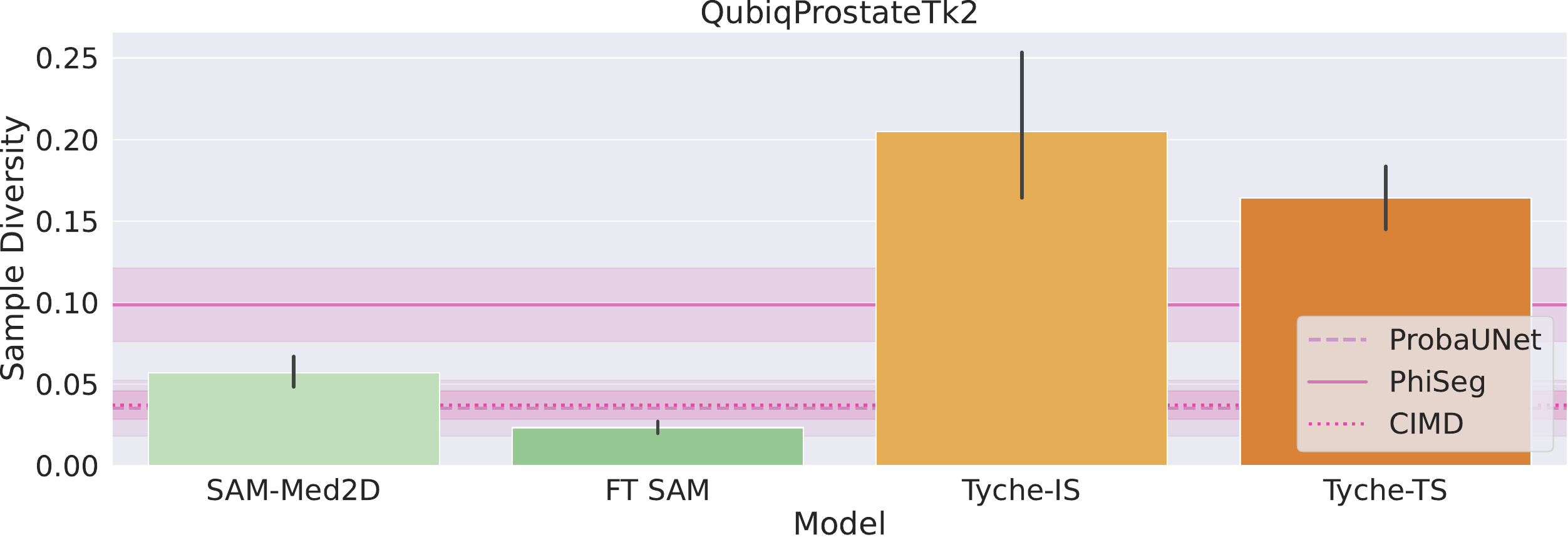}
    \includegraphics[width=0.45\textwidth]{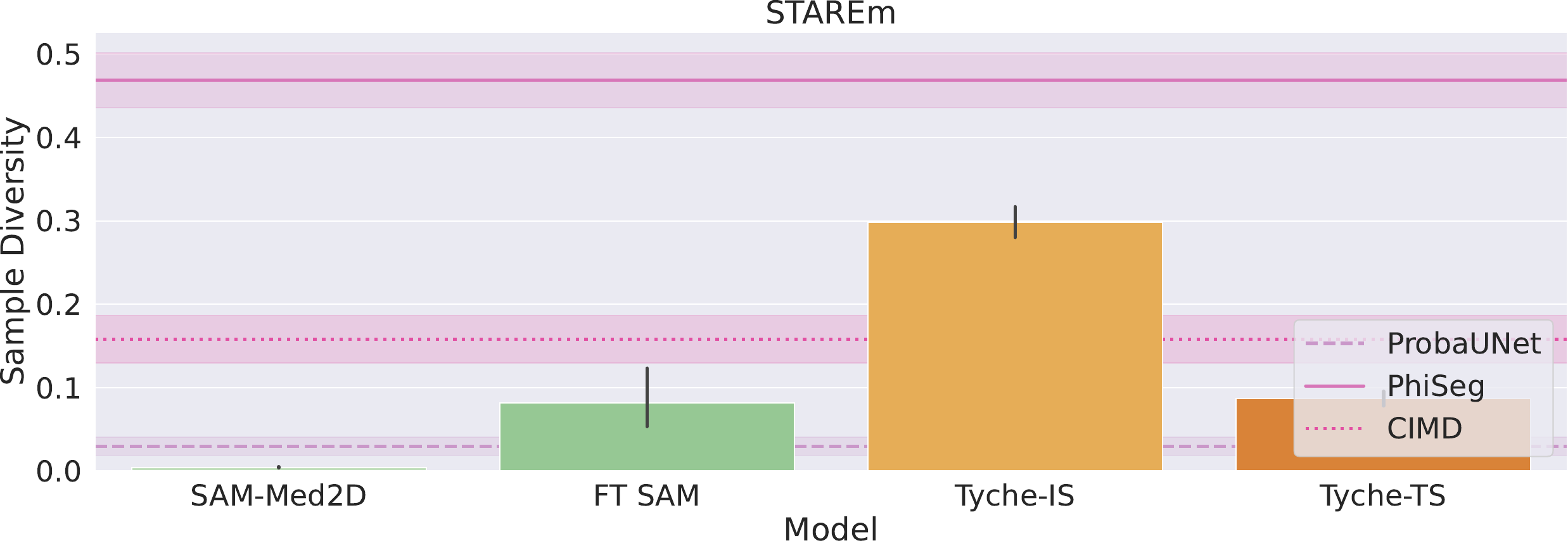}
    \caption{\textbf{Sample Diversity for Multi-Annotator Datasets.} Top to bottom: Hippocampus, LIDC-IDRI, Prostate Task 1, Prostate Task 2 and STARE. \emph{Tyche} performs well across datasets. We only show methods with diversity greater than 0. (Higher is better.)}
    \label{fig:MR_SD}
\end{figure}

\subpara{Sample Diversity}
Figure \ref{fig:MR_SD} shows sample diversity for the multi-annotator datasets. We only show the sample diversity for the methods outputing more than one segmentation candidate. For UniverSeg, SegGPT and SENet, the sample diversity is trivially 0. 

\newpage
\newpage
\subsection{Additional Visualizations}
We show additional visualization for \emph{Tyche} frameworks as well as all the benchmarks. 

\paragraph{Single-Annotator Data}

\begin{figure*}
    \centering
    \includegraphics[width=\textwidth]{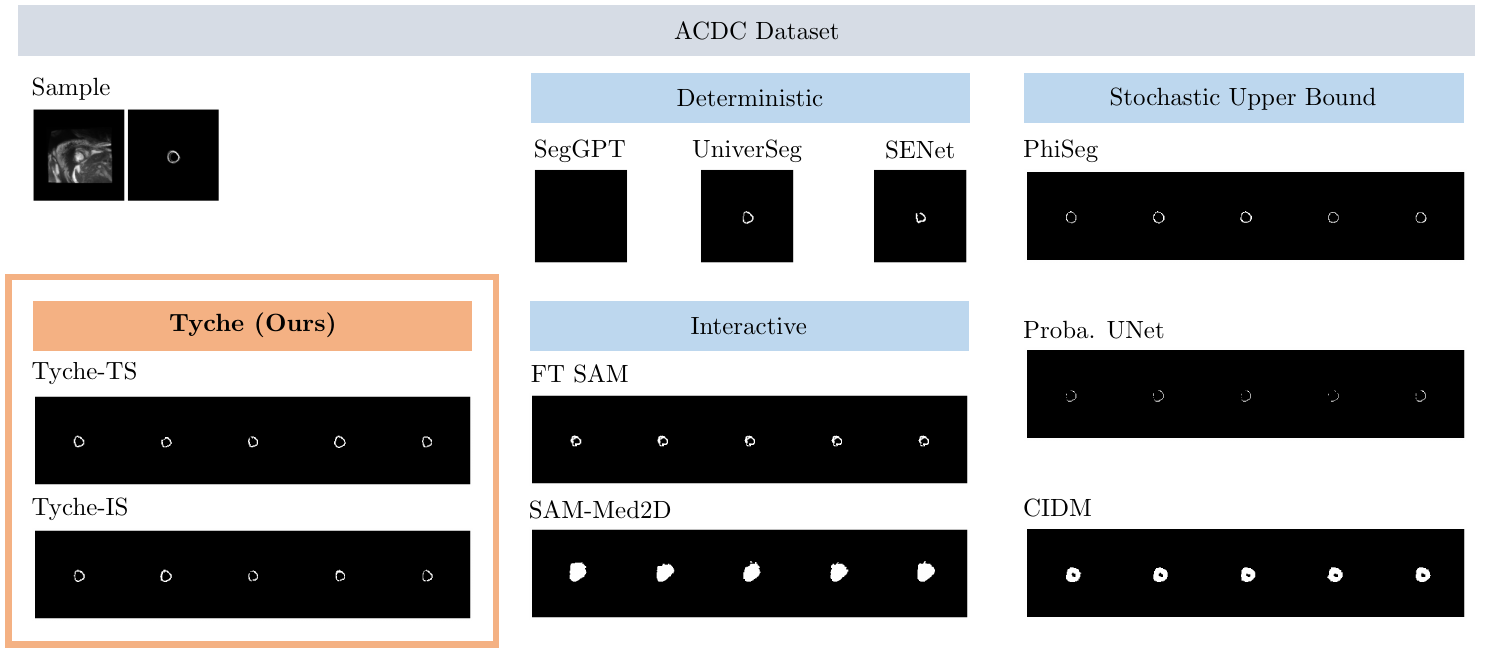}
    \caption{\textbf{Example Prediction for ACDC.}}
    \label{fig:sup:viz_ACDC}
\end{figure*}

\begin{figure*}
    \centering
    \includegraphics[width=\textwidth]{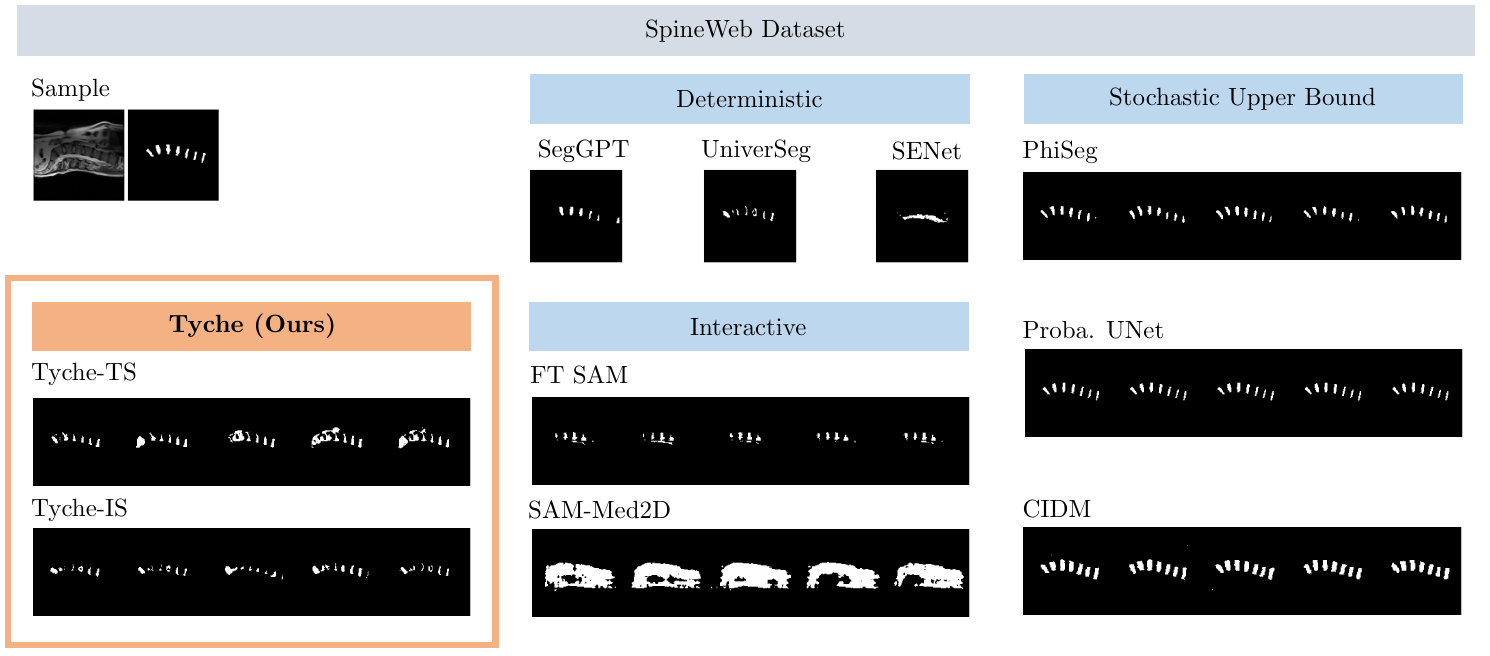}
    \caption{\textbf{Example Prediction for SpineWeb.}}
    \label{fig:sup:viz_SpineWeb}
\end{figure*}

\begin{figure*}
    \centering
    \includegraphics[width=\textwidth]{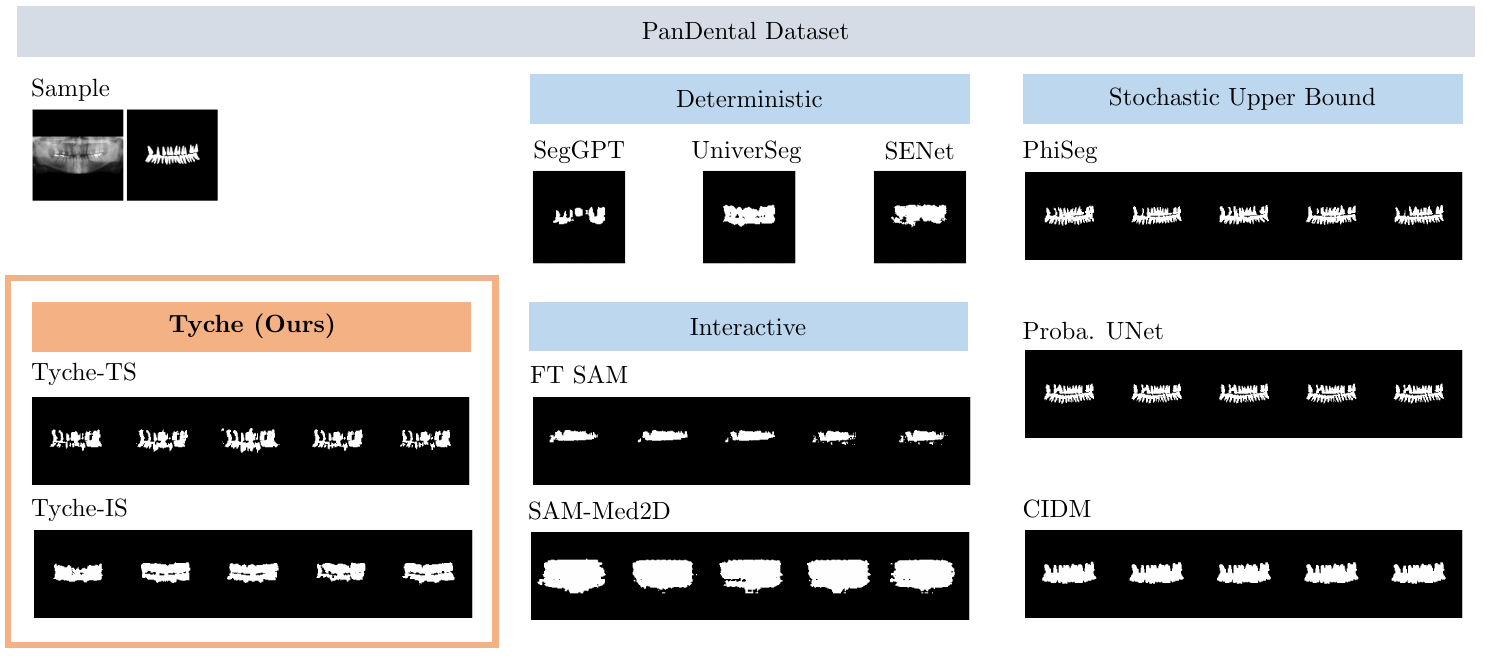}
    \caption{\textbf{Example Prediction for PanDental, task 1.}}
    \label{fig:sup:viz_PanDental1}
\end{figure*}

\begin{figure*}
    \centering
    \includegraphics[width=\textwidth]{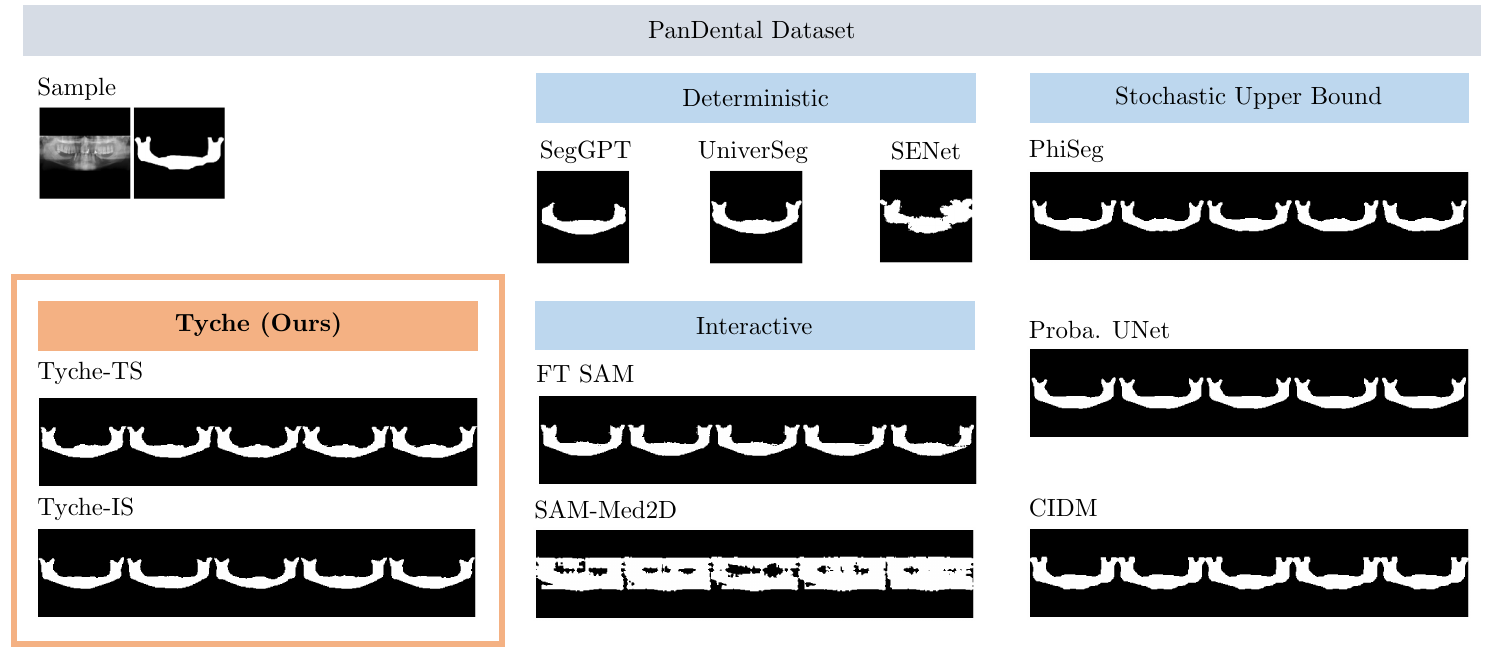}
    \caption{\textbf{Example Prediction for PanDental, task 2.}}
    \label{fig:sup:viz_PanDental2}
\end{figure*}

\begin{figure*}
    \centering
    \includegraphics[width=\textwidth]{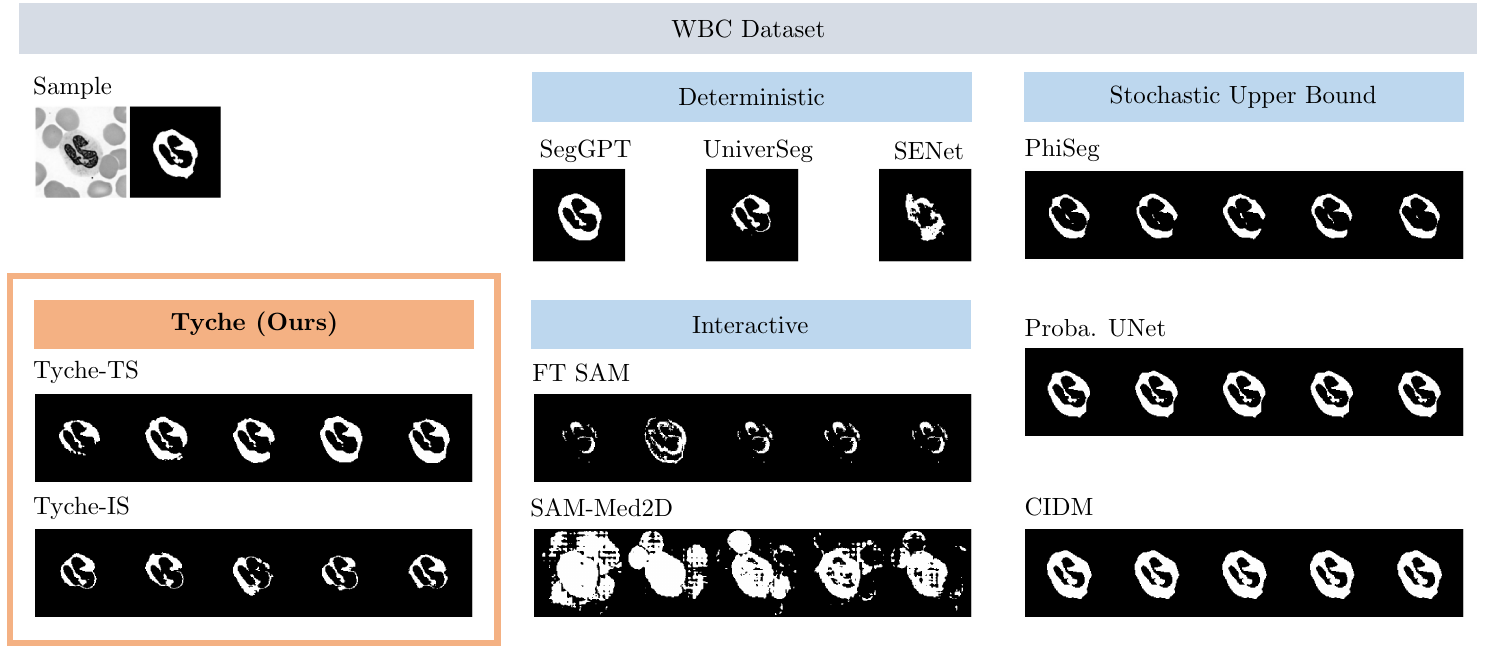}
    \caption{\textbf{Example Prediction for WBC.}}
    \label{fig:sup:viz_WBC}
\end{figure*}

Visualizations for ACDC are shown Figure \ref{fig:sup:viz_ACDC}. We show two PanDental examples Figure \ref{fig:sup:viz_PanDental1} and Figure \ref{fig:sup:viz_PanDental2}, one for each task. We show an example for SpineWeb Figure \ref{fig:sup:viz_SpineWeb} and one for WBC Figure \ref{fig:sup:viz_WBC}.
\label{sup:sr:results}


\begin{figure*}
    \centering
    \includegraphics[width=\textwidth]{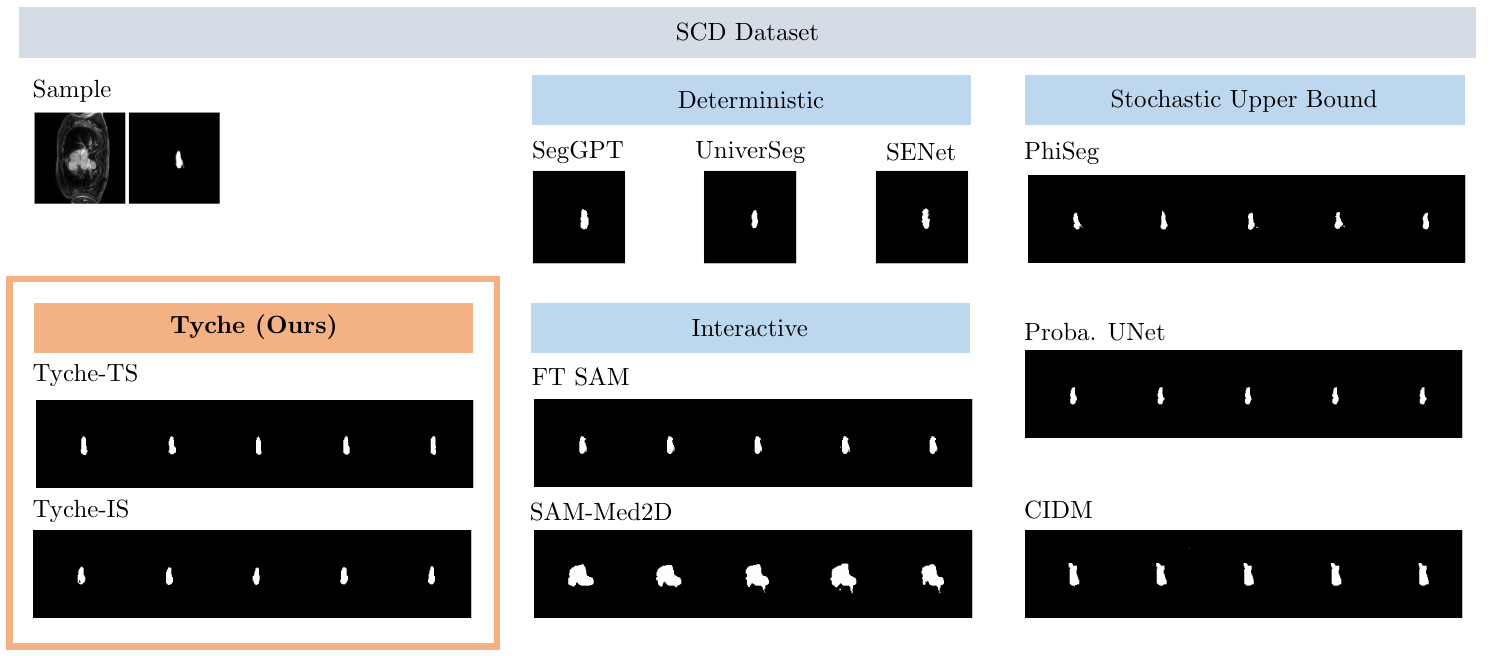}
    \caption{\textbf{Example Prediction for SCD.}}
    \label{fig:sup:viz_SCD}
\end{figure*}

\begin{figure*}
    \centering
    \includegraphics[width=\textwidth]{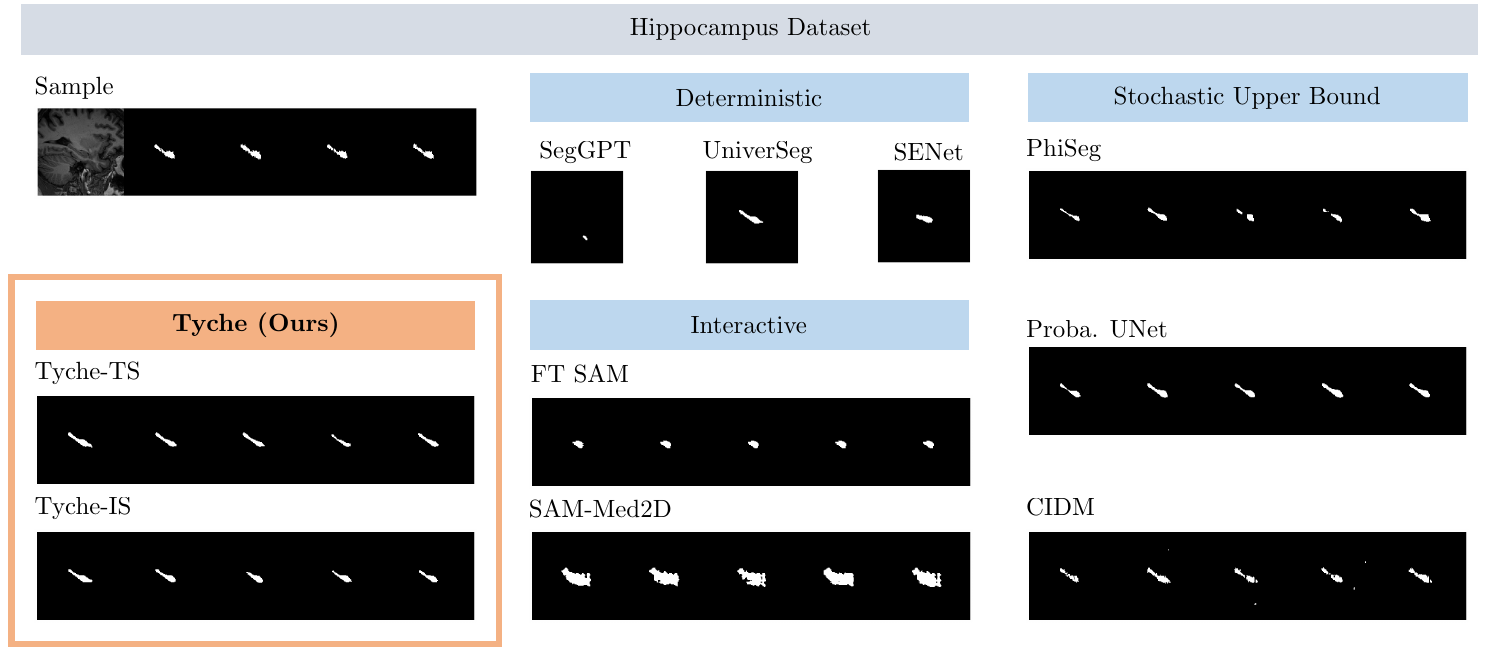}
    \caption{\textbf{Example Prediction for Hippocampus.}}
    \label{fig:sup:viz_Hippo}
\end{figure*}

\begin{figure*}
    \centering
    \includegraphics[width=\textwidth]{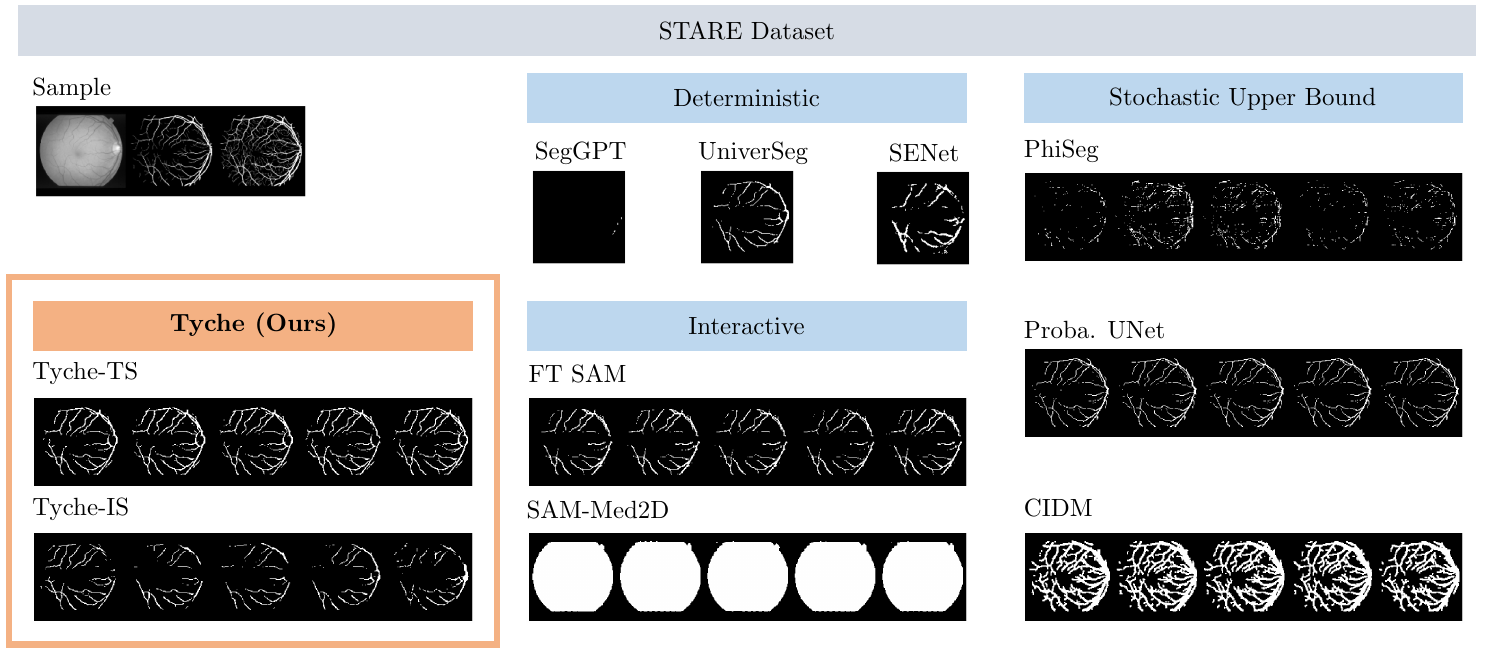}
    \caption{\textbf{Example Prediction for STARE.}}
    \label{fig:sup:viz_STARE}
\end{figure*}

\begin{figure*}
    \centering
    \includegraphics[width=\textwidth]{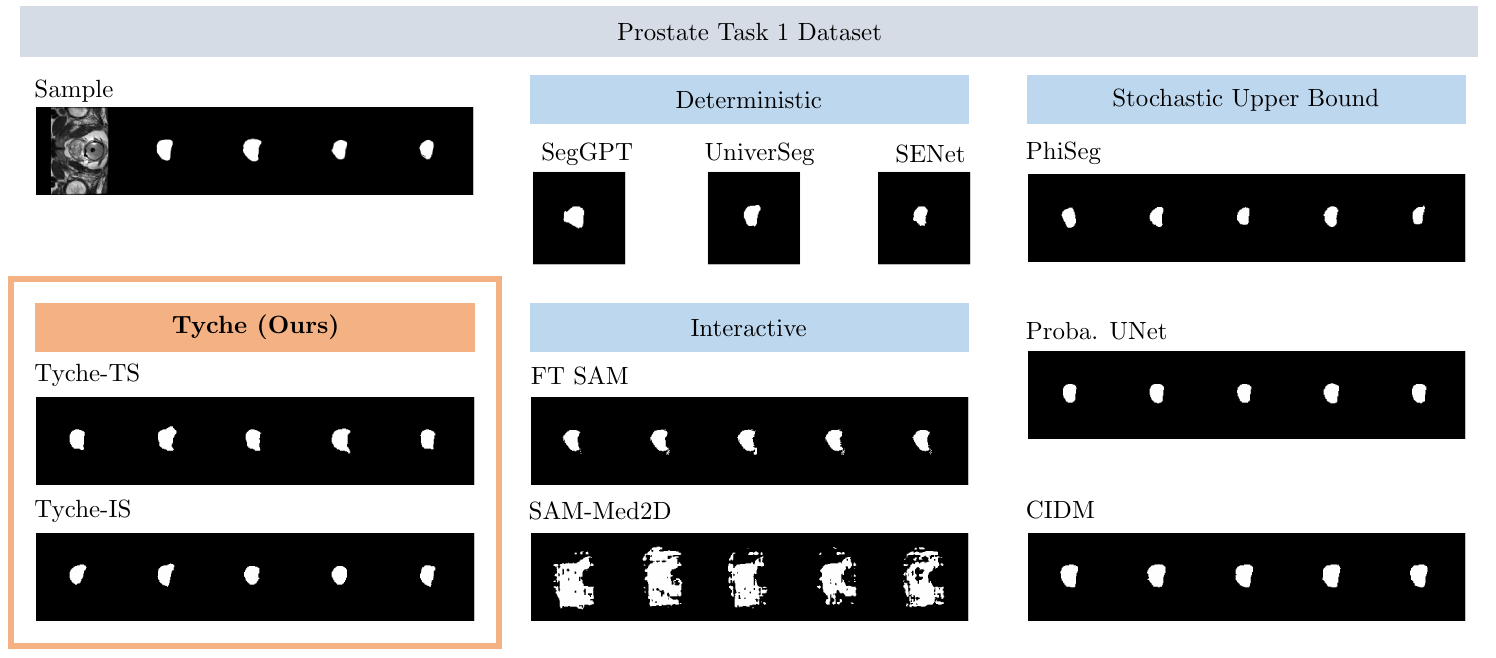}
    \caption{\textbf{Example Prediction for Prostate Task 1.}}
    \label{fig:sup:viz_Pr1}
\end{figure*}

\begin{figure*}
    \centering
    \includegraphics[width=\textwidth]{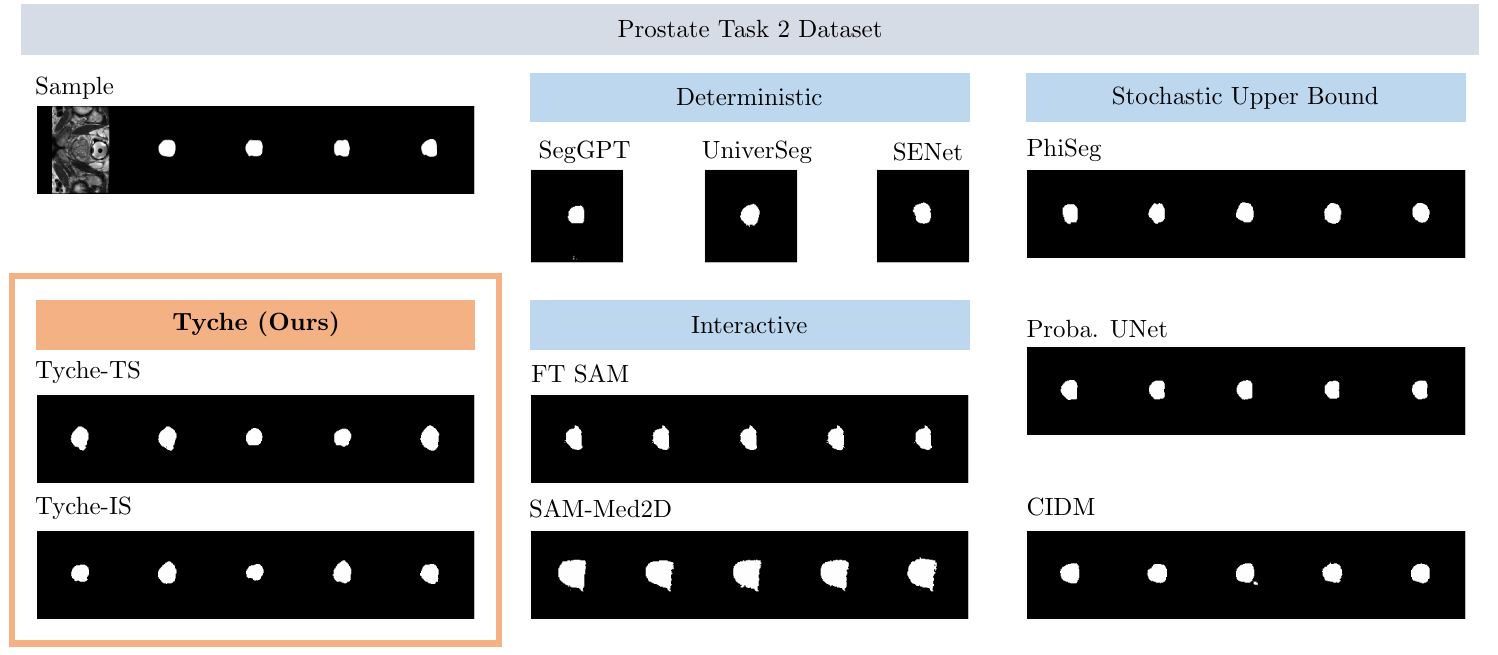}
    \caption{\textbf{Example Prediction for Prostate Task 2.}}
    \label{fig:sup:viz_Pr2}
\end{figure*}

\begin{figure*}
    \centering
    \includegraphics[width=\textwidth]{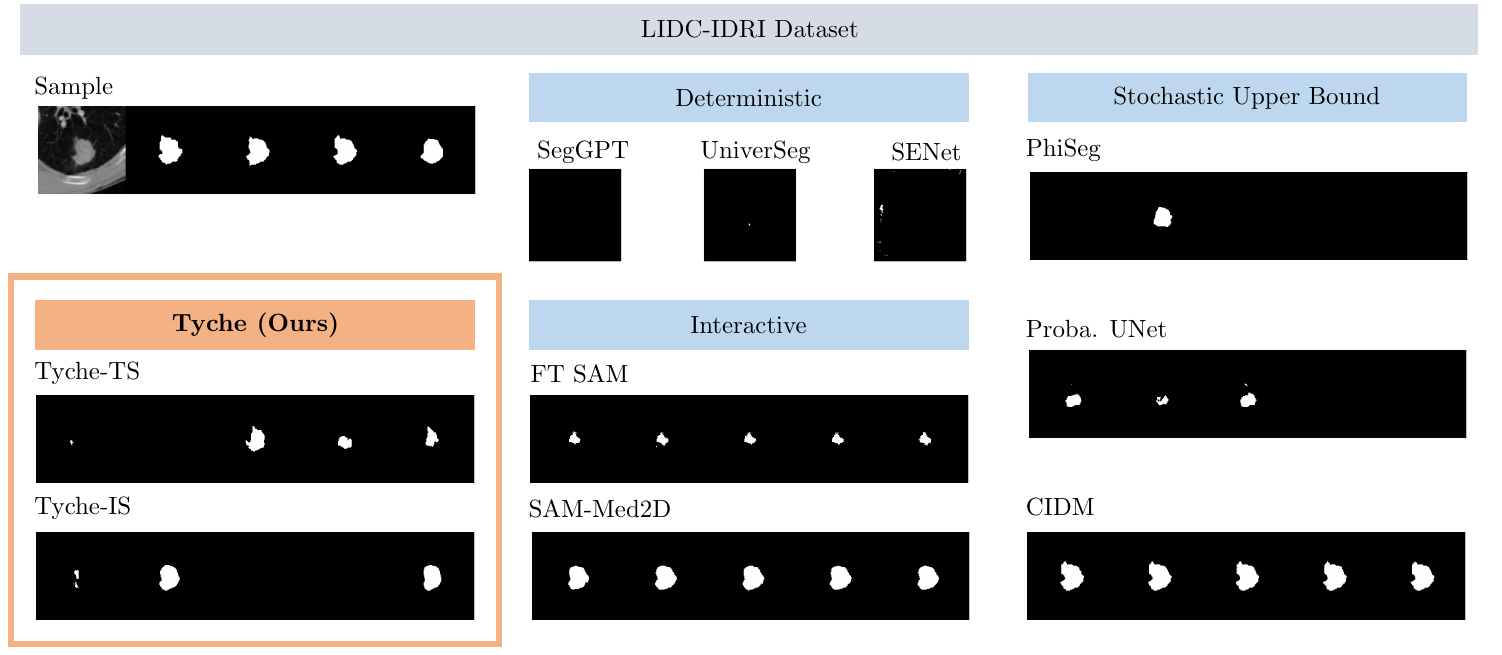}
    \caption{\textbf{Example Prediction for LIDC-IDRI.}}
    \label{fig:sup:viz_lidc}
\end{figure*}

\begin{figure*}
    \centering
    \includegraphics[width=\textwidth]{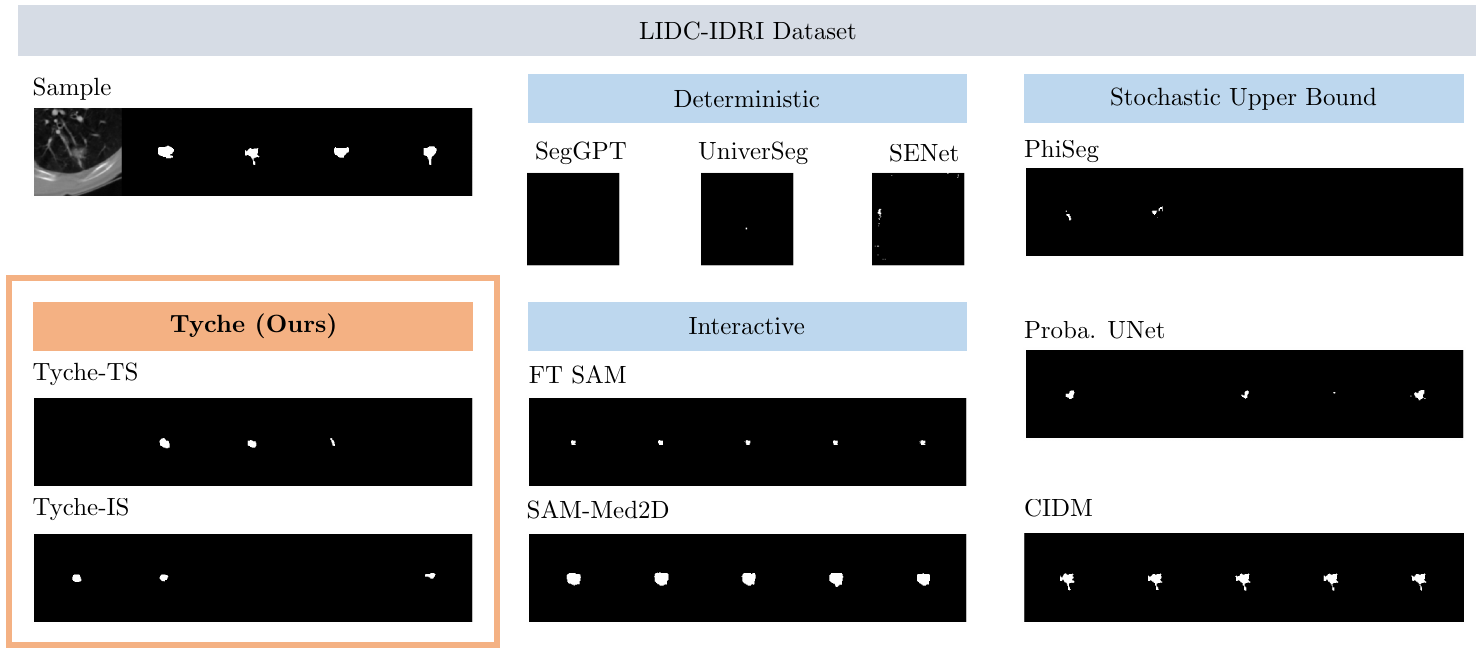}
    \caption{\textbf{Example Prediction for LIDC-IDRI.}}
    \label{fig:sup:viz_lidc2}
\end{figure*}

\newpage
\paragraph{Multi-Annotator Data}
We show an example prediction for the Hippocampus data Figure \ref{fig:sup:viz_Hippo} and one example prediction for STARE in Figure \ref{fig:sup:viz_STARE}.  We provide visualizations for each Prostate task Figures \ref{fig:sup:viz_Pr1} and \ref{fig:sup:viz_Pr2} respectively. Finally, we give two example visualizations for LIDC-IDRI Figures \ref{fig:sup:viz_lidc} and \ref{fig:sup:viz_lidc2}.

 \begin{table*}[t]
    
\rowcolors{2}{white}{gray!15}
\begin{tabular}{p{3cm}p{7.5cm}p{1.5cm}p{3cm}}

    \textbf{Dataset Name }   & \textbf{Description} & \textbf{\# of Scans} & \textbf{Image Modalities} \\ 
    \toprule
     ACDC~\cite{ACDC} & {Left  and right ventricular endocardium} & 99 & cine-MRI \\
     AMOS~\cite{AMOS} & Abdominal organ segmentation & 240 & CT, MRI \\
     BBBC003~\cite{BBBC003} & Mouse embryos & 15 & Microscopy 
     \\
     BBBC038~\cite{BBBC038} & Nuclei images & 670 & Microscopy 
     \\
     BUID~\cite{BUID} & Breast tumors & 647 & Ultrasound 
     \\
     BrainDev.~\cite{gousias2012magnetic, BrainDevFetal, BrainDevelopment, serag2012construction} & Adult and Neonatal Brain Atlases & 53 & multi-modal MRI \\
     BRATS~\cite{BRATS, bakas2017advancing, menze2014multimodal} & Brain tumors & 6,096 & multi-modal MRI \\
     BTCV~\cite{BTCV} & Abdominal Organs & 30 & CT  \\
     BUS~\cite{Bus} & Breast tumor & 163 & Ultrasound \\
     CAMUS~\cite{CAMUS}  & Four-chamber and Apical two-chamber heart & 500 & Ultrasound\\
     CDemris~\cite{cDemris}  & Human Left Atrial Wall & 60 & CMR \\
     CHAOS~\cite{Chaos_1, Chaos_2}  & Abdominal organs (liver, kidneys, spleen) & 40 & CT, T2-weighted MRI \\
     CheXplanation~\cite{CheXplanation} & Chest X-Ray observations & 170 & X-Ray
     \\
     CT-ORG\cite{CT_ORG} & Abdominal organ segmentation (overlap with LiTS) & 140 & CT 
     \\
     DRIVE~\cite{DRIVE} & Blood vessels in retinal images & 20 & Optical camera\\
     EOphtha~\cite{EOphtha} & Eye Microaneurysms and Diabetic Retinopathy & 102 & Optical camera\\
     FeTA~\cite{FeTA} & Fetal brain structures & 80 & Fetal MRI \\
     FetoPlac~\cite{FetoPlac} & Placenta vessel & 6 & Fetoscopic optical camera\\
     HMC-QU~\cite{HMC-QU, kiranyaz2020left} & 4-chamber (A4C) and apical 2-chamber (A2C) left  wall & 292 & Ultrasound \\
     HipXRay~\cite{HipXRay} & Ilium and femur & 140  &  X-Ray\\
     I2CVB~\cite{I2CVB} & Prostate (peripheral zone, central gland) & 19 & T2-weighted MRI \\
     IDRID~\cite{IDRID} & Diabetic Retinopathy & 54 & Optical camera \\
     ISLES~\cite{ISLES} & Ischemic stroke lesion & 180 & multi-modal MRI \\
     KiTS~\cite{KiTS} & Kidney and kidney tumor & 210 & CT \\
     LGGFlair~\cite{buda2019association, LGGFlair} & TCIA lower-grade glioma brain tumor & 110 & MRI \\
     LiTS~\cite{LiTS} & Liver Tumor & 131 & CT \\
     LUNA~\cite{LUNA} & Lungs & 888 & CT \\
     MCIC~\cite{MCIC} & Multi-site Brain regions of Schizophrenic patients & 390 & T1-weighted MRI\\
     MSD~\cite{MSD} & Collection of 10 Medical Segmentation Datasets & 3,225 & CT, multi-modal MRI\\
     NCI-ISBI~\cite{NCI-ISBI} & Prostate & 30 & T2-weighted MRI \\
     OASIS~\cite{OASIS-proc, OASIS-data} & Brain anatomy & 414 & T1-weighted MRI \\
     OCTA500~\cite{OCTA500} & Retinal vascular & 500 & OCT/OCTA \\
     PanDental~\cite{PanDental} & Mandible and Teeth & 215 & X-Ray \\
     PAXRay~\cite{PAXRay} & Thoracic organs & 880 & X-Ray \\
     PROMISE12~\cite{Promise12} & Prostate & 37 & T2-weighted MRI\\
     PPMI~\cite{PPMI} & Brain regions of Parkinson patients & 1,130 & T1-weighted MRI\\
     ROSE~\cite{Rose} & Retinal vessel & 117 & OCT/OCTA \\
     SCD~\cite{SCD} & Sunnybrook Cardiac Multi-Dataset Collection & 100 & cine-MRI \\
     SegTHOR~\cite{SegTHOR} & Thoracic organs (heart, trachea, esophagus) & 40 & CT \\
     SpineWeb~\cite{SpineWeb} & Vertebrae & 15 & T2-weighted MRI  \\
     ToothSeg~\cite{ToothSeg} & Individual teeth & 598 & X-Ray
     \\
     WBC~\cite{WBC} & White blood cell and nucleus & 400 & Microscopy \\
     WMH~\cite{WMH} & White matter hyper-intensities & 60 & multi-modal MRI \\
     WORD~\cite{Word} & Organ segmentation & 120 & CT \\
     \bottomrule
\end{tabular}
     \caption{\textbf{Collection of datasets in MegaMedical 2.0}. The entry number of scans is the number of unique (subject, modality) pairs for each dataset.}
     \label{tab:sup:megamedical2}
 \end{table*}

\begin{table*}[t]
    
\rowcolors{2}{white}{gray!15}
\begin{tabular}{p{3cm}p{7.5cm}p{1.5cm}p{3cm}}
    \textbf{Dataset Name }   & \textbf{Description} & \textbf{\# of Scans} & \textbf{Image Modalities} \\ 
    \toprule 
     LIDC-IDRI~\cite{armato2011lung} & Lung Nodules & 1018 & CT \\
     QUBIQ~\cite{qubiq} & Brain, kidney, pancreas and prostate & 209 & MRI T1, Multimodal MRI, CT\\
     STARE~\cite{STARE} & Blood vessels in retinal images & 20 & Optical camera \\
     \bottomrule
    \end{tabular}
     \caption{\textbf{Multi-Annotator Data}. The entry number of scans is the number of unique (subject, modality) pairs for each dataset.}
     \label{tab:sup:MAData}
 \end{table*}

\end{document}